%Paper: hep-ph/9304249
%From: bern@guinness.ias.edu (Zvi Bern)
%Date: Tue, 13 Apr 93 18:25:24 EDT

%%%%%%%%%%%%%%%%%%%%%%%%%%%%%%%%%%%%%%%%%%%%%%%%%%%%%%%%%%%%%%%%%%%%%%%%
% STRING-BASED PERTURBATIVE METHODS FOR GAUGE THEORIES by Zvi Bern.    %
% Lectures presented at TASI 1992, 66 pages, TeX twice for proper      %
% labeling, uuencoded compressed tar figures obtained with uufiles     %
% included at end.  Figures and instructions can be found after        %
% `THE UUENCODED FIGURES' after text.  All necessary macros included.  %
%%%%%%%%%%%%%%%%%%%%%%%%%%%%%%%%%%%%%%%%%%%%%%%%%%%%%%%%%%%%%%%%%%%%%%%%

\magnification=\magstep1
%%%%%%%%%%%%%%%%%%%%%%%%%%%%%%%%%%%%%%%%%%%%%%%%%%%%%%%%%%%%%%%%%
% various macros
%\input header

\newbox\SlashedBox
\def\slashed#1{\setbox\SlashedBox=\hbox{#1}
\hbox to 0pt{\hbox to 1\wd\SlashedBox{\hfil/\hfil}\hss}#1}
\def\hboxtosizeof#1#2{\setbox\SlashedBox=\hbox{#1}
\hbox to 1\wd\SlashedBox{#2}}

% The following is necessary so that we can get a partial slash
% inside a math display... sigh.
\def\mathslashed#1{\setbox\SlashedBox=\hbox{$#1$}
\hbox to 0pt{\hbox to 1\wd\SlashedBox{\hfil/\hfil}\hss}#1}

\def\ifsmall{\iffalse}  % default is unreduced.
\def\titlepagefont{}  % default is ordinary font.

% the ps: landscape must be the first special command in order
% to get the first page in landscape mode -- so we go through some
% contortions to define TeXgraphics in the default case.
\def\DefineTeXgraphics{%
\special{ps::[global] /TeXgraphics { } def}}  % No need to do anything

\def\today{\ifcase\month\or January\or February\or March\or April\or May
\or June\or July\or August\or September\or October\or November\or
December\fi\space\number\day, \number\year}
\def\eatPrefix19{}
\def\Year{\expandafter\eatPrefix\the\year}
\newcount\hours \newcount\minutes
\def\monthname{\ifcase\month\or
January\or February\or March\or April\or May\or June\or July\or
August\or September\or October\or November\or December\fi}
\def\shortmonthname{\ifcase\month\or
Jan\or Feb\or Mar\or Apr\or May\or Jun\or Jul\or
Aug\or Sep\or Oct\or Nov\or Dec\fi}

\def\TimeStamp{\hours\the\time\divide\hours by60%
\minutes -\the\time\divide\minutes by60\multiply\minutes by60%
\advance\minutes by\the\time%
${\rm \shortmonthname}\cdot\if\day<10{}0\fi\the\day\cdot\the\year%
\qquad\the\hours:\if\minutes<10{}0\fi\the\minutes$}

%\DefineTeXgraphics}

% restores pagenumbers

%\def\draft{\centerline{\it Preliminary Draft}\vskip 0.4in}

\newif\ifdraftmode
\newif\ifleftlabels  % Labels in left margins as well, for European-size paper

% Stolen from harvmac.tex 04/08/92
%       use \nolabels to get rid of eqn, ref, and fig labels in draft mode
\def\nolabels{\def\wrlabeL##1{}\def\eqlabeL##1{}\def\reflabeL##1{}}
\def\writelabels{\def\wrlabeL##1{\leavevmode\vadjust{\rlap{\smash%
{\line{{\escapechar=` \hfill\rlap{\sevenrm\hskip.03in\string##1}}}}}}}%
\def\eqlabeL##1{{\escapechar-1\rlap{\sevenrm\hskip.05in\string##1}}}%
\def\reflabeL##1{\noexpand\rlap{\noexpand\sevenrm[\string##1]}}}
\def\writeleftlabels{\def\wrlabeL##1{\leavevmode\vadjust{\rlap{\smash%
{\line{{\escapechar=` \hfill\rlap{\sevenrm\hskip.03in\string##1}}}}}}}%
\def\eqlabeL##1{{\escapechar-1%
\rlap{\sixrm\hskip.05in\string##1}%
\llap{\sevenrm\string##1\hskip.03in\hbox to \hsize{}}}}%
\def\reflabeL##1{\noexpand\rlap{\noexpand\sevenrm[\string##1]}}}
\nolabels

\newdimen\fullhsize
\newdimen\hstitle
\hstitle=\hsize % default
\newdimen\hsbody
\hsbody=\hsize % default
\newdimen\hbodyoffset
\hbodyoffset=\hoffset % default
\newbox\leftpage
\def\abstract#1{#1}
\def\rotated{\special{ps: landscape}
\magnification=1000  % This line must come before we change vsize,
                     % since \magnification sets it to a fixed value.
\baselineskip=14pt
\global\hstitle=9truein\global\hsbody=4.75truein
\global\vsize=7truein\global\voffset=-.31truein
\global\hoffset=-0.54in\global\hbodyoffset=-.54truein
\global\fullhsize=10truein
\def\DefineTeXgraphics{%
\special{ps::[global]
/TeXgraphics {currentpoint translate 0.7 0.7 scale
              -80 0.72 mul -1000 0.72 mul translate} def}}
 % 0.7 is slightly less than the ratio of horizontal sizes: 4.75 to 6.5
\let\lr=L
\def\ifsmall{\iftrue}
\def\titlepagefont{\twelvepoint}
\trueseventeenpoint
\def\almostshipout##1{\if L\lr \count1=1
      \global\setbox\leftpage=##1 \global\let\lr=R
   \else \count1=2
      \shipout\vbox{\hbox to\fullhsize{\box\leftpage\hfil##1}}
      \global\let\lr=L\fi}

\output={\ifnum\count0=1 %%% This is the HUTP version
 \shipout\vbox{\hbox to \fullhsize{\hfill\pagebody\hfill}}\advancepageno
 \else
 \almostshipout{\leftline{\vbox{\pagebody\makefootline}}}\advancepageno
 \fi}

\def\abstract##1{{\leftskip=1.5in\rightskip=1.5in ##1\par}} }

% Messages on lines by themselves
\def\linemessage#1{\immediate\write16{#1}}

% tagged sec numbers
\global\newcount\secno \global\secno=0
\global\newcount\appno \global\appno=0
\global\newcount\meqno \global\meqno=1
\global\newcount\subsecno \global\subsecno=0
% and figure numbers
\global\newcount\figno \global\figno=0

\newif\ifAnyCounterChanged
% If we are comparing numbers, there's no special problem.
% But if we are comparing roman numerals, we must be careful, because
% stuff read in from the aux file would be made up of ordinary
% characters (category code = 11), whereas \romannumeral generates
% characters with category code = 12..., so the stuff from the
% current run won't appear equal to the previous definition, as far
% as \warnIfChanged is concerned.
% To get around this, we have a macro \makeNormal, which converts
% letters `ivxlcdmIVXLCDM' to normal letters, no matter what their category
% code.  The macro has the convoluted form it does, with aftergroup's & all,
% to avoid blowing up TeX...
% The macro is used below in makeNormalizedRomappno, by which means we
% define the appendix counters to be strings containing vanilla versions
% of the letters... Sigh
\let\terminator=\relax
% The string to be normalized must not contain { and } tokens...
\def\normalize#1{\ifx#1\terminator\let\next=\relax\else%
\if#1i\aftergroup i\else\if#1v\aftergroup v\else\if#1x\aftergroup x%
\else\if#1l\aftergroup l\else\if#1c\aftergroup c\else%
\if#1m\aftergroup m\else%
\if#1I\aftergroup I\else\if#1V\aftergroup V\else\if#1X\aftergroup X%
\else\if#1L\aftergroup L\else\if#1C\aftergroup C\else%
\if#1M\aftergroup M\else\aftergroup#1\fi\fi\fi\fi\fi\fi\fi\fi\fi\fi\fi\fi%
\let\next=\normalize\fi%
\next}
% makes #1 a normalized version of #2...
\def\makeNormal#1#2{\def\doNormalDef{\edef#1}\begingroup%
\aftergroup\doNormalDef\aftergroup{\normalize#2\terminator\aftergroup}%
\endgroup}
% makes a normalized version of its argument:

\def\warnIfChanged#1#2{%
\ifundef#1% skip it
\else\begingroup%
\edef\oldDefinitionOfCounter{#1}\edef\newDefinitionOfCounter{#2}%
%\message{old: \oldDefinitionOfCounter}%
%\message{new: \newDefinitionOfCounter}%
\ifx\oldDefinitionOfCounter\newDefinitionOfCounter%
\else%
\linemessage{Warning: definition of \noexpand#1 has changed.}%
\global\AnyCounterChangedtrue\fi\endgroup\fi}

\def\Section#1{\global\advance\secno by1\relax\global\meqno=1%
\global\subsecno=0%
\bigbreak\bigskip% (combination \goodbreak\bigskip\bigskip)
\centerline{\twelvepoint \bf %
\the\secno. #1}%
\par\nobreak\medskip\nobreak}
\def\tagsection#1{%
\warnIfChanged#1{\the\secno}%
\xdef#1{\the\secno}%
\ifWritingAuxFile\immediate\write\auxfile{\noexpand\xdef\noexpand#1{#1}}\fi%
}
\def\section{\Section}
\def\Subsection#1{\global\advance\subsecno by1\relax\medskip %
\leftline{\bf\the\secno.\the\subsecno\ #1}%
\par\nobreak\smallskip\nobreak}
\def\tagsubsection#1{%
\warnIfChanged#1{\the\secno.\the\subsecno}%
\xdef#1{\the\secno.\the\subsecno}%
\ifWritingAuxFile\immediate\write\auxfile{\noexpand\xdef\noexpand#1{#1}}\fi%
}

\def\subsection{\Subsection}

\def\romappno{\uppercase\expandafter{\romannumeral\appno}}
\def\makeNormalizedRomappno{%
\expandafter\makeNormal\expandafter\normalizedromappno%
\expandafter{\romannumeral\appno}%
\edef\normalizedromappno{\uppercase{\normalizedromappno}}}
\def\Appendix#1{\global\advance\appno by1\relax\global\meqno=1\global\secno=0
\bigbreak\bigskip% (combination \goodbreak\bigskip\bigskip)
\centerline{\twelvepoint \bf Appendix %
\romappno. #1}%
\par\nobreak\medskip\nobreak}
\def\tagappendix#1{\makeNormalizedRomappno%
\warnIfChanged#1{\normalizedromappno}%
\xdef#1{\normalizedromappno}%
\ifWritingAuxFile\immediate\write\auxfile{\noexpand\xdef\noexpand#1{#1}}\fi%
}
\def\appendix{\Appendix}

\def\eqn#1{\makeNormalizedRomappno%
\ifnum\secno>0%
  \warnIfChanged#1{\the\secno.\the\meqno}%
  \eqno(\the\secno.\the\meqno)\xdef#1{\the\secno.\the\meqno}%
     \global\advance\meqno by1
\else\ifnum\appno>0%
  \warnIfChanged#1{\normalizedromappno.\the\meqno}%
  \eqno({\rm\romappno}.\the\meqno)%
      \xdef#1{\normalizedromappno.\the\meqno}%
     \global\advance\meqno by1
\else%
  \warnIfChanged#1{\the\meqno}%
  \eqno(\the\meqno)\xdef#1{\the\meqno}%
     \global\advance\meqno by1
\fi\fi%
\eqlabeL#1%
\ifWritingAuxFile\immediate\write\auxfile{\noexpand\xdef\noexpand#1{#1}}\fi%
}
\def\defeqn#1{\makeNormalizedRomappno%
\ifnum\secno>0%
  \warnIfChanged#1{\the\secno.\the\meqno}%
  \xdef#1{\the\secno.\the\meqno}%
     \global\advance\meqno by1
\else\ifnum\appno>0%
  \warnIfChanged#1{\normalizedromappno.\the\meqno}%
  \xdef#1{\normalizedromappno.\the\meqno}%
     \global\advance\meqno by1
\else%
  \warnIfChanged#1{\the\meqno}%
  \xdef#1{\the\meqno}%
     \global\advance\meqno by1
\fi\fi%
\eqlabeL#1%
\ifWritingAuxFile\immediate\write\auxfile{\noexpand\xdef\noexpand#1{#1}}\fi%
}
\def\anoneqn{\makeNormalizedRomappno%
\ifnum\secno>0
  \eqno(\the\secno.\the\meqno)%
     \global\advance\meqno by1
\else\ifnum\appno>0
  \eqno({\rm\normalizedromappno}.\the\meqno)%
     \global\advance\meqno by1
\else
  \eqno(\the\meqno)%
     \global\advance\meqno by1
\fi\fi%
}
\def\mfig#1#2{\global\advance\figno by1%
\relax#1\the\figno%
\warnIfChanged#2{\the\figno}%
\edef#2{\the\figno}%
\reflabeL#2%
\ifWritingAuxFile\immediate\write\auxfile{\noexpand\xdef\noexpand#2{#2}}\fi%
}

\def\fig#1{\mfig{fig.~}#1}

\catcode`@=11 % borrow the private macros of PLAIN (with care)

%\font\titlefont=cmr10 at 16pt
\font\ninerm=cmr9
\font\eightrm=cmr8
\font\sixrm=cmr6

\def\loadtrueseventeenpoint{
 \font\seventeenrm=cmr10 at 17.28truept
 \font\seventeeni=cmmi10 at 17.28truept
 \font\seventeenbf=cmbx10 at 17.28truept
 \font\seventeenit=cmti10 at 17.28truept
 \font\seventeensl=cmsl10 at 17.28truept
 \font\seventeensy=cmsy10 at 17.28truept
}
\def\loadfourteenpoint{
\font\fourteenrm=cmr10 at 14.4pt
\font\fourteeni=cmmi10 at 14.4pt
\font\fourteenit=cmti10 at 14.4pt
\font\fourteensl=cmsl10 at 14.4pt
\font\fourteensy=cmsy10 at 14.4pt
\font\fourteenbf=cmbx10 at 14.4pt
}
\def\loadtruetwelvepoint{
\font\twelverm=cmr10 at 12truept
\font\twelvei=cmmi10 at 12truept
\font\twelveit=cmti10 at 12truept
\font\twelvesl=cmsl10 at 12truept
\font\twelvesy=cmsy10 at 12truept
\font\twelvebf=cmbx10 at 12truept
}

\font\ninei=cmmi9
\font\eighti=cmmi8
\font\sixi=cmmi6
\skewchar\ninei='177 \skewchar\eighti='177 \skewchar\sixi='177

\font\ninesy=cmsy9
\font\eightsy=cmsy8
\font\sixsy=cmsy6
\skewchar\ninesy='60 \skewchar\eightsy='60 \skewchar\sixsy='60

\font\ninebf=cmbx9
\font\eightbf=cmbx8
\font\sixbf=cmbx6

\font\ninett=cmtt9
\font\eighttt=cmtt8

\hyphenchar\tentt=-1 % inhibit hyphenation in typewriter type
\hyphenchar\ninett=-1
\hyphenchar\eighttt=-1

\font\ninesl=cmsl9
\font\eightsl=cmsl8

\font\nineit=cmti9
\font\eightit=cmti8

 % unslanted text italic

\newskip\ttglue
\def\tenpoint{\def\rm{\fam0\tenrm}%
  \textfont0=\tenrm \scriptfont0=\sevenrm \scriptscriptfont0=\fiverm
  \textfont1=\teni \scriptfont1=\seveni \scriptscriptfont1=\fivei
  \textfont2=\tensy \scriptfont2=\sevensy \scriptscriptfont2=\fivesy
  \textfont3=\tenex \scriptfont3=\tenex \scriptscriptfont3=\tenex
  \def\it{\fam\itfam\tenit}\textfont\itfam=\tenit
  \def\sl{\fam\slfam\tensl}\textfont\slfam=\tensl
  \def\bf{\fam\bffam\tenbf}\textfont\bffam=\tenbf \scriptfont\bffam=\sevenbf
  \scriptscriptfont\bffam=\fivebf
  \normalbaselineskip=12pt
  \let\sc=\eightrm
  \let\big=\tenbig
  \setbox\strutbox=\hbox{\vrule height8.5pt depth3.5pt width\z@}%
  \normalbaselines\rm}

\def\twelvepoint{\def\rm{\fam0\twelverm}%
  \textfont0=\twelverm \scriptfont0=\ninerm \scriptscriptfont0=\sevenrm
  \textfont1=\twelvei \scriptfont1=\ninei \scriptscriptfont1=\seveni
  \textfont2=\twelvesy \scriptfont2=\ninesy \scriptscriptfont2=\sevensy
  \textfont3=\tenex \scriptfont3=\tenex \scriptscriptfont3=\tenex
  \def\it{\fam\itfam\twelveit}\textfont\itfam=\twelveit
  \def\sl{\fam\slfam\twelvesl}\textfont\slfam=\twelvesl
  \def\bf{\fam\bffam\twelvebf}\textfont\bffam=\twelvebf
  \scriptfont\bffam=\ninebf
  \scriptscriptfont\bffam=\sevenbf
  \normalbaselineskip=12pt
  \let\sc=\eightrm
  \let\big=\tenbig
  \setbox\strutbox=\hbox{\vrule height8.5pt depth3.5pt width\z@}%
  \normalbaselines\rm}

\def\fourteenpoint{\def\rm{\fam0\fourteenrm}%
  \textfont0=\fourteenrm \scriptfont0=\tenrm \scriptscriptfont0=\sevenrm
  \textfont1=\fourteeni \scriptfont1=\teni \scriptscriptfont1=\seveni
  \textfont2=\fourteensy \scriptfont2=\tensy \scriptscriptfont2=\sevensy
  \textfont3=\tenex \scriptfont3=\tenex \scriptscriptfont3=\tenex
  \def\it{\fam\itfam\fourteenit}\textfont\itfam=\fourteenit
  \def\sl{\fam\slfam\fourteensl}\textfont\slfam=\fourteensl
  \def\bf{\fam\bffam\fourteenbf}\textfont\bffam=\fourteenbf%
  \scriptfont\bffam=\tenbf
  \scriptscriptfont\bffam=\sevenbf
  \normalbaselineskip=17pt
  \let\sc=\elevenrm
  \let\big=\tenbig
  \setbox\strutbox=\hbox{\vrule height8.5pt depth3.5pt width\z@}%
  \normalbaselines\rm}

\def\seventeenpoint{\def\rm{\fam0\seventeenrm}%
  \textfont0=\seventeenrm \scriptfont0=\fourteenrm \scriptscriptfont0=\tenrm
  \textfont1=\seventeeni \scriptfont1=\fourteeni \scriptscriptfont1=\teni
  \textfont2=\seventeensy \scriptfont2=\fourteensy \scriptscriptfont2=\tensy
  \textfont3=\tenex \scriptfont3=\tenex \scriptscriptfont3=\tenex
  \def\it{\fam\itfam\seventeenit}\textfont\itfam=\seventeenit
  \def\sl{\fam\slfam\seventeensl}\textfont\slfam=\seventeensl
  \def\bf{\fam\bffam\seventeenbf}\textfont\bffam=\seventeenbf%
  \scriptfont\bffam=\fourteenbf
  \scriptscriptfont\bffam=\twelvebf
  \normalbaselineskip=21pt
  \let\sc=\fourteenrm
  \let\big=\tenbig
  \setbox\strutbox=\hbox{\vrule height 12pt depth 6pt width\z@}%
  \normalbaselines\rm}

\def\ninepoint{\def\rm{\fam0\ninerm}%
  \textfont0=\ninerm \scriptfont0=\sixrm \scriptscriptfont0=\fiverm
  \textfont1=\ninei \scriptfont1=\sixi \scriptscriptfont1=\fivei
  \textfont2=\ninesy \scriptfont2=\sixsy \scriptscriptfont2=\fivesy
  \textfont3=\tenex \scriptfont3=\tenex \scriptscriptfont3=\tenex
  \def\it{\fam\itfam\nineit}\textfont\itfam=\nineit
  \def\sl{\fam\slfam\ninesl}\textfont\slfam=\ninesl
  \def\bf{\fam\bffam\ninebf}\textfont\bffam=\ninebf \scriptfont\bffam=\sixbf
  \scriptscriptfont\bffam=\fivebf
  \normalbaselineskip=11pt
  \let\sc=\sevenrm
  \let\big=\ninebig
  \setbox\strutbox=\hbox{\vrule height8pt depth3pt width\z@}%
  \normalbaselines\rm}

\def\eightpoint{\def\rm{\fam0\eightrm}%
  \textfont0=\eightrm \scriptfont0=\sixrm \scriptscriptfont0=\fiverm%
  \textfont1=\eighti \scriptfont1=\sixi \scriptscriptfont1=\fivei%
  \textfont2=\eightsy \scriptfont2=\sixsy \scriptscriptfont2=\fivesy%
  \textfont3=\tenex \scriptfont3=\tenex \scriptscriptfont3=\tenex%
  \def\it{\fam\itfam\eightit}\textfont\itfam=\eightit%
  \def\sl{\fam\slfam\eightsl}\textfont\slfam=\eightsl%
  \def\bf{\fam\bffam\eightbf}\textfont\bffam=\eightbf \scriptfont\bffam=\sixbf%
  \scriptscriptfont\bffam=\fivebf%
  \normalbaselineskip=9pt%
  \let\sc=\sixrm%
  \let\big=\eightbig%
  \setbox\strutbox=\hbox{\vrule height7pt depth2pt width\z@}%
  \normalbaselines\rm}

 % use after $ in ninepoint sections
\def\tenbig#1{{\hbox{$\left#1\vbox to8.5pt{}\right.\n@space$}}}
\def\ninebig#1{{\hbox{$\textfont0=\tenrm\textfont2=\tensy
  \left#1\vbox to7.25pt{}\right.\n@space$}}}
\def\eightbig#1{{\hbox{$\textfont0=\ninerm\textfont2=\ninesy
  \left#1\vbox to6.5pt{}\right.\n@space$}}}

% Page layout
%\newinsert\footins
\def\footnote#1{\edef\@sf{\spacefactor\the\spacefactor}#1\@sf
      \insert\footins\bgroup\eightpoint
      \interlinepenalty100 \let\par=\endgraf
        \leftskip=\z@skip \rightskip=\z@skip
        \splittopskip=10pt plus 1pt minus 1pt \floatingpenalty=20000
        \smallskip\item{#1}\bgroup\strut\aftergroup\@foot\let\next}
\skip\footins=12pt plus 2pt minus 4pt % space added when footnote is present
%\count\footins=1000 % footnote magnification factor (1 to 1)
\dimen\footins=30pc % maximum footnotes per page

\newinsert\margin
\dimen\margin=\maxdimen
%\count\margin=0 \skip\margin=0pt % marginal inserts take up no space

\loadtruetwelvepoint % At FNAL...
\loadtrueseventeenpoint
\catcode`\@=\active
\catcode`@=12  % No longer.
\catcode`\"=\active

% \use\cs
% puts in the expansion of `\cs' if it's defined, the literal "\cs" otherwise.
\def\eatOne#1{}
\def\ifundef#1{\expandafter\ifx%
\csname\expandafter\eatOne\string#1\endcsname\relax}
\def\notTrue{\iffalse}\def\isTrue{\iftrue}
\def\ifdef#1{{\ifundef#1%
\aftergroup\notTrue\else\aftergroup\isTrue\fi}}
\def\use#1{\ifundef#1\linemessage{Warning: \string#1 is undefined.}%
{\tt \string#1}\else#1\fi}

%     \ref\label{text}
% generates a number, assigns it to \label, generates an entry.
% To list the refs on a separate page,  \listrefs
% \nref does the same without generating any text at the reference
% point

\global\newcount\refno \global\refno=1
\newwrite\rfile
\newlinechar=`\^^J
\def\ref#1#2{\the\refno\nref#1{#2}}
\def\nref#1#2{\xdef#1{\the\refno}%
\ifnum\refno=1\immediate\openout\rfile=refs.tmp\fi%
\immediate\write\rfile{\noexpand\item{[\noexpand#1]\ }#2.}%
\global\advance\refno by1}
% To start a long reference...
\def\lref#1#2{\the\refno\xdef#1{\the\refno}%
\ifnum\refno=1\immediate\openout\rfile=refs.tmp\fi%
\immediate\write\rfile{\noexpand\item{[\noexpand#1]\ }#2\semi}%
\global\advance\refno by1}
% To continue a long reference...
\def\cref#1{\immediate\write\rfile{#1\semi}}
% To end a long reference...

\def\semi{;\hfil\noexpand\break}

\def\inputAuxIfPresent#1{\immediate\openin1=#1
\ifeof1\message{No file \auxfileName; I'll create one.
}\else\closein1\relax\input\auxfileName\fi%
}
% For references, some journal names

% An .aux file --- for forward references...
\newif\ifWritingAuxFile
\newwrite\auxfile
\def\SetUpAuxFile{%
\xdef\auxfileName{\jobname.aux}%
% Read it in if it exists
\inputAuxIfPresent{\auxfileName}%
% Now write a new one.
\WritingAuxFiletrue%
\immediate\openout\auxfile=\auxfileName}

% Some generally useful notation
\def\L{\left(}\def\R{\right)}
\def\LP{\left.}\def\RP{\right.}
\def\LB{\left[}\def\RB{\right]}

\def\LV{\left|}\def\RV{\right|}

% Warn about changed counters...
\def\bye{\par\vfill\supereject%
\ifAnyCounterChanged\linemessage{
Some counters have changed.  Re-run tex to fix them up.}\fi%
\end}

%%%%%%%%%%%%%%%%%%%%%%%%%%%%%%%%%%%%%%%%%%%%%%%%%%%%%%%%%%%%
\SetUpAuxFile

\def\L{\left(}
\def\R{\right)}

\def\c{\mskip 1mu\cdot\mskip 1mu }
\def\Tr{\mathop{\rm Tr}\nolimits}
\def\Gr{\mathop{\rm Gr}\nolimits}
\def\sign{\mathop{\rm sign}\nolimits}
\def\Re{\mathop{\rm Re}\nolimits}

\def\nub{{\overline \nu}}
\def\si{\sigma}

\def\eps{\epsilon}

\def\LP{\left.}\def\RP{\right.}

\def\pol{\varepsilon}

\def\bdry#1#2{{\left[\vphantom{{0\atop0}}\smash{{#1\atop#2}}\right]}}

\def\dl^#1_#2{\delta^{#1}{}_{#2}}

\def\Gf#1#2{G_F\!\bdry{#1}{#2}\!}

\def\Gbd{\dot G_B}
\def\Gbdd{\ddot G_B}

\def\Ord{{\cal O}}

\def\A#1{{\cal A}_{#1}}

\catcode`@=11  % Make @ letter.
\def\meqalign#1{\,\vcenter{\openup1\jot\m@th
   \ialign{\strut\hfil$\displaystyle{##}$ && $\displaystyle{{}##}$\hfil
             \crcr#1\crcr}}\,}
\catcode`@=12  % No longer.

\def\n#1#2{\nu_{#1#2}}
\def\nb#1#2{\nub_{#1#2}}

% counters

% General parameters
\baselineskip 15pt
\overfullrule 0.5pt

\def\si{\sigma}

\def\Tr{\mathop{\rm Tr}\nolimits}

\def\A#1{{\cal A}_{#1}}

\def\pol{\varepsilon}

\def\c{\,\cdot\,}

\def\Re{\mathop{\rm Re}}
\def\L{\left(}\def\R{\right)}
\def\LP{\left.}\def\RP{\right.}
\def\spa#1.#2{\left\langle#1\,#2\right\rangle}
\def\spb#1.#2{\left[#1\,#2\right]}
\def\lor#1.#2{\left(#1\,#2\right)}
\def\sand#1.#2.#3{%
\left\langle\smash{#1}{\vphantom1}^{-}\right|{#2}%
\left|\smash{#3}{\vphantom1}^{-}\right\rangle}
\def\sandp#1.#2.#3{%
\left\langle\smash{#1}{\vphantom1}^{-}\right|{#2}%
\left|\smash{#3}{\vphantom1}^{+}\right\rangle}
\def\sandpp#1.#2.#3{%
\left\langle\smash{#1}{\vphantom1}^{+}\right|{#2}%
\left|\smash{#3}{\vphantom1}^{+}\right\rangle}
\catcode`@=11  % Make @ letter.
\def\meqalign#1{\,\vcenter{\openup1\jot\m@th
   \ialign{\strut\hfil$\displaystyle{##}$ && $\displaystyle{{}##}$\hfil
             \crcr#1\crcr}}\,}
\catcode`@=12  % No longer.

%%%%%%%%%%%%%%%%%%%%%%%%%%%%%%%%%%%%%%%%%%%%%%%%%%%%%%%%%%%%%%%%%%%
%
\newread\epsffilein    % file to \read
\newif\ifepsffileok    % continue looking for the bounding box?
\newif\ifepsfbbfound   % success?
\newif\ifepsfverbose   % report what you're making?
\newdimen\epsfxsize    % horizontal size after scaling
\newdimen\epsfysize    % vertical size after scaling
\newdimen\epsftsize    % horizontal size before scaling
\newdimen\epsfrsize    % vertical size before scaling
\newdimen\epsftmp      % register for arithmetic manipulation
\newdimen\pspoints     % conversion factor
\pspoints=1bp          % Adobe points are `big'
\epsfxsize=0pt         % Default value, means `use natural size'
\epsfysize=0pt         % ditto
\def\epsfbox#1{\global\def\epsfllx{72}\global\def\epsflly{72}%
   \global\def\epsfurx{540}\global\def\epsfury{720}%
   \def\lbracket{[}\def\testit{#1}\ifx\testit\lbracket
   \let\next=\epsfgetlitbb\else\let\next=\epsfnormal\fi\next{#1}}%
\def\epsfgetlitbb#1#2 #3 #4 #5]#6{\epsfgrab #2 #3 #4 #5 .\\%
   \epsfsetgraph{#6}}%
\def\epsfnormal#1{\epsfgetbb{#1}\epsfsetgraph{#1}}%
\def\epsfgetbb#1{%
%
%   The first thing we need to do is to open the
%   PostScript file, if possible.
%
\openin\epsffilein=#1
\ifeof\epsffilein\errmessage{I couldn't open #1, will ignore it}\else
%
%   Okay, we got it. Now we'll scan lines until we find one that doesn't
%   start with %. We're looking for the bounding box comment.
%
   {\epsffileoktrue \chardef\other=12
    \def\do##1{\catcode`##1=\other}\dospecials \catcode`\ =10
    \loop
       \read\epsffilein to \epsffileline
       \ifeof\epsffilein\epsffileokfalse\else
%
%   We check to see if the first character is a % sign;
%   if not, we stop reading (unless the line was entirely blank);
%   if so, we look further and stop only if the line begins with
%   `%%BoundingBox:'.
%
          \expandafter\epsfaux\epsffileline:. \\%
       \fi
   \ifepsffileok\repeat
   \ifepsfbbfound\else
    \ifepsfverbose\message{No bounding box comment in #1; using defaults}\fi\fi
   }\closein\epsffilein\fi}%
%
%   Now we have to calculate the scale and offset values to use.
%   First we compute the natural sizes.
%
\def\epsfclipstring{}% do we clip or not?  If so,
\def\epsfsetgraph#1{%
   \epsfrsize=\epsfury\pspoints
   \advance\epsfrsize by-\epsflly\pspoints
   \epsftsize=\epsfurx\pspoints
   \advance\epsftsize by-\epsfllx\pspoints
%
%   If `epsfxsize' is 0, we default to the natural size of the picture.
%   Otherwise we scale the graph to be \epsfxsize wide.
%
   \epsfxsize\epsfsize\epsftsize\epsfrsize
   \ifnum\epsfxsize=0 \ifnum\epsfysize=0
      \epsfxsize=\epsftsize \epsfysize=\epsfrsize
      \epsfrsize=0pt
%
%   We have a sticky problem here:  TeX doesn't do floating point arithmetic!
%   Our goal is to compute y = rx/t. The following loop does this reasonably
%   fast, with an error of at most about 16 sp (about 1/4000 pt).
%
     \else\epsftmp=\epsftsize \divide\epsftmp\epsfrsize
       \epsfxsize=\epsfysize \multiply\epsfxsize\epsftmp
       \multiply\epsftmp\epsfrsize \advance\epsftsize-\epsftmp
       \epsftmp=\epsfysize
       \loop \advance\epsftsize\epsftsize \divide\epsftmp 2
       \ifnum\epsftmp>0
          \ifnum\epsftsize<\epsfrsize\else
             \advance\epsftsize-\epsfrsize \advance\epsfxsize\epsftmp \fi
       \repeat
       \epsfrsize=0pt
     \fi
   \else \ifnum\epsfysize=0
     \epsftmp=\epsfrsize \divide\epsftmp\epsftsize
     \epsfysize=\epsfxsize \multiply\epsfysize\epsftmp
     \multiply\epsftmp\epsftsize \advance\epsfrsize-\epsftmp
     \epsftmp=\epsfxsize
     \loop \advance\epsfrsize\epsfrsize \divide\epsftmp 2
     \ifnum\epsftmp>0
        \ifnum\epsfrsize<\epsftsize\else
           \advance\epsfrsize-\epsftsize \advance\epsfysize\epsftmp \fi
     \repeat
     \epsfrsize=0pt
    \else
     \epsfrsize=\epsfysize
    \fi
   \fi
%
%  Finally, we make the vbox and stick in a \special that dvips can parse.
%
   \ifepsfverbose\message{#1: width=\the\epsfxsize, height=\the\epsfysize}\fi
   \epsftmp=10\epsfxsize \divide\epsftmp\pspoints
   \vbox to\epsfysize{\vfil\hbox to\epsfxsize{%
      \ifnum\epsfrsize=0\relax
        \includegraphics{#1}%
      \else
        \epsfrsize=10\epsfysize \divide\epsfrsize\pspoints
        \includegraphics{#1}%
      \fi
      \hfil}}%
\global\epsfxsize=0pt\global\epsfysize=0pt}%
%
%   We still need to define the tricky \epsfaux macro. This requires
%   a couple of magic constants for comparison purposes.
%
{\catcode`\%=12 \global\let\epsfpercent=%\global\def\epsfbblit{%BoundingBox}}%
%
%   So we're ready to check for `%BoundingBox:' and to grab the
%   values if they are found.
%
\long\def\epsfaux#1#2:#3\\{\ifx#1\epsfpercent
   \def\testit{#2}\ifx\testit\epsfbblit
      \epsfgrab #3 . . . \\%
      \epsffileokfalse
      \global\epsfbbfoundtrue
   \fi\else\ifx#1\par\else\epsffileokfalse\fi\fi}%
%
%   Here we grab the values and stuff them in the appropriate definitions.
%
\def\epsfempty{}%
\def\epsfgrab #1 #2 #3 #4 #5\\{%
\global\def\epsfllx{#1}\ifx\epsfllx\epsfempty
      \epsfgrab #2 #3 #4 #5 .\\\else
   \global\def\epsflly{#2}%
   \global\def\epsfurx{#3}\global\def\epsfury{#4}\fi}%
%
%   We default the epsfsize macro.
%
\def\epsfsize#1#2{\epsfxsize}
%
%   Finally, another definition for compatibility with older macros.
%

%%%%%%%%%%%%%%%%%%%%%%%%%%%%%%%%%%%%%%%%%%%%%%%%%%%%%%%%%%%%%%%%%%

%\epsfverbosetrue

\def\ref#1#2{\nref#1{#2}}
\overfullrule 0pt
\hfuzz 40pt
\hsize 6. truein
\vsize 8.5 truein
%\nopagenumbers

\loadfourteenpoint
\newcount\eqncount
\newcount\sectcount
\eqncount=0
\sectcount=0
\def\secta{\global\advance\sectcount by1
\eqncount=0}

\def\equn{
\global\advance\eqncount by1
\eqno{(\the\sectcount.\the\eqncount)}        }
\def\put#1{\global\edef#1{(\the\sectcount.\the\eqncount)}     }

\def\section#1{\global\advance\secno by1\relax\global\meqno=1%
\global\subsecno=0%
\bigbreak\bigskip% (combination \goodbreak\bigskip\bigskip)
\noindent{\twelvepoint \bf %
\the\secno. #1}%
\par\nobreak\medskip\nobreak}

\def\eps{\epsilon}
\def\pol{\varepsilon}
\def\epsh{{[\eps]}}
\def\delmeps{\delta_{(-\eps)}}

\def\c{\,\cdot\,}

\def\Re{\mathop{\rm Re}}

\def\eps{\epsilon}
\def\as#1{a_{\sigma(#1)}}
\def\ks#1{k_{\sigma(#1)}}
\def\ps#1{\pol_{\sigma(#1)}}

\def\N{{(4\pi)^{\eps/2} \over (16\pi^2)} }
\def\x#1#2{x_{#1 #2}}
\def\half{{\textstyle{1\over 2}}}
\def\Gammait{{\mit \Gamma}}
\def\Lambdait{{\mit \Lambda}}
\def\HV{HV}
\def\CDR{CDR}
\def\FDH{FDH}
\chardef\hyphen=45
\def\floorHalfN{\left\lfloor n/2\right\rfloor}

\def\helicitiesA#1#2#3#4{1^{#1{}},2^{#2},3^{#3},4^{#4}}
\def\soq{(s)}
\def\toq{(t)}
\def\moq{(\mu^2)}

\def\ThetaHat{{\widehat\Theta}}
\def\lQ{l_Q}

%%%%%%%%%%%%%%%%%%%%%%%%%%%%%%%%%%%%%%%%%%%%%%%%%%%%%%%%%%%%%%%%%%%%%%%%%

\ref\Short{Z. Bern and D.A.\ Kosower, Phys.\ Rev.\ Lett.\ 66:1669 (1991)}

\ref\Long{Z. Bern and D.A.\ Kosower, 379:451 (1992)}

\ref\Pascos{Z. Bern and D.A.\ Kosower, in {\it Proceedings of the PASCOS-91
Symposium}, eds.\ P. Nath and S. Reucroft}

\ref\AmplLet{Z. Bern, L. Dixon and D.A. Kosower, CERN-TH.6803/93}

\ref\Future{Z. Bern, L. Dixon and D.A. Kosower, in preparation}

\ref\Ellis{R.K.\ Ellis and J.C.\ Sexton, Nucl.\ Phys.\ B269:445 (1986)}

\ref\SpinorHelicity{%
F.\ A.\ Berends, R.\ Kleiss, P.\ De Causmaecker, R.\ Gastmans and T.\ T.\ Wu,
        Phys.\ Lett.\ 103B:124 (1981)\semi
P.\ De Causmaeker, R.\ Gastmans,  W.\ Troost and  T.\ T.\ Wu,
Nucl. Phys. B206:53 (1982)\semi
R.\ Kleiss and W.\ J.\ Stirling,
   Nucl.\ Phys.\ B262:235 (1985)\semi
   J.\ F.\ Gunion and Z.\ Kunszt, Phys.\ Lett.\ 161B:333 (1985)\semi
 R.\ Gastmans and T.T.\ Wu,
{\it The Ubiquitous Photon: Helicity Method for QED and QCD} (Clarendon Press)
(1990)}

\ref\XZC{Z.\ Xu, D.-H.\ Zhang and L. Chang, Nucl.\ Phys.\ B291:392 (1987)}

\ref\TreeColor{F.A.\ Berends and W.T.\ Giele,
Nucl.\ Phys.\ B294:700 (1987)\semi
M.\ Mangano and S.J.\ Parke, Nucl.\ Phys.
B299:673 (1988)\semi
M.\ Mangano, Nucl.\ Phys.\ B309:461 (1988)}

\ref\MPX{%
M.\ Mangano, S. Parke and Z.\ Xu, Nucl.\ Phys.\ B298:653 (1988)}

\ref\Susy{S.J. Parke and T. Taylor, Phys. Lett. B157:81 (1985)}

\ref\Recursive{F.A. Berends and W.T. Giele, Nucl.\ Phys.\ B306:759 (1988)}

\ref\ManganoReview{M. Mangano and S.J. Parke, Phys.\ Rep.\ 200:301 (1991)}

\ref\KLN{D.A. Kosower, B.-H.\ Lee and V.P. Nair, Phys.\ Lett.\
201B:85 (1988)}

\ref\LightconeRecurrence{D.A. Kosower, Nucl.\ Phys.\ B335:23(1990)}

\ref\Mapping{Z. Bern and D.C. Dunbar,  Nucl.\ Phys.\ B379:562 (1992)}

\ref\Scherk{J. Scherk, Nucl.\ Phys.\ {B31} (1971) 222\semi
A. Neveu and J. Scherk, Nucl.\ Phys.\ {B36} (1972) 155}

\ref\GSB{M.\ B.\ Green, J.\ H.\ Schwarz and L.\ Brink, Nucl.\ Phys.\
B198:472 (1982)}

\ref\Minahan{J.\ Minahan, Nucl.\ Phys.\ B298:36 (1988)}

\ref\Heterotic{D.J. Gross, Jeffrey A. Harvey, Emil Martinec and Ryan Rohm,
Nucl.\ Phys.\ B256:253 (1985); Nucl.\ Phys.\ B267:75 (1986)}

\ref\KLT{
L.\ Dixon, J.\ Harvey, C.\ Vafa and E.\ Witten, Nucl.\ Phys.\
{B261}:678 (1985), {B274}:285 (1986)\semi
K.S.\ Narain, Phys.\ Lett.\ {169B}:41 (1986)\semi
W.\ Lerche, D.\ Lust and A.N.\ Schellekens, Nucl.\ Phys.\
{B287}:477 (1987)\semi
H.\ Kawai, D.C.\ Lewellen and S.-H.H.\ Tye, Nucl.\ Phys.\
B288:1 (1987)\semi
K.S.\ Narain, M.H.\ Sarmadi and C.\ Vafa,
Nucl.\ Phys.\ {B288}:551 (1987)\semi
I. Antoniadis, C.P.\ Bachas and C.\ Kounnas, Nucl.\ Phys.\
{B289}:87 (1987)}

\ref\Beta{Z.\ Bern and D.A.\ Kosower, Phys.\ Rev.\ D38:1888 (1988)\semi
Z.\ Bern and D. A.\ Kosower, in proceedings, {\it Perspectives in String
Theory}, eds. P.\ Di Vecchia and J. L.\ Petersen, Copenhagen 1987}

\ref\Bosonic{Z. Bern, Phys.\ Lett.\ 296B:85 (1992)}

\ref\LightByLight{R. Karplus and M. Neuman, Phys.\ Rev.\ 80:380 (1950);
Phys.\ Rev.\ 83:776 (1951)\semi
J.M. Jauch and F. Rohrlich, {\it The Theory of Photons and Electrons}
(Addison-Weseley) (1959)\semi
Hung Cheng, Er-Cheng Tsai and Xiquan Zhu, Phys.\ Rev.\ D26:922 (1982)}

\ref\Gravity{Z. Bern, D.C. Dunbar and T. Shimada, in preparation}

\ref\GN{J.L.\ Gervais and A. Neveu, Nucl.\ Phys.\ B46:381 (1972)}

\ref\Matt{M.J.\ Strassler,  Nucl.\ Phys.\ B385:145 (1992)}

\ref\FirstQuantized{L. Brink, P. Di Vecchia and P. Howe, Phys.\ Lett.\ 65B
Nucl.\ Phys. \ B118 (1977) 76\semi
F.A.\ Berezin and M.S.\ Marinov, JETP Lett.\ 21:320 (1975)\semi
R.Casalbouni, Nouvo Cimento 33A:389 (1976)\semi
M.B.\ Halpern and  P. Senjanovic, Phys.\ Rev.\ D15:1655 (1977)\semi
M.B.\ Halpern, A. Jevicki and P. Senjanovic, Phys.\ Rev.\ D16:2476 (1977)\semi
E.S.\ Fradkin and A.A.\ Tseytlin, Phys.\ Lett.\ 158B:316,1985;
163B:123 (1985); Nucl.\ Phys.\ B261:1 (1985)}

\ref\QCDTexts{Ta-Pei Cheng and Ling-Fong Li, {\it Gauge Theory
of Elementary Particle Physics}, (Clarendon Press) (1984)\semi
V.D. Barger and R.J.N. Philips, {\it  Collider Physics} (Addison-Wesley)
(1987)\semi
R.D. Field, {\it Applications of Perturbative QCD} (Addison-Wesley) (1989)}

\ref\EKS{F. Aversa, P. Chiappetta, M. Greco and J.P. Guillet,
Phys.\ Lett.\ 210B:225 (1988); Nucl.\ Phys.\ B327:105 (1989); Phys.\ Lett.\
B211:465 (1988)\semi
S.D. Ellis, Z. Kunszt and D.E. Soper,
Phys.\ Rev.\ Lett.\ 62:726 (1989);  Phys.\ Rev.\ Lett.\ 64:2121 (1990)}

\ref\GieleGlover{W.T. Giele and E.W.N. Glover, Phys.\ Rev.\ D46:1980 (1992)}

\ref\Jets{W.T. Giele, E.W.N. Glover and D.A. Kosower, Fermilab-pub-92-213-T,
Fermilab-conf-92-230-T}

\ref\UAData{T. Hansl-Kozanecka, SLAC Summer Institute Lectures, 1991}

\ref\LEPResults{S. Bethke and S. Catani, CERN-TH-6484-92}

\ref\LEPCompute{R.K. Ellis, D.A. Ross and A.E. Terrano,
Phys.\ Rev.\ Lett.\ 45:1226 (1980); Nucl.\ Phys.\ B178:421 (1981)}

\ref\Fermion{Z. Bern, L. Dixon and D.A. Kosower, in progress\semi
L. Dixon and Y. Shadmi, in progress}

\ref\Yehudai{R.K. Ellis, W. Giele and E. Yehudai, in progress}

\ref\ChanPaton{J.E.\ Paton and H.M.\ Chan, Nucl.\ Phys.\ B10:516 (1969)}

\ref\KobaNielsen{Z.\ Koba and H.B.\ Nielsen, Nucl.\ Phys.\ B12:517 (1969)}

\ref\Private{D.A.\ Kosower, private communication}

\ref\ParkeTaylor{S.J. Parke and T.R. Taylor, Phys.\ Rev.\ Lett.\ 56:2459,1986}

\ref\Grisaru{M.T. Grisaru, H.N. Pendelton and P. van Nieuwenhuizen, Phys. Rev.
D15:996 (1977)\semi
M.T. Grisaru and H.N. Pendelton, Nucl.\ Phys.\ B124:81 (1977)}

\ref\BasicSusy{J. Wess and J. Bagger, {\it SuperSymmetry and
SuperGravity} (Princeton University Press) (1983)\semi
S.J. Gates, M.T. Grisaru, M. Rocek and W. Siegel,
{\it Superspace or One Thousand and One Lessons in Supersymmetry}
(Benjamin/Cummings) (1983)}

\ref\SchwarzReview{J.H.\ Schwarz, Phys. Reports 89:223 (1982)}

\ref\typeII{L.\ Dixon, V.\ Kaplunovsky and C. Vafa, Nucl.\ Phys.\
B294:43 (1987)\semi
H.\ Kawai, D.C.\ Lewellen and S.-H. H.\ Tye, Phys.\ Lett.\ 191B:63 (1987)}

\ref\OpenString{Z.\ Bern and D.C.\ Dunbar,  Phys.\ Rev.\ Lett.\ 64:827 (1990);
          Phys.\ Lett.\ 242B:175 (1990)}

\ref\Tseytlin{R.R.\ Metsaev and A.A.\ Tseytlin, Nucl.\ Phys.\ B298:109 (1988)}

\ref\Polyakov{A.M.\ Polyakov, Phys.\ Lett.\ 103B:207 (1981); 103B:211 (1981)}

\ref\GSW{M.B.\ Green, J.H.\ Schwarz,
and E.\ Witten, {\it Superstring Theory} (Cambridge University
Press) (1987)}

\ref\Lovelace{C. Lovelace, Phys.\ Lett.\ 34B:500 (1971)}

\ref\Color{Z. Bern and D.A.\ Kosower, Nucl.\ Phys.\ B362:389 (1991)}

\ref\BjD{J.D. Bjorken, Stanford Ph.D. thesis (1958)\semi
J.D. Bjorken and S.D. Drell, {\it Relativistic Quantum Fields}
(McGraw-Hill, 1965)\semi
J. Mathews, Phys.\ Rev. 113:381 (1959)\semi
S. Coleman and R. Norton, Nuovo Cimento 38:438 (1965)\semi
C.S. Lam and J.P. Lebrun, Nuovo Cimento 59A:397 (1969)}

\ref\Lam{C.S. Lam,  MCGILL-92-32;  MCGILL-92-53}

\ref\Korner{
G.A. Schuler, S. Sakakibara and J.G. Korner, Phys.\ Lett.\ 194B:125
(1987)\semi
J.G. Korner and P. Sieben, Nucl.\ Phys.\ B363:65 (1991)}

\ref\KSpinor{D.A. Kosower, Phys.\ Lett.\ B254:439 (1991)}

\ref\HV{G. 't\ Hooft and M. Veltman, Nucl.\ Phys.\ B44:189 (1972)}

\ref\Siegel{W. Siegel, Phys.\ Lett.\ 84B:193 (1979)}

\ref\WaveFunction{Z. Bern and D.A.\ Kosower, Nucl.\ Phys.\ B321:605 (1989)\semi
Z. Bern, D.A.\ Kosower and K. Roland, Nucl.\ Phys.\ B334:309 (1990)}

\ref\Rozhanskii{L.V.\ Rozhanskii, Int.\ J.\ Mod.\ Phys.\ A5:571 (1990)}

\ref\QCDReview{J.C.\ Collins, D.E.\ Soper,
and G.\ Sterman, in {\it Perturbative Quantum Chromodynamics},
ed.\ A.H.\ Mueller (World Scientific, 1989)\semi
R. K. Ellis, private communication}

\ref\Integrals{Z. Bern, L. Dixon and D.A. Kosower, SLAC--PUB--5947;
SLAC--PUB--6001, Phys. Lett. B, to appear\semi
R.K.\ Ellis, W. Giele and E. Yehudai, in progress}

\ref\LeeNauenberg{T. Kinoshita, J. Math.\ Phys.\ 3:650 (1962)\semi
T.D. Lee and M. Nauenberg, Phys.\ Rev.\ 133:B1549 (1964)}

\ref\BerendsGrav{F.A. Berends, W.T.\ Giele and H. Kuijf,
Phys. Lett.\ {211B}:91 (1988)}

\ref\KLTtree{ H. Kawai, D.C.\ Lewellen and S.-H.H.\ Tye,
Nucl.\ Phys.\ {B269}:1 (1986)}

\ref\Background{G. 't Hooft,
{\it in} Acta Universitatis Wratislavensis no.\
38, 12th Winter School of Theoretical Physics in Karpacz; {\it
Functional and Probabilistic Methods in Quantum Field Theory},
Vol. 1 (1975)\semi
B.S.\ DeWitt, {\it in} Quantum gravity II, eds. C. Isham, R.\ Penrose and
D.\ Sciama (Oxford, 1981)\semi
L.F.\ Abbott, Nucl.\ Phys.\ B185:189 (1981)}

\ref\AGS{L.F. Abbott, M.T. Grisaru and R.K. Schaefer,
Nucl.\ Phys.\ B229:372 (1983)}

\ref\Unpublished{Z. Bern, unpublished}

\ref\Roland{K. Roland, Phys.\ Lett.\ B289:148 (1992)\semi
            Z. Bern, D.A. Kosower and K. Roland, in progress}

%%%%%%%%%%%%%%%%%%%%%%%%%%%%%%%%%%%%%%%%%%%%%%%%%%%%%%%%%%%%%%%%%%%%%%%%%%%%%

\noindent
hep-ph/9304249 \hfill UCLA/93/TEP/5 \break

%{String-Based Perturbative Methods for Gauge Theories

\vskip 1.5 cm

\baselineskip 15 pt
\centerline{\bf STRING-BASED PERTURBATIVE METHODS}
\vskip .1 cm
\centerline{\bf FOR GAUGE THEORIES%}
\footnote{${}^*$}%
{Lectures presented at the Theoretical Advanced Summer Institute (TASI) 1992,
Boulder, CO.} }

\vskip 3. cm
\centerline{ Zvi Bern}
\centerline{\it Department of Physics}
\centerline{\it UCLA}
\centerline{\it Los Angeles, CA 90024}

%\vskip 1 cm

\vskip 2.5 truecm

\vskip 1. cm
{ \narrower\smallskip
\centerline{\bf Abstract}
\vskip .1 cm
%\ninerm\baselineskip 10 pt
Recent progress in the computation of one-loop gluon amplitudes is
reviewed.  These methods were originally derived from superstring
theory and are significantly more efficient than conventional Feynman
rules.  With these methods, explicit computations
can be performed beyond those achieved by traditional methods.
\smallskip}

%\vskip 0.3 truecm

\vfill
\break

%%%%%%%%%%%%%%%%%%%%%%%%%%%%%%%%%%%%%%%%%%%%%%%%%%%%%%%%%%%%%%%%%%%%%%%%%%%%%
%\hfill UCLA/93/TEP/5 \break

%{String-Based Perturbative Methods for Gauge Theories

\baselineskip 12 pt
\centerline{\bf STRING-BASED PERTURBATIVE METHODS}
\vskip .1 cm
\centerline{\bf FOR GAUGE THEORIES}%
%\footnote{${}^*$}%
%{Lectures presented at the Theoretical Advanced Summer Institute (TASI) 1992,
%Boulder, CO.} }

\vskip .6 cm
\centerline{ Zvi Bern}
\centerline{\it Department of Physics, UCLA}
\centerline{\it Los Angeles, CA 90024}

%\vskip 1 cm

%\vskip 2.0 truecm
\baselineskip12pt

\vskip .6 cm
{ \narrower\smallskip
\centerline{\bf Abstract}
\ninerm\baselineskip 10 pt
Recent progress in the computation of one-loop gluon amplitudes is
reviewed.  These methods were originally derived from superstring
theory and are significantly more efficient than conventional Feynman
rules.  With these methods, explicit computations
can be performed beyond those achieved by traditional methods.
\smallskip}

%\vskip 0.3 truecm

%\vfill\break

%\vskip .3 cm
\baselineskip 12 pt
\section{Introduction and Overview}
\tagsection\IntroductionSection

The ability to uncover new physics at accelerators relies to a large
extent on subtracting known physics; in particular, QCD loop
corrections provide a significant background but are in general quite
formidable to calculate.  Intermediate expressions can be many
thousands of times larger in size than final expressions.  This
explosion of terms has been one of the major obstacles preventing
computations required by experiment from being performed.  In these
lectures new techniques which bypass much of the algebra associated
with one-loop Feynman diagram computations in gauge theories
[\use\Short --\use\Future] are discussed.  Many of the ingredients
that make up the new techniques are directly motivated by string
theory.

As an example of the power of the new technique, the Ellis and Sexton
[\use\Ellis] computation of the next-to-leading order contributions to
the $2 g \rightarrow 2 g$ cross-section required 108 diagrammatic
interferences; with the new methods only two relatively simple
diagrams are required.  Furthermore, with the new string-based
techniques the one-loop five-gluon amplitudes have been
computed yielding a compact form [\use\AmplLet]. These amplitudes have
not been obtained with traditional techniques.

Recent years have also seen substantial progress in improving the
situation in tree-level calculations.  Tree-level matrix elements have
been essential for checks of QCD processes and for estimates of QCD
backgrounds to new physics searches.  Four ideas which have
contributed to improvements in calculational ability are the spinor
helicity method for gluon polarization vectors
[\use\SpinorHelicity,\XZC], the color decomposition
[\use\TreeColor,\use\MPX], supersymmetry identities [\use\Susy], and
the Berends and Giele recurrence relations [\use\Recursive]. (A review
of these ideas can be found in ref.~[\use\ManganoReview].)  The tree-level
color decomposition [\use\MPX,\use\KLN] and recurrence relations
[\use\LightconeRecurrence] emerge quite naturally from string
theories.  The first three developments have also played an important
role at loop level.  (Recursion relations of the Berends and Giele
type may very well play an important role at loop-level in the
future.)

The one-loop method which is discussed in these lecture notes was
originally derived from string theory.  Although based on string
theory it has been summarized in terms of simple rules which require
no knowledge of string theory [\Long,\Pascos]; the structure of these
rules can also be understood from conventional field theory through
a particular gauge choice and organization of the amplitude
[\use\Mapping].  The structure of string amplitudes indicates why it
might be advantageous to organize a field theory computation according
to string theory: a string amplitude contains all field theory
diagrams in a single compact `master formula'.  This may be compared
to conventional field theory where the various Feynman diagrams bear
little relationship to each other.  Since string theories contain
gauge theories in the infinite string tension limit
[\use\Scherk,\use\GSB,\use\Minahan] and have a simpler organization of
the amplitudes than field theories, one might expect a string-based
calculation of the amplitude to be more efficient than a traditional
Feynman diagram calculation.
The original derivation of the new
one-loop method [\use\Long] was based on the field theory limit of an
appropriate heterotic string [\use\Heterotic,\use\KLT] amplitude.
Although the appropriate four-dimensional heterotic string
construction turns out to be fairly intricate [\use\Beta], the
consistency of the string guarantees that no extraneous problems
enter.  Once the correctness of the method is understood much simpler
string constructions, such as the bosonic one in reference [\use\Bosonic],
suffice. By organizing the various contributions which survive in the
field theory limit diagrammatic rules for computing amplitudes were
derived.

One way to quantify the gain in efficiency over traditional methods is
by comparing the calculation of the virtual corrections to
one-loop gluon scattering $2g
\rightarrow 2 g$ to the calculation of light-by-light scattering
$2\gamma \rightarrow 2 \gamma$.  Using modern spinor
helicity methods [\use\SpinorHelicity,\use\XZC] the light-by-light
computation is already far simpler than the traditional computation
[\use\LightByLight].  The power of the string-based methods is such
that the gluon calculation is only a bit more difficult than the
already much simplified photon computation.  This can be contrasted to
conventional field theory where the complexity of the non-abelian
gluon Feynman vertex as compared to the photon vertex implies that a
gluon scattering calculation should be significantly more complicated
than a photon scattering calculation.  Perhaps even more remarkably,
with string-based methods the growth in the complexity of a graviton
scattering computation as compared to a photon scattering computation is
relatively inconsequential as compared to conventional field theory
expectations; in particular, a one-loop graviton-by-graviton
scattering computation using string theory is only moderately more
complicated than the already much simplified light-by-light scattering
computation [\use\Gravity].

Is string theory `required' for field theory calculations?  The answer
is both yes and no.  To develop and extend the methods string theory
has been crucial and can be expected to continue to be useful.  To
actually evaluate amplitudes there is no need to turn to string
theory. The main role of string theory is to provide a principle for
discovering compact representations for field theory amplitudes. As
yet, there is no corresponding principle in conventional field theory.
In particular, given the string-based rules for the one-loop $n$-gluon
amplitudes and the understanding of these rules in conventional field
theory [\use\Mapping], there does not appear to be a clear way to
extend the rules to multi-loops, or gravity without referring back to
string theory to at least some extent.  It is, however, possible to
formulate a conventional field theory framework for obtaining much of
the efficiency of the string-based method by working backwards from
the string-based rules.  At one loop the main field theory ideas which
can be used to improve the efficiency of a calculation and are
inspired from string theory are the use of the background field gauge
for the loop part of diagrams, the non-linear Gervais and Neveu gauge
[\use\GN] for the tree parts of diagrams, color ordering of vertices,
systematic organization of the vertex algebra and a second order
formalism for fermions which helps make supersymmetry relations
manifest.  The spinor helicity method [\use\SpinorHelicity,\use\XZC]
is also natural within this framework.

These field theory ideas can be applied more generally to gauge theory
amplitude calculations which involve non-abelian vertices
[\use\Mapping].  In the future, extensions of the rules to include
external fermions, weak interactions and multi-loops can be expected,
but in the meantime, at least some of above ideas can be directly
applied to any Feynman diagram computation in non-abelian gauge
theory.

A complementary field theory approach [\use\Matt] for understanding
the string-based methods in a field theory context is through a first
quantized formalism [\use\FirstQuantized].  Its main advantage is that
it is simpler than dealing with string theory and is useful for
gaining an understanding of the string-based loop substitution rules.
With this approach one obtains a description of the one-loop effective
action, although as yet it does not provide a satisfactory
description of scattering amplitudes nor of the tree parts of
diagrams.  It might, however, provide an alternative path for
extensions to multi-loops.

These lectures are organized as follows: first the basic motivation
from experiment for wanting to compute loop diagrams with large
multiplicities is explained in Section~\use\QCDSection.  Such Feynman
diagram calculations are generally quite formidable although
important for new physics searches at colliders.  In Section
\use\TreeSection, tree-level techniques which carry over to loop level
are reviewed; the three methods are the color decomposition, spinor
helicity techniques, and supersymmetry identities.  Since the
loop-level method is based on string theory, a review of some of the
relevant string ideas is given in Section~\use\StringSection.
Although lacking complete string consistency, the bosonic string is
used as a basis of discussion because of its simplicity as compared to
a fully consistent four-dimensional heterotic string.  By taking the
field theory limit of an appropriately constructed string theory,
gauge theory amplitudes can be recovered.  These amplitudes are
organized in a particularly compact way.  In
Section~\use\TreeLoopSection\ the modifications that are needed when
applying the tree-level methods of Section~\use\TreeSection\ to loop
level is discussed.  One form of string-based rules is presented in
Section~\use\RulesSection.  These rules are then applied, in
Section~\use\ExampleSection, to a specific calculation of a one-loop
gluon helicity amplitude that would be rather difficult to evaluate by
traditional Feynman diagram methods, but is rather easy in the
string-based method. Results are also presented for one-loop four- and
five-gluon helicity amplitudes; these amplitudes were first calculated
with the string-based methods.  An amusing example of a four-graviton
calculation is also presented that would be exceedingly difficult to
evaluate via traditional Feynman diagrams but is rather simple using
string theory.  The basic structure of the string-based rules can be
understood by a particular organization of field theory; this is
explained in Section~\use\FieldTheorySection.  At tree-level, the
non-linear Gervais-Neveu gauge makes the match to string theory less
obscure and its rather simple diagrammatic structure partly explains
how string theory can avoid many of the large cancellations inherent
in conventional field theory.  At loop-level a different gauge choice
is required to make the structure of string-based rules less obscure:
background field Feynman gauge.  The background field method is
briefly reviewed.  The generic structure of the one-loop effective
action implied by string theory is then given followed by a discussion
of how one would apply string motivated field theory ideas to more
general calculations.  Section~\use\FieldTheorySection\ then concludes
with a discussion of the first quantized approach.  Finally in
Section~\use\ConclusionSection\ a summary and outlook for the future
is given.

\section{One-Loop Perturbative QCD}
\tagsection\QCDSection

\noindent
{\it \QCDSection.1 Requirements by Experiments}
\vskip .1 cm

The fundamental question of perturbative QCD is whether new physics is
hiding in the QCD background.  The QCD background generally swamps new
physics signals.  As an example, in \fig\BackgroundFig\ the
number of events is plotted against the two-jet invariant mass.  At
approximately 90 GeV one might expect to see peaks from $W$ and $Z$
production; as seen in the figure these peaks are swamped by the QCD
background.  Another example is $t$ quark searches at Fermilab
which must deal with significant QCD backgrounds.  In general, to find
new physics it is important to subtract the QCD background.  The more
precisely the subtraction can be done, the more likely that new physics
can be identified at colliders.

One of the characteristics of many of the interesting events at
accelerators are jets (which are collimated bunches of hadrons heading
out from the interaction). A key ingredient that enters into the
theoretical computation of jets are Feynman diagrams which describe
the partons.  Other essential ingredients which make up the
computation are the structure functions describing the initial state
partons and the final state hadronization process. Further details can
be found in standard textbooks [\use\QCDTexts].  (In general the
conversion of the matrix elements into physical scattering processes
that can be compared to experiment is nontrivial because of
complexities associated with soft and collinear divergences
[\use\EKS]; Giele and Glover [\use\GieleGlover,\use\Jets] have,
however, constructed a convenient formalism for performing
that step.)   These lectures will deal with only the Feynman diagram
part of the computation.

%\vskip .3 cm
\centerline{\epsfxsize 5.2 truein \epsfbox{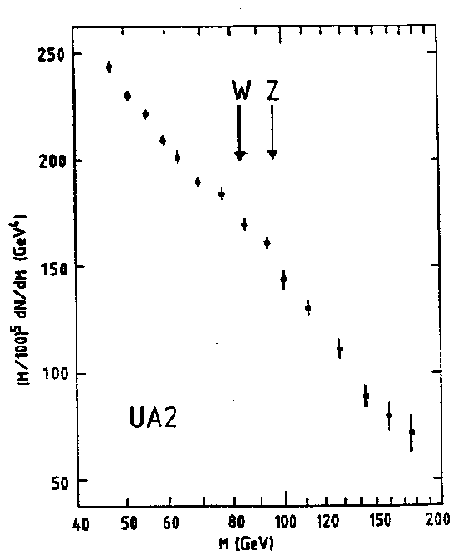}}
\nobreak
\vskip -.2 cm\nobreak
{ \narrower\smallskip \ninerm
\noindent{\ninebf Fig.~\BackgroundFig:} An example of the QCD background:
the $Z$ and $W$ peaks are completely swamped.  (Data from UA2 collaboration.)
\smallskip}

\vskip .4 cm

In Feynman diagram computations one starts with the Born or
tree diagrams, since these give the leading order contribution
and are the simplest to compute.  The tree
level diagrams form the cornerstone for extracting physics from
colliders [\use\SpinorHelicity,\TreeColor,\use\ManganoReview].  The
tree-level computations, however, miss essential physics
[\use\EKS,\use\GieleGlover].  There are three basic problems:

\item
{1)} The tree-level jet cone angle dependence is wrong.  The physical
origin of the jet cone angle dependence is that when two jets are
nearby they could either be counted as a single jet or as two jets
depending on the jet cone definitions.  Depending on how a given jet
is counted the apparent cross-section will change. In \fig\ConeSizeFig\
an example of the difference between the tree-level predictions and loop
level predictions is shown for $W + 1$ jet production.

\vskip .4 cm
\centerline{\epsfxsize 6.0 truein \epsfbox{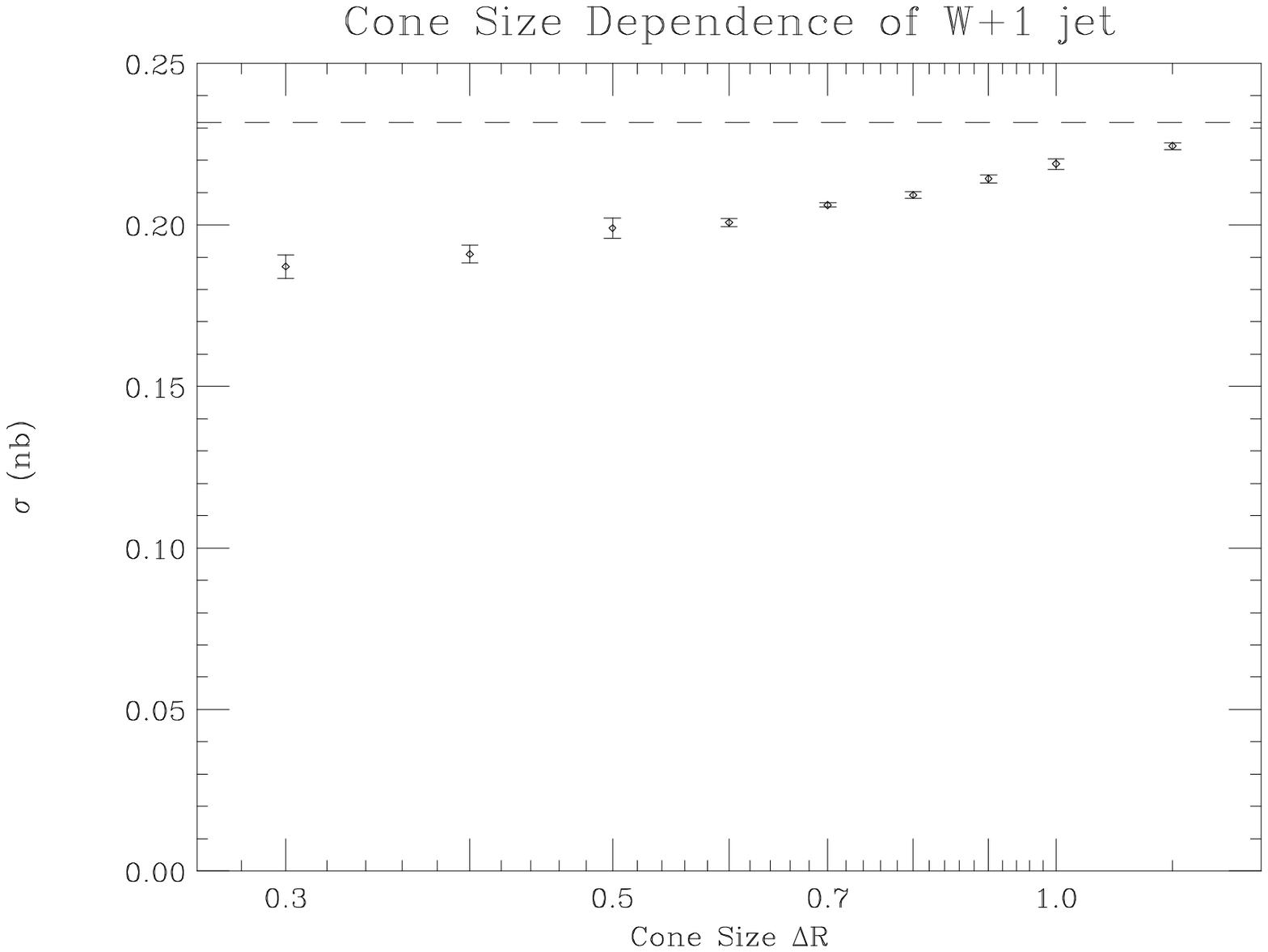}}
\nobreak
\vskip -8.3 cm
{\baselineskip 10 pt
\narrower\smallskip\noindent \ninerm \nobreak
{\ninebf Fig.~\ConeSizeFig :}  An example of the
dependence of the cross-section on the jet cone-size. The dotted line
is the tree-level result and shows no dependence on the cone size.
The data points are theoretical results
including one-loop corrections and exhibit a significant cone-size dependence.
(From ref.~[\use\Jets].)
\smallskip}

\vskip .4 cm

\item
{2)} The tree-level renormalization scale dependence is wrong.
Physical quantities should not depend on the renormalization scale.
At tree level the dependence on the scale enters due to the coupling constant
sitting in front of the amplitude; the precise value which should be
chosen for the scale (or equivalently, the value of the coupling
constant) is not clear, leading to variations in the predicted
cross-section of more than fifty percent for sensible choices of the scale,
as indicated in
\fig\RenormScaleFig.  This leads to the commonly quoted large theoretical
uncertainty.
As indicated in fig.~\RenormScaleFig\ the one-loop
corrections tend to reduce the scale dependence to about five or ten
percent.

\item{3)} In QCD there are large infrared logarithms which in general cannot
be neglected.  Such logarithms are not accounted for by tree
calculations and are related to the incorrect cone angle dependence.

\noindent
To a large extent one-loop corrections fix these problems.   This
provides the basic motivation for performing loop level QCD
computations.

\vskip .5 cm
\centerline{\epsfxsize 6.0 truein \epsfbox{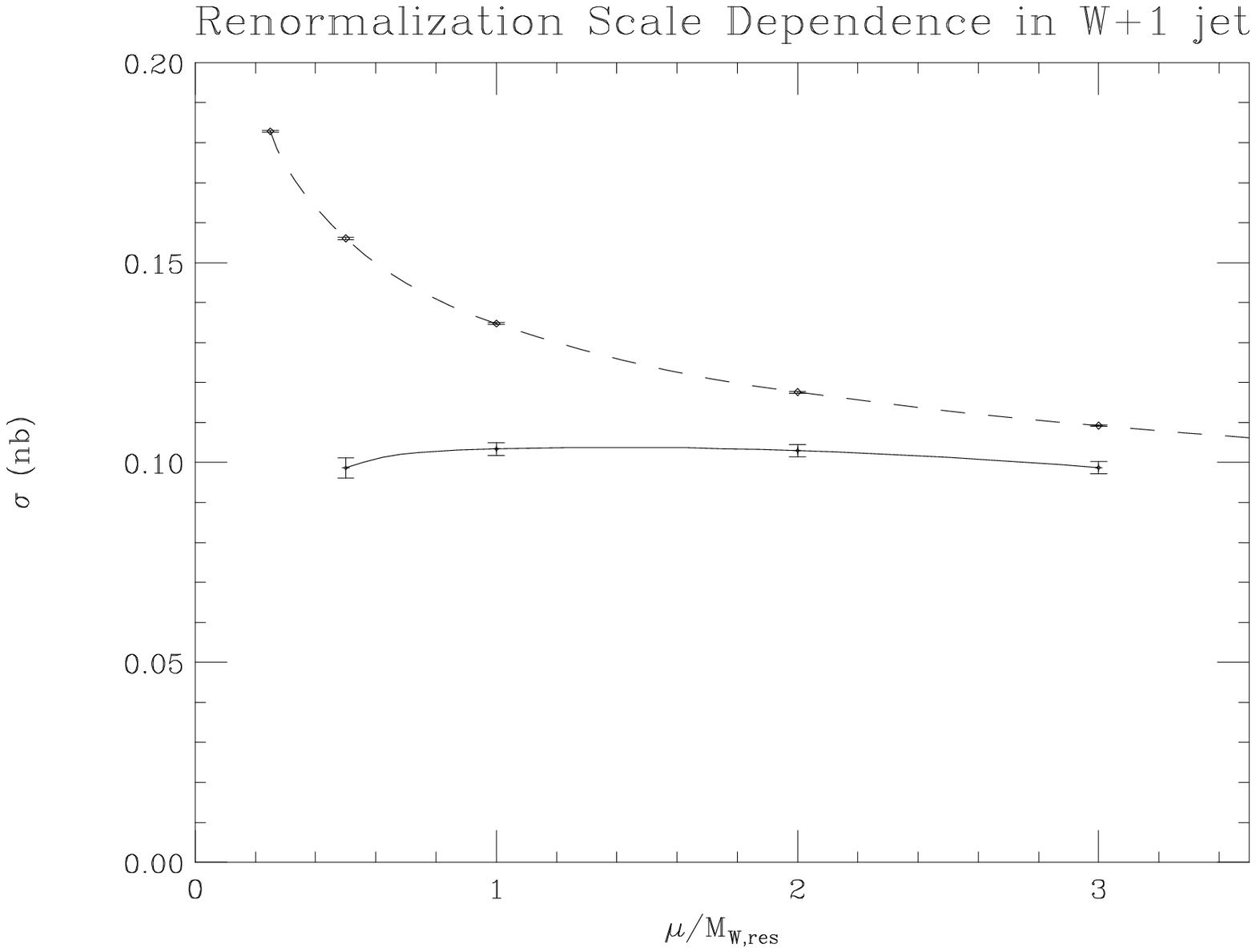}}
\nobreak
\vskip -8.2 cm \nobreak
{\baselineskip 10 pt \narrower\smallskip\noindent \ninerm {\ninebf
Fig.~\RenormScaleFig :} The renormalization scale dependence of a
cross-section at tree-level and with one-loop corrections.  The dotted
line is the tree result which exhibits a strong scale dependence while
the solid line includes one-loop corrections and exhibits a much
weaker dependence over a wide scale. (From ref.~[\use\Jets].)
\smallskip}

\vskip .4 cm

As one example of a relevant loop computation, experimenters at
Fermilab would like to measure the strong coupling constant $\alpha_s$
and its running.  At Fermilab one would be able to measure the
coupling constant at 250 GeV which is at an energy well beyond what
can be currently achieved at LEP.  (At these higher energies jets are
easier to resolve.) A good way to obtain $\alpha_s$ is from the ratio
of the three-jet cross section to the two-jet cross section.  Roughly
speaking this quantity is proportional to $\alpha_s$.  Good data with
approximately twenty percent experimental errors exist since 1985
from the UA1 and UA2 experiments at CERN for $\alpha_s$ at various
mass scales [\use\UAData].  Unfortunately, because of the lack of all
the required one-loop calculations of the three-jet cross-section, the
theoretical uncertainties associated with this quantity are on the
order of a hundred percent.  This situation may be compared to LEP
which quotes $\alpha_s$ at the mass of the $Z$ with a total
theoretical and experimental uncertainty of about ten percent
[\use\LEPResults].  One reason why the relevant theoretical
computations have been performed for LEP, but not for hadron machines,
is that diagrams involving initial state electrons instead of gluons
and other partons are easier to compute; the relevant loop corrections
have been given in refs.~[\use\LEPCompute].  (The conversion of the
matrix elements into physical quantities is also more complicated
[\use\EKS,\use\GieleGlover] for hadron scattering.)  In terms of the
diagrams, the one-loop three-jet calculation for LEP requires box
diagrams at worst while the corresponding calculation for hadron
colliders requires pentagon diagrams.  (Pentagon diagrams are
generally many orders of magnitude more complicated to evaluate than
box diagrams).

Using the string-based methods discussed in these lectures the first
computation of the one-loop gluon matrix elements required for the
three-jet cross-section has been performed [\use\AmplLet].  The quark
contributions have not been computed but progress has been made with
string-based methods [\use\Fermion].  From a traditional field theory
point of view these are generally easier to compute than the gluon
contributions.

Other examples of computations which are relevant for current
experiments but have not been performed as yet are:

\item
{1)} One-loop corrections to $W$ + $n$-jet production at hadron
colliders with $n\ge 2$, where $W\rightarrow \ell \nu$.  This forms a
background to $t$ quark searches at Fermilab.

\item
{2)} One-loop corrections to $Z$ + $n$-jet production at hadron
colliders with $n\ge 2$ where $Z\rightarrow\bar\nu\nu$. This forms a
background to missing transverse energy searches for new physics at
Fermilab.

\item
{3)} One-loop corrections to $Z \rightarrow 4$ jets [\use\Yehudai].
This would be be useful for measurement of $\alpha_s$ from the four-
to three-jet ratio at LEP.  The calculation is also equivalent to
a large extent to the calculation of one-loop corrections to $W,Z$ + 2
jet production at Fermilab.

\item
{4)} Two-loop corrections to $Z\rightarrow 3$ jets.  LEP is currently
sensitive to these corrections.

\item
{5)} Two-loop corrections to two-jet production at Fermilab.  The
only way to decisively prove that one-loop corrections are adequate is to
calculate the two-loop corrections and show that they are unimportant.

Experimenters need theorists to perform the loop computations
associated with these processes, so why haven't theorists calculated
them?

\vskip .2 cm
\noindent{\it \use\QCDSection.2 Difficulty of Loop Computations}
\vskip .1 cm

Another way to phrase the above question is: why are perturbative
QCD computations so complicated?  The answer is that there are
too many Feynman diagrams and each Feynman diagram is too complicated,
especially those diagrams which contain gluons.  Such diagrams are
important at high energies.
An underlying cause of the complexity is that
the non-abelian vertices which are
given in \fig\FeynmanVertFig\ are relatively complicated. Since the vertices
each contain six terms, one encounters a rapidly growing number of
terms as one sews together vertices with propagators to form Feynman diagrams.

\vskip .4 cm
{\baselineskip 14 pt
\epsfxsize 1.1 truein \epsfbox{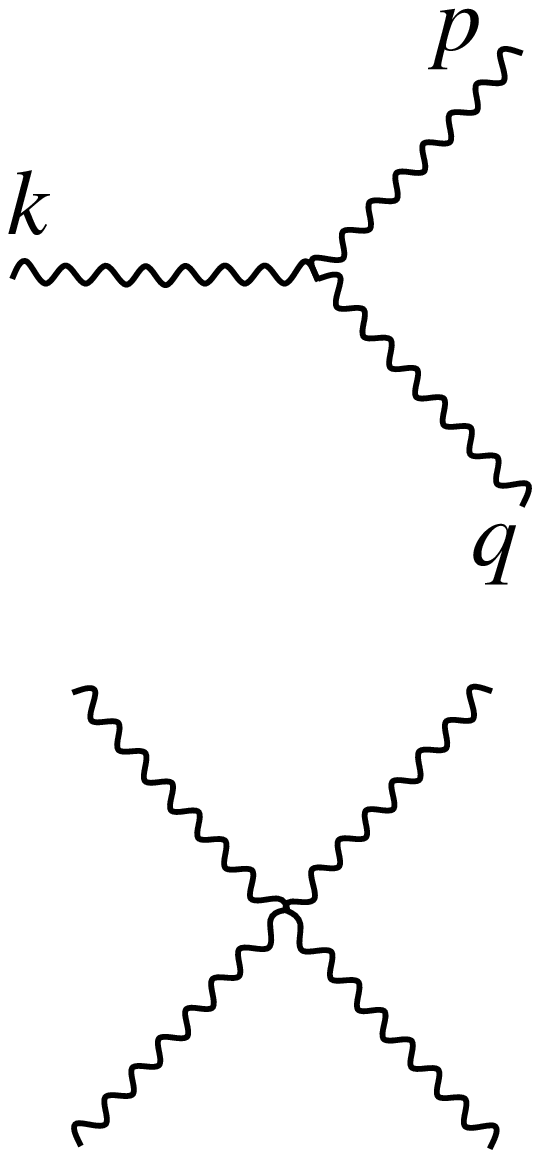}
\nobreak
\vskip - 5.9 truecm
\noindent \hskip 2.7 truecm $\nu$ \hskip .06 truecm $b$\par\nobreak
\vskip .6 truecm \nobreak
\noindent\hskip .1 truecm  $\mu$ \hskip .06 truecm $a$\par\nobreak
\noindent\hskip 2.7 truecm $\rho$ \hskip .06 truecm $c$\par\nobreak
\vskip -1.4 truecm\nobreak
\noindent \hskip 3.1 truecm
$ = - g f^{abc} \Bigr( \eta_{\mu\nu} (k - p)_{\rho}
                  + \eta_{\nu\rho} (p - q)_{\mu}
                  +\eta_{\rho\mu} (q - k)_{\nu} \Bigr) $ \par\nobreak
\vskip 5. truecm \par\nobreak
\vskip -3.6 truecm \nobreak
\noindent \hskip .07 truecm $a$ \hskip .2 truecm $\mu$
\hskip 1.05 truecm $\nu$ \hskip .2 truecm $b$\par\nobreak
\vskip 5. truecm \nobreak
\vskip -3.1 truecm \nobreak
\noindent \hskip .07 truecm $d$ \hskip .2 truecm $\lambda$
\hskip 1.05 truecm $\rho$ \hskip .2 truecm $c$ \par\nobreak
\vskip 4.1 truecm \nobreak
\vskip - 7.1 truecm \nobreak
$$
= \left\{
\eqalign{-i g^2 [ & f^{abe} f^{ecd}
(\eta_{\mu\rho}\eta_{\nu\lambda} - \eta_{\mu\lambda}\eta_{\nu\rho}) \cr
\null + \null & f^{ade}f^{ebc} (\eta_{\mu\nu} \eta_{\rho\lambda} -
\eta_{\mu\rho}\eta_{\nu\lambda}) \cr
\null + \null & f^{ace} f^{ebd} (\eta_{\mu\nu} \eta_{\rho\lambda} -
\eta_{\mu\lambda} \eta_{\nu\rho} ) ] \cr} \right.
$$
\vskip 7.1 truecm\nobreak
\vskip -6.2 truecm \nobreak
{\baselineskip 10 pt \narrower\smallskip\noindent \ninerm\nobreak
{\ninebf Fig. \FeynmanVertFig:} The conventional three- and
four-point Feynman vertices. \narrower}
}
\vskip .6 cm

As a simple example consider the pentagon diagram one
would encounter in a brute force three-jet computation.  A naive count
of the number of terms gives about $6^5$ terms.  (This count
is slightly reduced by the use of on-shell conditions but
increased by observing that each internal momentum is a sum of
momenta.)  Each term is associated with an integral which
evaluates to an expression on the order of a page in length.
This means that one is faced with about $10^4$ pages of algebra
for this single diagram.   As bad as this situation might seem, it is
actually much worse because of the structure of the results.
After evaluating the integrals and summing over diagrams
one obtains expressions of the form
$$
{N_1 \over D_1} + {N_2\over D_2} + \cdots
\anoneqn
$$
where the $N_i$ and $D_i$ are the numerators and denominators one
encounters when performing the integrals.  In general the denominators
contain spurious singularities which cancel only after putting large
numbers of terms on a common denominator; this unfortunately causes an
explosion of terms in the numerators.  It is therefore not too
surprising that the three-jet computation, which involves pentagon
diagrams, has not yet been performed
with the traditional methods employed, for example, by Ellis and
Sexton [\use\Ellis] in their two-jet computation.

The basic observation for being able to improve on conventional computations
is that Feynman diagram computations always involve large cancellations
amongst the various terms.  Anyone who has done a
Feynman diagram computation has undoubtedly asked themselves why
vasts amounts of algebra are required when answers tend to be
quite small.  A nice example of this is the four-gluon helicity amplitude
$$
A_{4;1}^{\rm 1 - loop} (1^-, 2^+, 3^+, 4^+) = - {i \over 48 \pi^2}
{\spb2.4{}^2 u \over \spb1.2 \spa2.3 \spa3.4 \spb4.1 }
\anoneqn
$$
where the plus and minus signs associated with each leg denote the
helicity, the various brackets refer to the spinor helicity notation
discussed in the next section, and $u= 2k_1\c k_3$ is a Mandelstam
variable.  (This amplitude has been color decomposed, which will be
discussed in Sections~\TreeSection\ and \TreeLoopSection.)  Although
this expression fits on a line, a brute force computation performed in
the conventional way would start with expressions containing about
$10^4$ terms. Clearly there is considerable room for improving Feynman
diagram computations at one loop.

\section{Tree Level Methods}
\tagsection\TreeSection

The tree-level techniques have been already been reviewed in the
article of Mangano and Parke [\use\ManganoReview] so here only those
techniques that have been carried over to loop level will be
discussed. The three important tree-level tools which have been carried
over to loop level are color decomposition [\use\TreeColor],
spinor-helicity techniques [\use\SpinorHelicity,\use\XZC], and supersymmetry
identities [\use\Susy].

\vskip .2 cm
\noindent
{\it \use\TreeSection.1 The Color Decomposition}
\vskip .1 cm

In terms of ordinary Feynman rules the notion of color ordering is fairly
simple to implement. The Yang-Mills structure constants are rewritten in
terms of fundamental representation matrices
$$
f^{abc} = -{i\over\sqrt2} \Tr\L \LB T^a, T^b\RB T^c\R
\eqn\FundConvert
$$
where the normalization of the
generators is $\Tr(T^a T^b) = \delta^{ab}$.
The color ordered gluon Feynman rules for ordinary Feynman gauge
are depicted in \fig\FeynmanColorFig.  These rules are obtained from
ordinary Feynman rules given in fig.~\FeynmanVertFig\
by restricting attention to a given color ordering.
By using eq.~\FundConvert\ and extracting the coefficient of $\Tr(T^a T^b T^c)$
the color ordered three-vertex is obtained.  The same type of analysis
leads to the color ordered four-vertex which is given
by the coefficient of $\Tr(T^a T^b T^c T^d)$.  With these rules one computes
a partial amplitude corresponding to a single color trace term;  the
diagrams should be drawn in a planar fashion with the external legs
following the ordering of the color trace under consideration.
The full tree-level amplitude can then be reconstructed from the
partial amplitudes by multiplying by
the associated color trace and summing over all non-cyclic
permutations
$$
\A{n}(\{k_i,\pol_i,a_i\}) =
g^{n-2} \sum_{\sigma\in S_n/Z_n} \Tr(T^{\as1}\cdots T^{\as{n}})
A_n(\ks1,\ps1;\ldots;\ks{n},\ps{n})
\eqn\TreeAmplitudeDecomposition
$$
where $k_i$, $\pol_i$, and $a_i$ are respectively the momentum,
polarization vector, and color index of the $i$-th external
gluon.  $S_n/Z_n$ is the set of non-cyclic
permutations of $\{1,\ldots, n\}$.  Note that the partial amplitudes
have been defined with the powers of the coupling constant removed.

\vskip .3 cm
{\baselineskip 14 pt
{\epsfxsize 1.1 truein \epsfbox{feynmanvertfig.ps} }
\nobreak
\vskip - 5.9 truecm \nobreak
\noindent \hskip 2.7 truecm $\nu$\par\nobreak
\vskip .6 truecm \nobreak
\noindent\hskip .1 truecm  $\mu$\par\nobreak
\noindent\hskip 2.7 truecm $\rho$\par\nobreak
\vskip -1.7 truecm\nobreak
\noindent \hskip 3.1 truecm
% V^{\rm Feynman}_{\mu \nu\rho} (k, p , q)
$ \displaystyle  = {i\over\sqrt{2}} \Bigr( \eta_{\mu\nu} (k -p)_{\rho}
                  + \eta_{\nu\rho} (p -q)_{\mu}
                  + \eta_{\rho\mu} (q -k)_{\nu} \Bigr) $ \par\nobreak
\vskip 4.9 truecm \nobreak
\vskip -3.6 truecm \nobreak
\noindent \hskip .07 truecm $a$ \hskip .2 truecm $\mu$
\hskip 1.05 truecm $\nu$ \hskip .2 truecm $b$\par\nobreak
\vskip 5. truecm \nobreak
\vskip -3.2 truecm \nobreak
\noindent \hskip .07 truecm $d$ \hskip .2 truecm $\lambda$
\hskip 1.05 truecm $\rho$ \hskip .2 truecm $c$ \par\nobreak
\vskip 4.1 truecm \nobreak
\vskip - 6.0 truecm \nobreak
% V^{\rm Feynman}_{\mu \nu\rho\lambda}
\noindent \hskip 3.1 truecm
$\displaystyle = i \eta_{\mu\rho} \eta_{\nu \lambda} -
   {i\over 2} ( \eta_{\mu \nu} \eta_{\rho\lambda} +
\eta_{\mu\lambda} \eta_{\nu\rho})$ \par\nobreak
\vskip 6.0 truecm\nobreak
\vskip -4.8 truecm \nobreak
{\baselineskip 10 pt\narrower\smallskip\noindent\ninerm
{\ninebf Fig.~\FeynmanColorFig :} The color ordered Feynman gauge vertices
for obtaining the partial amplitudes $A_n$.
\smallskip}
}

\vskip .6 cm

The immediate advantage of rewriting Feynman rules in this way is that
fewer diagrams contribute.  As a simple example with conventional
Feynman diagrams one would have a total of four conventional Feynman
diagrams, depicted in \fig\FourPointTreeFig\ for the four-point
tree amplitude.  With color ordered Feynman rules one would compute the
partial amplitude $A_4(1,2,3,4)$ associated with the color trace
$\Tr(T^{a_1} T^{a_2} T^{a_3} T^{a_4})$ and
would not need
to include diagram~\use\FourPointTreeFig{c}, since the ordering of the
legs do not follow the ordering of the color trace.
It is a simple exercise at the four-point level to verify that
these color ordered rules reproduce the results obtained from
conventional Feynman rules.

\vskip .6 cm
\centerline{\epsfxsize 3.8 truein \epsfbox{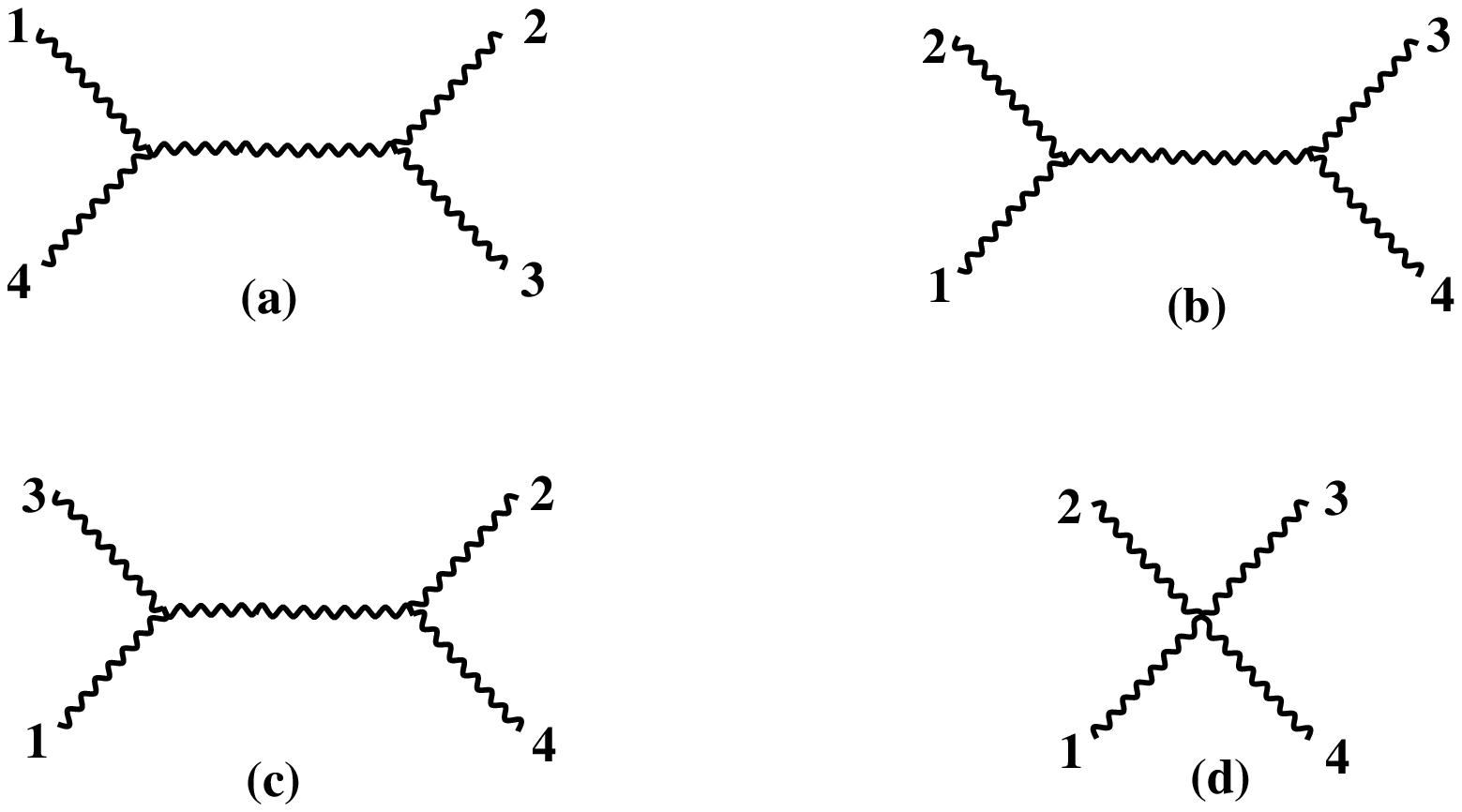} }

\nobreak
\vskip .2 cm \nobreak
{\baselineskip 10 pt\narrower\smallskip\noindent\ninerm
{\ninebf Fig.~\FourPointTreeFig:}
The four-point Feynman diagrams.  Color ordered Feynman rules do not
include diagram (c) for $A_4(1,2,3,4)$.
\smallskip}

\vskip .6 cm

The color decomposition (\use\TreeAmplitudeDecomposition) follows
from string theory.  In an open string theory, the full
on-shell amplitude for the scattering of $n$ massless vector mesons
can be written as the sum over non-cyclic permutations of the external
legs of Chan-Paton factors [\use\ChanPaton] times Koba-Nielsen partial
amplitudes [\use\KobaNielsen]
$$
\eqalign{
\A{n}^{\rm string}(\{k_i,\pol_i,a_i\})  & =
g^{n-2} \hskip -.15 cm  \sum_{\sigma\in S_n/Z_n} \hskip -.2 cm
\Tr(T^{\as1}T^{\as2}\cdots T^{\as{n}}) \cr
\null & \hskip 2 cm \times
A_n^{\rm KN}(\ks1,\ps1;\ldots;\ks{n},\ps{n}) \; . \cr }
\eqn\StringTreeAmplitude
$$
By taking the infinite string tension limit, the string amplitudes
reduce to field theory amplitudes yielding the field theory color
decomposition (\use\TreeAmplitudeDecomposition).  For tree amplitudes
where all the external legs are gluons, the matter content is
irrelevant, since the matter fields cannot appear as internal lines.
Thus one can use the open bosonic string, the simplest of all string
constructions, at tree level.  The decomposition of the string
amplitude leads immediately to the decomposition of on-shell $n$-gluon
amplitudes (\use\TreeAmplitudeDecomposition).

This color decomposition is actually quite a bit more useful than just a
reduction in the number of diagrams that must be considered. Two
additional advantages are the ability to use $U(N)$ color matrices
instead of $SU(N)$ matrices and certain
identities satisfied by the partial amplitudes.
The partial amplitudes $A_n$ possess a number of nice
properties that follow immediately from the properties of
the Koba-Nielsen amplitudes.  Each is gauge invariant on shell,
that is invariant
under the substitution $\pol_i\rightarrow \pol_i + \lambda k_i$
for each leg independently.  It is also invariant under cyclic
permutation of its arguments, and satisfies a reflection
identity,
$$A_n(n,\ldots,1) = (-1)^n A_n(1,\ldots,n)\,,\eqn\ReflectionIdentity$$
where the notation
$A_n(1,\ldots,n) = A_n(k_1,\pol_1;\ldots;k_{n},\pol_{n})$ is used.
The $U(1)$ gauge boson is an integral part of the string theory
(its presence is necessary for unitarity), but in the infinite-tension
limit, it must decouple from $SU(N)$ gauge boson amplitudes;
thus in field theory
$$
\A{n}(\{k_i,\pol_i,a_i\}_{i=1}^{n-1};k_n,\pol_n,a_{U(1)}) = 0 \; .
\eqn\DecouplingTreeBase
$$
This can be used to derive a decoupling identity, simply by extracting
the coefficient of
$\Tr(T^{a_1}\cdots T^{a_{n-1}})$, which is
$$
\sum_{\sigma\in Z_{n-1}} A_n(\si(1),\ldots,\si({n-1}),n) = 0.
\eqn\TreeDecouplingIdentity
$$
(This identity can also be derived starting with the twist operator in
open string theory.
Mangano, Parke, and Xu [\use\MPX,\use\Recursive,\use\KLN] term the
identity a dual Ward identity.)  Substituting additional photons
for gluons leads to equations which are linearly dependent on
equation~(\TreeDecouplingIdentity).

An advantage of using a $U(N)$ gauge group instead of an $SU(N)$ gauge
group in the color decomposition is that the $U(N)$ Fierz identities
\defeqn\Fierz
$$
\eqalignno{
\Tr(T^a X) \Tr(T^a Y) &= \Tr(X Y) &
(\Fierz{a}) \cr
\Tr(T^a X T^a Y) & = \Tr(X) \Tr(Y) &
(\Fierz{b}) \cr }
$$
are simpler than their $SU(N)$ counterparts.  This is useful when
squaring and summing over colors in order to obtain the cross-section.

In summary, the color ordered Feynman rules lead to significant
simplifications as compared to conventional Feynman rules.
However, the real power of color ordering occurs when coupled
with other ideas.

\vskip .2 cm
\noindent
{\it \use\TreeSection.2 Spinor Helicity Techniques}
\vskip .1 cm

The spinor helicity method [\use\SpinorHelicity,\use\XZC] involves a
rewriting of gluon (or photon) polarization vectors in terms of spinor
inner products. At first sight the point of this rewriting may not be
clear, but with a few simple examples its power becomes evident.  This
technique implicitly makes use of clever on-shell gauge
transformations in order to make large numbers of terms vanish in a
given computation.

In the formalism of Xu, Zhang and Chang a gluon polarization vector
is written as
$$
\pol^{(+)}_\mu (k;q) =
{\sand{q}.{\gamma_\mu}.k
\over \sqrt2 \langle q^- | k^+ \rangle },\hskip 1cm
\pol^{(-)}_\mu (k;q) =
{\sandpp{q}.{\gamma_\mu}.k
\over \sqrt{2} \langle k^+ | q^- \rangle},
\anoneqn
$$
where $|k^{\pm} \rangle $ is a Weyl spinor, with plus and minus helicities,
$k$ is the on-shell momentum of the gluon and
$q$ is an arbitrary reference momentum satisfying $q^2 =0$,
$k\cdot q \not = 0$.  These polarization vectors satisfy
the conditions for circular polarization
$$
k \cdot \pol^{(\pm)}_\mu (k;q) = 0 \; ,
\hskip 1 cm
(\pol^{(\pm)})^2 = 0 \; ,
\hskip 1 cm
\pol^{(+)} \cdot \pol^{(-)} = -1
\anoneqn
$$
and are therefore sensible definitions for helicities.
The convention that all momenta are outgoing is used; the effect of
this is to flip helicity notation on an incoming line.

It is convenient to define
abbreviations for the various spinor products and the Lorentz product,
$$\eqalign{
\spa{j}.l &= \spa{k_j}.{k_l} = \langle {k_j}^{-} | {k_l}^{+}\rangle \cr
\spb{j}.l &= \spb{k_j}.{k_l} = \langle {k_j}^{+} | {k_l}^{-}\rangle \cr
\lor{j}.l &= \spa{j}.{l}\spb{l}.j = 2 k_j \cdot k_l\;.\cr
}\anoneqn$$
The spinor products are antisymmetric,
$$\spa{j}.l = -\spa{l}.j,\hskip 1cm
\spb{j}.l = -\spb{l}.j
\anoneqn$$
and can be evaluated explicitly using,
$$\eqalign{
\spa{k_1}.{k_2} &=
  \sqrt{\L k_1^t-k_1^z\R\,\L k_2^t+k_2^z\R}
      \exp(i \mathop{\rm atan}(k_1^y/k_1^x))
  - (1\leftrightarrow2)\cr
 &= \sqrt{ {k_2^t+k_2^z\over k_1^t+k_1^z} }\,\L k_1^x + i k_1^y\R
  - (1\leftrightarrow2)\cr
\spb{k_1}.{k_2} &= \mathop{\rm sign}(k_1^t k_2^t)\L\spa{k_2}.{k_1}\R^*.
}\anoneqn$$

Gauge-invariant quantities are independent of the choice of reference
momentum $q$, because changing $q$ just corresponds to a gauge transformation
[\use\XZC]
$$
\pol^{(+)}_\mu(k;q') = \pol^{(+)}_\mu(k;q)
+ {\sqrt{2}\spa{q}.{q'}\over\spa{q}.k\spa{q'}.k}\, k_\mu
\eqn\SpinorGaugeTrans
$$
which follows from the rearrangement or Schouten identity
$$
\spa1.2\spa3.4 = \spa1.4\spa3.2 + \spa1.3\spa2.4\;.
\anoneqn
$$
The Fierz identity,
$$
\sand1.{\gamma^\mu}.2\,\sandpp3.{\gamma_\mu}.4 =
2 \spa1.4\spb3.2
\eqn\FierzIdentity$$
and the fact that $\sand1.{\gamma^\mu}.2 = \sandpp2.{\gamma^\mu}.1$
can be used to evaluate dot products of polarization vectors.

Given the reference momenta, the various dot products
are simply

$$\eqalign{
&k_j\cdot\pol^{(+)}_l(k_l;q_l)
= {\spa{q_l}.j\spb{j}.l
\over \sqrt2\,\spa{q_l}.{l}} \; , \hskip 2.6 cm
k_j\cdot\pol^{(-)}_l(k_l;q_l)
= {\spb{q_l}.j\spa{j}.l
\over \sqrt2\,\spb{l}.{q_l}} \; ,  \cr
&\pol^{(-)}_j(k_j;q_j)\cdot\pol^{(-)}_l(k_l;q_l)
= {\spa{j}.{l}\,\spb{q_l}.{q_j}
\over \spb{j}.{q_j}\,\spb{l}.{q_l}} \; , \hskip 1. cm
\pol^{(+)}_j(k_j;q_j)\cdot\pol^{(+)}_l(k_l;q_l)
= {\spa{q_j}.{q_l}\,\spb{l}.{j}
\over \spa{q_j}.{j}\,\spa{q_l}.{l}} \; , \cr
& \hskip 4 cm
\pol^{(+)}_j(k_j;q_j)\cdot\pol^{(-)}_l(k_l;q_l)
= {\spa{q_j}.{l}\,\spb{q_l}.{j}
\over \spa{q_j}.{j}\,\spb{l}.{q_l}} \; .\cr
}\eqn\DotProducts
$$
In making a choice of reference momenta, it is useful to keep the properties
noted by Mangano~et~al.~[\use\MPX] in mind.  With the first argument
to a polarization vector denoting the momentum of the gluon, and the
second its reference momentum, these properties are
$$\eqalign{
q\cdot\pol^{(\pm)}(k;q) &= 0\cr
\pol_j^{(\pm)}(k_j;q)\cdot \pol_l^{(\pm)}(k_l;q) &= 0\cr
\pol_j^{(\mp)}(k_j;q)\cdot \pol_l^{(\pm)}(k_l;k_j) &= 0\cr
}\anoneqn
$$
so that it is desirable to choose the same reference momenta for all
gluons of a given helicity, and to take this momentum to be the momentum
of one of the opposite-helicity gluons.  This will greatly reduce
the number of non-vanishing $\pol_i\cdot\pol_j$ invariants.
It also turns out that
within the set of choices suggested by these properties, it is preferable
to choose a reference momentum that is cyclicly adjacent to the momentum
of the gluon.

As one simple example for the amplitude ${\cal A}(1^-,2^+,3^+,4^+)$,
consider reference momenta $(k_4,k_1,k_1,k_1)$ for the
legs (1,2,3,4) respectively, leading to the simplifications
$$
\eqalign{
\pol_i \cdot \pol_j = 0, & \hskip 1.5 cm
k_4\c \pol_1 = k_1 \c \pol_2 = k_1\cdot \pol_3 = k_1 \cdot \pol_4 = 0\cr
&k_3\cdot \pol_1 = - k_2 \cdot \pol_1 \; , \hskip 1.2 cm
k_4\cdot \pol_2 = -k_3 \cdot \pol_2 \; ,  \cr
&k_4\cdot \pol_3 = - k_2 \cdot \pol_3 \; , \hskip 1.2 cm
k_3\cdot \pol_4 = - k_2 \cdot \pol_4 \; . \cr}
\eqn\HelicitySimplifications
$$
The reason for using the spinor helicity method is now evident;
many of the dot products of polarization vectors amongst themselves and with
the external momenta simply vanish.  Since an amplitude consists
of sums of products of these dot products, with the spinor helicity method
many of the terms in an amplitude will also vanish with a
judicious choice of the reference momenta.

\vskip .6 cm
\centerline{\epsfxsize 5.7 truein \epsfbox{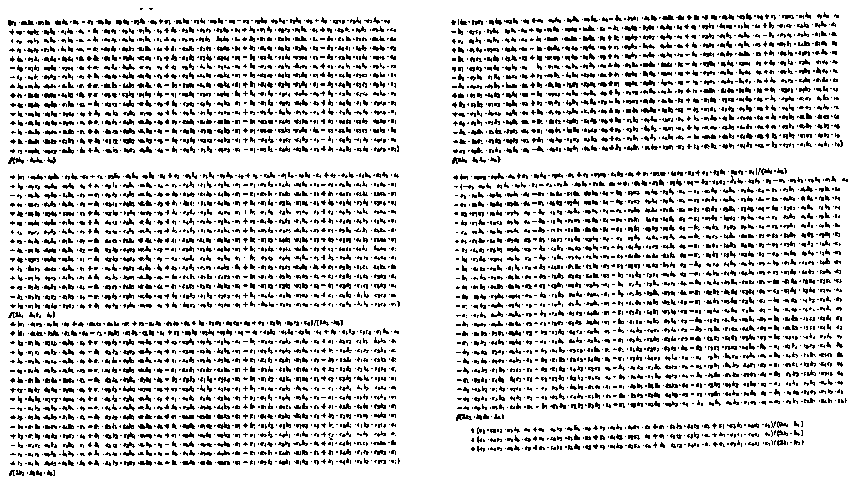}}
\nobreak
\vskip .2 cm \nobreak
{\baselineskip 10 pt \narrower\smallskip \noindent\ninerm {\ninebf
Fig.~\use\FiveGluonFig:} An unreadable form of the five-gluon tree
amplitude in terms of dot products of momentum and
polarization vectors to illustrate its complexity.
\smallskip}

\vskip .6 cm

The five-gluon tree amplitude provides a rather clear demonstration of the
power of the spinor helicity method.
In \fig\FiveGluonFig%
\footnote{$^\dagger$}{I thank D. Kosower for providing this figure.}
 the color ordered five-gluon tree
amplitude is presented in conventional but unreadable form to
illustrate the seeming complexity. The results in this figure
are written in terms
of dot products of polarization vectors and momenta.
Even if this figure were legible it would still be fairly painful to
use since one would need to square it and sum over helicities before
obtaining the cross-section.  Now consider the same expression using the
spinor helicity formalism:
$$
\eqalign{
& A_5^{\rm tree} (1^+, 2^+ , 3^+, 4^+, 5^+) = 0 \cr
& A_5^{\rm tree} (1^-, 2^+, 3^+, 4^+, 5^+) = 0 \cr
& A_5^{\rm tree} (1^+, \cdots, j^-, \cdots k^-, \cdots 5^+) =
i {\spa{j}.{k}^4 \over \spa1.2\spa2.3\spa3.4\spa4.5\spa5.1}  \cr}
\anoneqn
$$
where $j$ and $k$ are two negative helicity legs.  All other
configurations can be obtained from these basic ones by relabelings
and parity.
{}From this example it is clear that
a basic point behind using a helicity formalism is that results
which would otherwise need many pages can be expressed in a few lines.

As another example to illustrate the power of spinor helicity methods
consider the gluon helicity amplitudes $A_n^{\rm tree}(1^+, 2^+,
\cdots, n^+)$ and $A_n^{\rm tree}(1^-,2^+, \cdots, n^+)$.  Using
spinor helicity it is easy to argue that these amplitudes vanish
[\use\Private].  In the first case one chooses all the reference momenta
equal $q_i = q \not = k_j$, while in the second case one could choose
$q_1=k_2$ and $q_{i\not = 1} = k_1$. Thus choices exist for $A_n^{\rm
tree}(1^+, 2^+, \cdots, n^+)$ and $A_n^{\rm tree}(1^-,2^+, \cdots,
n^+)$ so that $\pol_i\c\pol_j=0$ for all $i,j$. It is then not
difficult to argue that all terms in a tree-level amplitude contain
at least one $\pol_i \c \pol_j$.  This can be obtained using
conventional Feynman diagrams in Feynman gauge.  First consider those
diagrams with only three-point vertices. These diagrams have $n$-legs
and $n-2$ vertices.  Since each vertex is linear in momenta (and there
are no other sources of momenta in the numerator in Feynman gauge) in
any given term at most $n-2$ momenta can contract with the
polarization vectors to form $\pol_i\c k_j$; this leaves two $\pol_i$
in every term which contract with one another to yield at least one
factor of $\pol_i\c \pol_j$.  The inclusion of four-point vertices
only helps this argument since a four-point vertex
implies an additional $\pol_i\c \pol_j$ in
every term.  Thus in one swoop infinitely many
tree-level Feynman diagrams have been evaluated with the result
$$
A_n^{\rm tree} (1^\pm, 2^+ , 3^+, \cdots, n^+) = 0 \; .
\eqn\SpinorZeroResult
$$
At loop level this argument does not work because diagrams exist
which have $n$ legs and $n$ vertices so all polarization vectors can
be simultaneously contracted into momenta; indeed, the corresponding
one-loop amplitudes do not vanish.

The other tree-level
helicity amplitudes do not vanish because reference momenta
cannot be chosen to make all $\pol_i\cdot \pol_j$ vanish
simultaneously.  There is
however, one other remarkably simple result and that is the Parke-Taylor
formula [\use\ParkeTaylor]
$$
A_n^{\rm tree}(1^+, 2^+, \cdots, j^-, \cdots, k^-, \cdots, n^+) =
i {\spa{j}.{k}^4 \over
\spa1.2 \spa2.3 \spa3.4 \cdots \spa n.1}
\anoneqn
$$
where all the legs are plus helicities except for legs $j$ and $k$
which are minus helicities.
This formula has been proven with the
Berends-Giele recursion relations [\use\Recursive,\use\LightconeRecurrence].

\vskip .2 cm
\noindent
{\it \use\TreeSection.3 Supersymmetry Identities}
\vskip .1 cm

Supersymmetry relates bosonic amplitude to fermionic ones.  Besides
the usual applications to model building, supersymmetry is a useful
computational tool in QCD calculations.  Although QCD is not a
supersymmetric theory, the supersymmetry identities [\use\Grisaru] do
provide general relationships between bosonic and fermionic
amplitudes.  Once diagrams containing either bosons or fermions are
calculated, information about the other case can be deduced from the
identities.  These relationships can then be applied to QCD
[\use\Susy] as we briefly review here.  We will follow the notation
and discussion of ref.~[\use\ManganoReview] in order to obtain
identities that will be useful later in these lecture notes.

The supersymmetry transformation turns bosons into fermions
and is given by
$$
[Q(\eta) , g^{\pm} (p)] = \mp \Gammait^{\pm}(p, \eta) \Lambdait^{\pm}(p)\; ,
\hskip 1.5 cm [Q(\eta), \Lambdait^{\pm}(p)] = \mp \Gammait^{\mp} (p, \eta)
g ^{\pm}(p) \; .
\anoneqn
$$
The supercharge is $Q(\eta)$ where $\eta$ is an arbitrary
anti-commuting parameter. The gluon field is $g$ while the
fermion gluino field is $\Lambdait$ and the $\pm$ superscripts denote
the helicity.   The coefficients $\Gammait$ are given by
$$
\Gammait^+ (p, \eta) = [\Gammait^- (p, \eta)]^* = {\bar \eta} u_- (p)
\anoneqn
$$
where $u_-(p)$ is a negative helicity spinor satisfying the massless
Dirac equation.  A convenient choice of the anticommuting parameter
is $ \bar\eta = \theta \bar u_+(k)$
where $\theta$ is a Grassmann parameter and
$k$ is an arbitrary null vector so that
$$
\Gammait^+ (p, \eta) =  \theta \langle k^+ | p^- \rangle = \theta \, [kp] \; .
\anoneqn
$$
The last expression is in terms of the compact spinor helicity notation.

Supersymmetry identities are obtained by using the
fact that in a supersymmetric theory the supercharge $Q$ annihilates the
vacuum [\use\BasicSusy].
The basic supersymmetric identity is then
$$
0 = \langle [Q, \prod_{i=1}^n \phi_i ] \rangle_0 =
\sum_{i=1}^n \langle \phi_1 \cdots [Q, \phi_i] \cdots \phi_n \rangle_0
\anoneqn
$$
where $\langle \cdots \rangle_0$ means the vacuum expectation value.

{}From here the specific supersymmetry identities can be derived.
Consider for example
$$
\eqalign{
0 & = \langle [ Q, \Lambdait_1^+ g_2^- g_3^+ \cdots g_n^+] \rangle_0 \cr
& = - \Gammait^- (p_1, k) A(g_1^+ , g_2^- , g_3^+\cdots g_n^+) -
\Gammait^- (p_2, k) A(\Lambdait_1^+ , \Lambdait_2^- , g_3^+,
\cdots,  g_n^+) \cr
\null & \hskip 1 cm
+ \sum_{i=3}^n \Gammait^+(p_i, k) A(\Lambdait_1^+, g_2^- , g_3^+, \cdots,
\Lambdait_i^+, \cdots g_n^+) \cr }
\anoneqn
$$
where $k$ is an arbitrary null momentum vector.
Since the gluon-fermion vertex conserves fermion helicity, all amplitudes
with like helicity (outgoing) fermions vanish so the third term containing
the sum drops out.
(Note that the notation is such that an incoming `$+$' has opposite
helicity as an outgoing `$+$'.)
Thus we obtain
$$
\eqalign{
0 & = \langle [ Q, \Lambdait_1^+ g_2^- g_3^+ \cdots g_n^+] \rangle_0 \cr
& = - \Gammait^- (p_1, k) A(g_1^+ , g_2^- , g_3^+\cdots g_n^+) -
\Gammait^- (p_2, k) A(\Lambdait_1^+ , \Lambdait_2^- , g_3^+,
\cdots,  g_n^+) \; . \cr }
\anoneqn
$$
By choosing $k=p_1$ or $k=p_2$ the two identities
$$
A_n^{\rm susy}(g_1^+ , g_2^- , g_3^+, \cdots g_n^+) =0
\eqn\SusyOneMinus
$$
and
$$
A_n^{\rm susy} (\Lambdait_1^+ , \Lambdait_2^- , g_3^+, \cdots,  g_n^+) = 0
\anoneqn
$$
are obtained.
Thus, in any space-time supersymmetric theory these amplitudes vanish
to all loop orders.
Other examples of supersymmetry identities are
$$
A_n^{\rm susy} (g_1^+, g_2^+ , \cdots , g_n^+) = 0
\eqn\SusyAllPlus
$$
and
$$
A_n^{\rm susy} (g_1^-, g_2^- , g_3^+, \cdots , g_n^+)
= {\spa1.2 \over \spa1.3}
A_n^{\rm susy}(g_1^-, \Lambdait_2^-, \Lambdait_3^+, g_4^+, \cdots, g_n^+) \; .
\eqn\SusyTwoMinus
$$

These identities can be immediately applied to tree-level QCD
computations.  At tree level, the $n$-gluon amplitudes are completely
independent of the matter content of a particular theory since by fermion
number conservation, fermion or scalar lines can never appear inside an
$n$-gluon diagram.  This means that the supersymmetry identities
(\use\SusyOneMinus) and (\use\SusyAllPlus) imply that
$$
A_n^{\rm qcd\ tree} (g_1^\pm, g_2^+ , \cdots , g_n^+) = 0 \; .
\anoneqn
$$
This agrees with the result obtained in eq.~(\use\SpinorZeroResult) through
spinor helicity methods.

A more complete discussion of applications of supersymmetry
identities to tree-level QCD computations can be found in
ref.~[\use\ManganoReview].

\section{Basic String Theory}
\tagsection\StringSection

The infinite tension limit of a string theory is a field theory
[\use\Scherk,\use\GSB,\use\Minahan].  In
order to use string theory as a computational tool, control of the
massless matter content of the string theory is required, because
colored massless matter particles can run around the loops.  It is
possible to build consistent heterotic string theories
[\use\Heterotic] whose infinite-tension limit is a non-abelian gauge
theory where one of the factors is an $SU(N)$ with no matter fields
[\use\Beta].  The technology needed for such a construction is the one
used to construct four-dimensional string models [\use\KLT]; the
formulation of Kawai, Lewellen and Tye is particularly simple,
although any of the other formulations can be used depending on one's
taste.  In the original derivation of the string-based rules
[\use\Long,\use\Pascos], it was essential to use a consistent string
in order to prevent extraneous problems from entering.  Without full
string consistency there would be no guarantee that the final results
obtained would be correct.  A heterotic string was used in the
original derivation of the string-based rules because bosonic
strings always contain unwanted massless scalars and tachyons, while
four-dimensional type~II [\use\SchwarzReview,\use\typeII]
and type~I [\use\OpenString] superstrings do not have a rich enough
variety of fully consistent models.

However, given that a consistent heterotic string derivation as
well as a conventional field theory understanding [\use\Mapping] now exists,
there is no longer a need to build fully consistent
strings as one can verify results either by comparing to the heterotic
construction or to field theory.  It turns out that any string
model will suffice; if the gauge group representation is not correct
or the number of flavors is not the desired one this can be fixed by
hand in the field theory limit.  The important information that string
theory supplies is the compact structure of the amplitude.

Bosonic string constructions are generally much simpler than super or
heterotic string constructions so that is what will be discussed here.
The open bosonic string discussed here is identical to the one used by
Metsaev and Tseytlin [\use\Tseytlin] to obtain the Yang-Mills
$\beta$-function from string theory.  This string is given by a naive
truncation of an oriented open bosonic string to four-dimensions.  In
this way all massless colored scalars arising from the dimensional
compactification are simply thrown away.  This string is inconsistent
as a fundamental string theory because of the naive truncation of the
spectrum.  Another technicality is that the string does contain a
tachyon, which might be worrisome; however, one can handle this with
the prescription that exponentially large terms due to the tachyon
should be dropped in the same way that exponentially small terms from
the higher mass states are dropped.  These potential difficulties are
of no concern in the field theory limit where the correctness of the
final results can be independently verified.  What is important here
is the basic structure that emerges from string theory without facing
the full technicalities of heterotic string constructions.

In general, an amplitude in string theory is evaluated by performing
the Polyakov surface integral [\use\Polyakov]
$$
A_n  \sim \int DX \exp\Bigl[ {1\over \alpha'}
\int d^2 \nu \; \partial_\alpha X^\mu \partial_\alpha X_\mu \Bigr]
V_1 V_2 \cdots V_n
\anoneqn
$$
where the $V_i \sim \pol_i \c \partial X e^{ik\cdot X}$ are the vertex
operators for external gluons.  At one-loop this path integral is
performed on a world-sheet annulus.
Since the world-sheet bosons are free,  Wick's theorem can be used
to evaluate the string $n$-gluon amplitude in terms of the two-point
correlation on the annulus
$$
\eqalign{
\langle X_{\mu}(\nu_1) X^{\nu}(\nu_2) \rangle
& = \delta_\mu{}^\nu G_B(\n 12) = - \delta_\mu{}^\nu
\Bigl[ \log\ |{2\sinh(\n 12) } | - { (\n12)^2 \over\tau }
- 4 q \sinh^2 (\n12) \Bigr] \cr
\null & \hskip 1cm
 + \Ord(q^2) \cr }
\anoneqn
$$
where $\tau=-\log(q)/2$ is the real modular parameter of the annulus,
$\nu_i$ represents the location of the vertex operator on the annulus
and $\n ij = \nu_i - \nu_j$.
(These parameters are $\pi/i$ times the conventional one in
refs.~[\use\SchwarzReview,\use\GSW].)
As discussed in ref.~[\use\Long], in the field theory limit
these parameters are proportional to
sums of Schwinger proper time parameters.
A repeated application of Wick's theorem to evaluate the surface integral
yields the string partial amplitude
$$
\eqalign{
A_{n;1} = & i {(4\pi)^{\eps/2} \over 16 \pi^2}
(\sqrt{2})^n (\alpha')^{n/2 - 2}
\int_0^\infty d\tau  \int \prod_{i=1}^{n-1} d \nu_i\theta(\nu_i-\nu_{i+1})\;
{\tau}^{-2+\eps/2} Z \cr
& \null \hskip .5 cm \times
\prod_{i<j}^n \exp \biggl\{
\alpha' k_i\c k_j G_B(\n ij)  +
\sqrt{\alpha'}(k_i\c\pol_j - k_j\c\pol_i ) \, \Gbd(\n ij) \cr
\null & \hskip 3 cm
- \pol_i\c\pol_j\, \Gbdd(\n ij) \biggr\}
\biggr|_{\rm multi-linear}
\cr}
\eqn\StringAmplitude
$$
where
$$
\Gbd(\nu) = {1\over 2} {\partial \over \partial \nu} G_B(\nu) \; ,
\hskip 2 cm
\Gbdd(\nu) = {1\over 4} {\partial^2 \over \partial \nu^2} G_B(\nu)
\anoneqn
$$
and $\nu_n$ is fixed at $\tau$.  The `multi-linear' signifies that
after expanding the exponential only terms which are linear in all $n$
polarizations vectors are to be kept. The
string oscillator contributions to the partition function are
$$
Z =  q^{-1}
\prod_{n=1}^\infty(1 - q^n)^{-2(1-\delta_R\eps/2)} \; .
\eqn\StringPartFunc
$$
Full consistency of the string demands that the dimension
$D=26$ [\use\Lovelace], but for the purposes of obtaining
field theory amplitudes $D=4-\eps$ where $\eps$ is the dimensional
regularization parameter necessary to handle infrared divergences; the
regularization parameter $\delta_R$, included in the string partition
function, determines the precise form of the regularization
[\use\Long]. (The regularization issues
will be discussed further in Sections~\use\TreeLoopSection\ and
\use\FieldTheorySection.)
In order to obtain a sensible field
theory limit, the leading $q^{-1}$ has been maintained by hand independent
of the number of dimensions.  (A fully consistent heterotic string
such as the one used in ref.~[\use\Long]
does not require any adjustments, such as this one.)
The field theory limit of the amplitude (\use\StringAmplitude) yields the pure
Yang-Mills contributions to the amplitude including Faddeev-Popov
ghosts.  The conventions have been adjusted so that in the field
theory limit the number of $\pi$'s and 2's which need to be shuffled
around are minimized.

Partial amplitudes associated with two color traces
are a bit different since the string vertex operators
are located on both boundaries of the annulus; examples can be
found in chapter 8 of ref.~[\use\GSW].

In order to take the infinite string tension limit $\alpha'
\rightarrow 0$ of the string amplitude (\use\StringAmplitude), it is
convenient to first integrate by parts on the string world-sheet in
order to remove all $\Gbdd$ from the kinematic factor
[\use\Long,\use\Pascos].  (The analysis of the field theory limit can
also be performed without the integration-by-parts step
[\use\Future] so it should not be taken as an essential ingredient to
the string-based method.)  In open string theory there are potential
boundary terms, but these can be removed by an appropriate analytic
continuation in external momenta since all the boundary terms contain
a factor of $|\nu_i -
\nu_j|^{-n - \alpha' k_i \c k_j} |_{\nu_i\rightarrow\nu_j} = 0$.
(One technicality is that the periodicity on the annulus under $\nu
\rightarrow \nu+\tau$ must be used to remove some of the surface
terms.)

As an example of the integration by parts procedure
consider the following term in three-point function
$$
\hskip - .2 cm
\eqalign{
& \int \prod_{i} d \nu_i \; \Gbdd(\nb12) \Gbd(\nb23 )
\exp\Bigl[\alpha' (k_1 \c k_2 G_B(\n12) + k_2 \c k_3 G_B(\n23)
+ k_1 \c k_3 G_B(\n13)) \Bigr] \cr
& \hskip 1 cm
\longrightarrow -\alpha'
\int \prod_{i} d \nu_i \; \Gbd(\nb12)\Gbd(\nb23 )\Bigl(k_1\c k_2 \Gbd(\n12)
+k_1\c k_3 \Gbd(\nb 13) \Bigr) \cr
& \hskip 2 cm \times
\exp\Bigl[\alpha' (k_1 \c k_2 G_B(\n12) + k_2 \c k_3 G_B(\n23)
+ k_1 \c k_3 G_B(\n13)) \Bigr]  \cr }
\eqn\IBPExample
$$
where the integration by parts was performed with respect to $\nu_1$.
In appendix B of ref.~[\use\Color] it was proven that
all $\Gbdd$'s can always be eliminated from the kinematic function,
by appropriate integration by parts.

In the field theory limit, the contributions to an integrated-by-parts
one-loop amplitude
can be classified in terms
of tree and loop parts.   The
tree parts are obtained by first extracting the massless
poles in the $S$-matrix before taking the field theory limit of the loop.
Examples of these kinematic poles are found in the regions where
$\nu_i \rightarrow \nu_j$ and are of the form
$$
\int d \nu_i {1\over \nu_{ij}^{1+ \alpha' k_i \c k_j}} \longrightarrow
- {1 \over \alpha' k_i \c k_j}  \hskip 1.5 cm (\alpha' \rightarrow 0) \; .
\anoneqn
$$
In general, the kinematic poles extracted in this way correspond to the poles
of a scalar $\phi^3$ diagram.

After kinematic poles have been extracted, the field theory limit of the
loop is needed.
This is obtained by taking $\tau, |\nu| \rightarrow \infty$ which corresponds
to squeezing the annulus down to a field theory loop.
The values of the Green functions in this limit are
$$
\eqalign{
\exp(G_B(\nu)) &\rightarrow
\exp\Bigl({\nu^2\over\tau}-|\nu| \Bigr)
\times \hbox{constant} \cr
\Gbd(\nu) & \rightarrow {\nu \over \tau} - \sign(\nu) (
{\textstyle{1\over 2}}+ e^{-2|\nu|} - q e^{2|\nu|}  ) \; .  \cr}
\anoneqn
$$
The exponentiated bosonic Green function was not expanded
beyond $\Ord(q^0)$; after carrying out the integration by parts procedure
the higher order terms do not contribute
since they carry too many explicit powers of $\alpha'$.  For
$\Gbd$, terms through $\Ord(q)$ should be kept due to the presence
of the overall $q^{-1}$ in the string amplitude (\StringAmplitude).

In the field theory limit two types of loop contributions are obtained
depending on whether a power of $q$ is extracted from the string
partition function or from the Green functions.  For the former
contribution one simply keeps the leading order contributions from the
bosonic Green functions. This type of contribution is described by the
bosonic zero-mode [\use\GSW] or loop momentum integral of the string
[\use\Mapping].
A product of $\Gbd$'s contains exponentially
growing and decaying terms as well as terms which are constant.
In general, when terms proportional to $q= e^{-2 \tau}$ are extracted
from a product of $\Gbd$ in order to cancel the overall $q^{-1}$,
a factor of the form
$$
\exp \Bigl[\Bigl( |x_k - x_l| - \sum |x_i - x_j| \Bigr) \tau \Bigr]
\anoneqn
$$
is obtained where $\x_i\equiv \nu_i / \tau$.
In order to avoid exponential suppression or growth as $\alpha' \rightarrow 0$
the sum must add up to cancel within the exponential exactly.
This will happen only if each $x_i$ which appears with a positive sign
also appears with a negative sign after expressing the absolute
values in terms of the $x_i$s directly.
The correct prescription for
dealing with exponentially growing terms due to the tachyon
is to simply drop them
in the same way that exponentially decaying terms are dropped. (The
exponential growth is an artifact of the Schwinger proper
time representation of tachyonic propagators.)

The result of collecting those terms where the exponential
terms completely cancel is that only those which form a {\it cycle}
of $\Gbd$'s, defined to be a product of $\Gbd$'s
with indices arranged in the form
$$
\Gbd(\n {i_1}{i_2}) \Gbd(\n {i_2}{i_3})  \cdots \Gbd(\n {i_m} {i_1}) \; ,
\anoneqn
$$
will not vanish.
Furthermore, the cyclic ordering of the indices must follow the same
ordering of the corresponding legs in the partial amplitude.

The superstring works in pretty much the same manner, except there are
now fermionic fields on the world-sheet.   A superstring is essential
in order to be able to include space-time fermions into the
string formalism.  Although all superstrings must maintain
world-sheet supersymmetry for their consistency, such
strings are not necessarily space-time supersymmetric.  In particular,
the string models of interest which have QCD-like spectra are not
space-time supersymmetric.

In the superstring the vertex operator is of the form
$$
V \sim \pol \c (\partial X + i \psi k \c \psi) e^{ik\c X}
\eqn\SuperVert
$$
where $X$ is the same world sheet bosonic field as in the bosonic string
and $\psi$ is a free fermionic field.  As for the bosonic string
these vertex operators are inserted into the Polyakov path integral
$$
A_{n}^{\rm superstring}\L \{k_i, \pol_i\}\R \sim
\int [DX][D\psi] \exp\LB-S\RB\,
 V\L k_1,\pol_1\R\cdots V\L k_n,\pol_n\R
\anoneqn
$$
in order to obtain amplitudes.  The contributions of the world-sheet
fermions can be computed by noting that they are free fields so
that Wick's theorem
can be used.  In this way, any product of fermion fields can be
evaluated from the basic two-point correlation function
$$
\langle \psi^\mu (\nu_1)  \psi^\sigma (\nu_2) \rangle_\beta^\alpha
= \delta^{\mu\sigma} \Gf{\alpha}{\beta} (\nu_1 - \nu_2)
\anoneqn
$$
where $\alpha$ and $\beta$ refer to the particular world sheet boundary
conditions. One of the features controlled by these boundary conditions
is whether the particles in the loop are bosons or fermions [\use\GSW].

It is a simple matter to verify from Wick's theorem and the vertex
operator (\use\SuperVert) that the fermionic
Green functions always arrange themselves into cycles.  For example
$$
\langle \psi^{\mu_1} (\nu_1)  \psi^{\sigma_1} (\nu_1)
\psi^{\mu_2} (\nu_2) \psi^{\sigma_2} (\nu_2)
\psi^{\mu_3} (\nu_3) \psi^{\sigma_3} (\nu_3)
\rangle \sim
G_F(\n12) G_F(\n23) G_F(\n31)
\anoneqn
$$
exhibits the cycle structure. These cycles are analogous to the
cycles of bosonic Green functions discussed above.

Since world-sheet supersymmetry in a superstring relates $\psi^\mu$ to
$X^\mu$ it turns out that it is possible to obtain the contributions
from the world-sheet fermions from the world-sheet bosons
[\use\Bosonic].  For an appropriate integration by parts the fermion
Green functions $G_F$ of a superstring satisfy the constraint that
after an appropriate integration by parts the superstring kinematic
expression vanishes after substituting $G_F^{i,j} \rightarrow
-\Gbd^{i,j}$.  In this way a precise match between the $G_F$
cycles and $\Gbd$ cycles can be made.
A relation between the bosonic and fermionic Green function
contributions to the amplitude is, of course, no surprise since this
is precisely the role of world-sheet supersymmetry.  (The relationship
between the $\Gbd$ and $G_F$ terms follows from the cancellation of
spurious $F_1$ formalism [\use\GSW] tachyon poles; in particular the
tadpole diagram with all legs pinched together should not have a
tachyon pole, so the pole contribution generally is the form of a
total derivative.  An appropriate integration-by-parts to remove the
total derivative then makes the matchup between $G_F$ and $\Gbd$ terms
manifest.)  By matching up the world-sheet fermions to the world-sheet
bosons in this way one can show that the results obtained from a
bosonic string match those of a superstring.  This is of particular
interest for the case where space-time fermions circulate in the loop
since this cannot be obtained directly from a bosonic string.  The
trick is thus to include $G_F$ superstring contributions as additional
$\Gbd$ contributions.  In this way rules can be constructed which
contain space-time fermions in the loop, but are based on the simpler
bosonic string kinematic master formula.  Rules obtained in this way
are presented in Section~\use\RulesSection .

\section{Tree Level Methods at Loop Level}
\tagsection\TreeLoopSection

The tree-level methods discussed in Section~\use\TreeSection\ carry over
to loop level; there are, however,
a number of differences between the situation
at loop and tree level as we discuss in this section.

\vskip .3 cm
\noindent
{\it {\use\TreeLoopSection.1 The Color Decomposition}}
\vskip .1 cm

\def\kr#1{k_{\rho(#1)}}

A detailed discussion of the one-loop color decomposition has been given
in ref.~[\use\Color].
The major difference between the tree  color decomposition
and one-loop decomposition is that at one
loop up to two color traces can appear in a given term.
In particular, the
$SU(N)$ four-point gluon amplitude can be written in the form,
\chardef\hyphen=45
$$
\eqalign{
\A{4}^{\rm one\hyphen loop} = & g^4
\sum_{\sigma \in S_4/Z_4}
N \Tr(T^{a_{\si(1)}}T^{a_{\si(2)}}T^{a_{\si(3)}}T^{a_\si{(4)}})
A_{4;1} (\si(1), \si(2), \si(3), \si(4)) \cr
& \null + \sum_{\sigma \in S_4/Z_2^3} \Tr(T^{a_{\si(1)}}T^{a_{\si(2)}})
\Tr(T^{a_{\si(3)}}T^{a_{\si(4)}})
A_{4;3} (\si(1), \si(2); \si(3), \si(4)) \; . \cr}
\eqn\FourAmplitude
$$
\def\fourperm#1#2#3#4{(#1\,#2\,#3\,#4)}%
The notation `$S_4/Z_4$' denotes the set of all permutations $S_4$ of
four objects, omitting the purely cyclic transformations
$(1\rightarrow2,2\rightarrow3,3\rightarrow4,4\rightarrow1)$, etc.
The notation `$S_4/Z_2^3$' refers again to the set of permutations
of four objects but with permutations considered equivalent (and
only one representative picked) if they exchange labels within a
single trace or exchange the two traces:
$S_4/Z_2^3 = \{\fourperm1234,\fourperm1324,\fourperm1423\}$.

More generally, the one-loop color decomposition for adjoint representation
states is given by
$$
\A{n}^{\rm 1\hyphen loop} = g^n
\sum_{j=1}^{\floorHalfN+1}
\sum_{\rho\in S_n/S_{n;j}} \Gr_{n;j}\L\rho(1),\ldots,\rho(n)\R
A_{n;j}\L\kr1,\pol_{\rho(1)};\ldots;\kr{n},\pol_{\rho(n)} \R
\eqn\LoopColorDecomposition
$$
\def\phstar{{\phantom{*}}}%
\def\LO{\rm LO}%
\def\NLO{\rm NLO}%
\def\Atree{A^{{\rm tree}\,*}}%
where $S_n/Z_n$ is the set of non-cyclic
permutations of $\{1,\ldots, n\}$; $\Gr_{n;j}$ denote the double-trace
structures
$$\eqalign{
\Gr_{n;1}\L 1,\ldots,n\R &= \Tr(1)\Tr(T^{a_1}\ldots T^{a_n})\cr
&= N\Tr(T^{a_1}\ldots T^{a_n})\cr
\Gr_{n;j}\L 1,\ldots,n\R &= \Tr(T^{a_1}\ldots T^{a_{j-1}})
\Tr(T^{a_j}\ldots T^{a_n}),\cr
}\anoneqn$$
and $S_{n;j}$ is the subset of the permutation group $S_n$ that leaves
the trace structure $\Gr_{n;j}$ invariant.  ($S_{n;1}$ is just the
set of cyclic permutations of $n$ objects, $Z_n$.)
For pure-glue amplitudes
in $SU(N)$, the partial amplitude
$A_{n;2}$ drops out
since its coefficient includes a trace over a
single $SU(N)$ generator, which vanishes identically.

The contribution from
fundamental representation states is a bit simpler and can be obtained
from the same partial amplitudes which were used for adjoint states
and is given by
$$
{\cal A}^{\rm fund}_n (\{ a_i, k_i, \pol_i\}) =
g^n \sum_{\sigma\in S_n/S_{n;1}} \Tr(T^{\as1} \cdots T^{\as{n}} )
A_{n;1}(\ks1,\ps1;\ldots;\ks{n},\ps{n}) \; .
\eqn\FundamentalColor
$$

The color decomposition (\use\LoopColorDecomposition) can be
heuristically understood from the open
bosonic string [\use\ChanPaton].
At one loop, the schematic form of the $n$-point amplitude
is
$$
\eqalign{
{\cal A}_n^{\rm string} &(\{ a_i, k_i, \eps_i\}) = \cr
&
\sum_{\sigma\in S_n/Z_n} N \Tr(T^{\as1} \cdots T^{\as{n}} )
A_1^{\rm string}(\ks1,\ps1;\ldots;\ks{n},\ps{n})\cr
\null &
+\sum_m \sum_{\sigma\in S_n/Z_m\times Z_{n-m}}
\hskip -2mm\Tr(T^{\as1} \cdots T^{\as{m}} )
\Tr(T^{\as{m+1}} \cdots T^{\as{n}} )\cr
\null & \hskip 2 cm  \times
A_2^{\rm string}(\ks1,\ps1;\ldots;\ks{n},\ps{n})
+{\cal O} (\alpha') \cr}
\eqn\OneloopDecomp
$$
where the first term appears when all gluons are attached
to a single string boundary as depicted in \fig\OpenStringFig{a}, while
the second term appears when
gluons are attached to both string boundaries as depicted in
fig.~\use\OpenStringFig{b}.
The higher order corrections in the inverse string tension $\alpha'$
(which do contain
terms with three or more non-trivial traces) arise from graviton
exchange.
Such contributions disappear in the gauge theory (or infinite-tension)
limit where the coupling to gravitons and other colorless states vanishes.

\break

\vskip .3 cm
\centerline{\epsfxsize 7. truecm \epsfbox{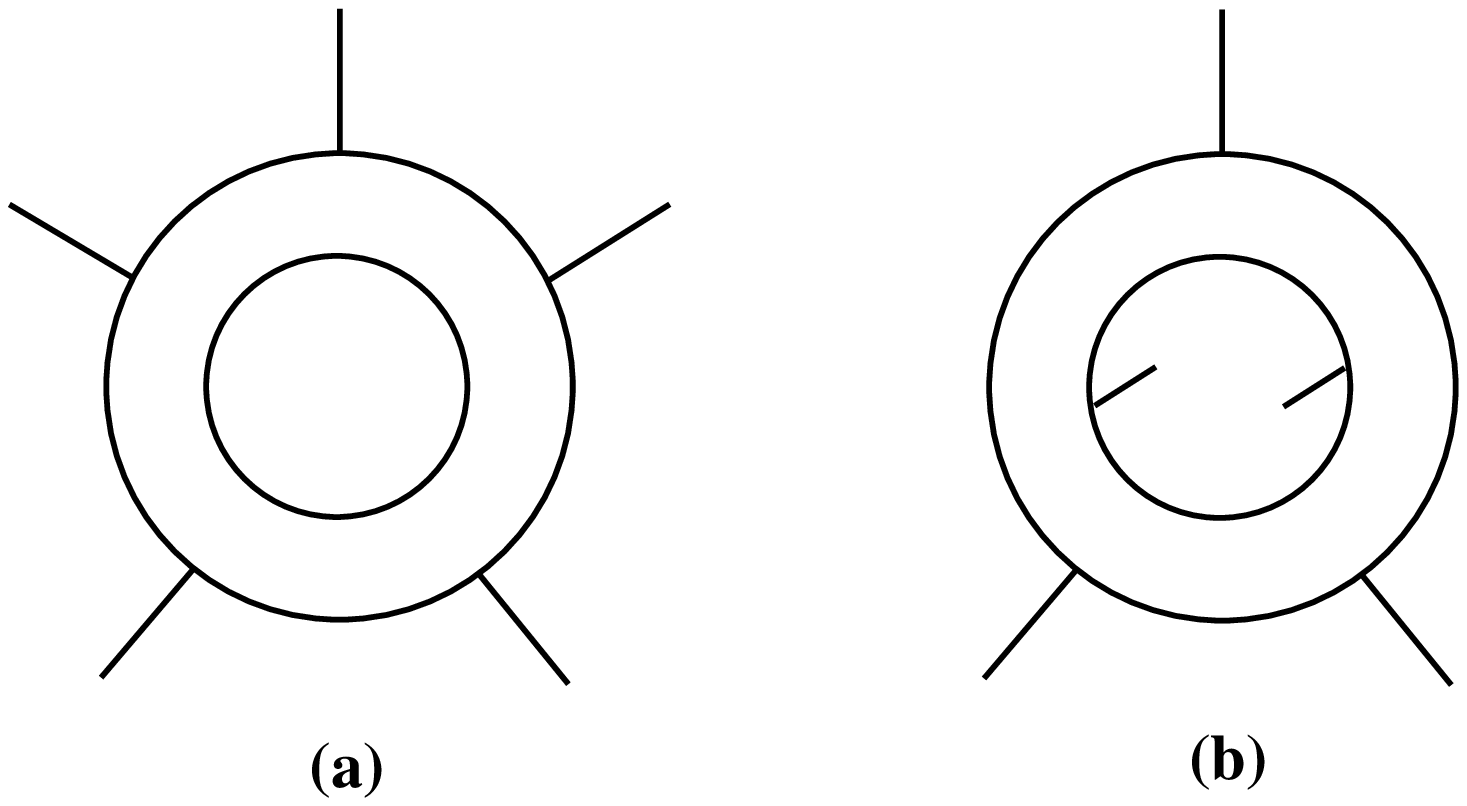} }
\nobreak
\vskip .1 cm
{\baselineskip 10 pt\narrower\smallskip\noindent\ninerm\nobreak
{\ninebf Fig.~\OpenStringFig:} The open string diagrams.
When the vertex operators are all attached to the same boundary
as in (a) a single color trace is obtained while if the vertex operators
are attached to both boundaries as in (b)
a product of two color traces is obtained.
\smallskip}

\vskip .6 cm

The one-loop partial amplitudes $A_{n;j}$ have properties analogous
to those of their tree-level counterparts: they are gauge-invariant
on-shell, satisfy a symmetry equation,
$$
\forall \sigma \in S_{n;j},\qquad
A_{n;j}(\si(1),\ldots,\si(n)) = A_{n;j}(1,\ldots, n)
\anoneqn$$
and a reflection identity,
$$
A_{n;j}(R_{n;j}(1,\ldots,n))
= (-1)^n A_{n;j}(1,\ldots,n)
\anoneqn$$
where
$$
R_{n;j}(i_1,\ldots,i_n)
= (i_{j-1},\ldots,i_1, i_n,\ldots,i_j) \; .
\anoneqn$$

In addition the partial amplitudes
satisfy a set of $U(1)$ decoupling equations which were
discussed in detail in ref.~[\use\Color].  In the case
of the four-point
function, these take the form
$$\eqalign{
A_{4;3} (1,2,3,4) &=
\sum_{\si\in S_4/Z_4} A_{4;1} (\si(1), \si(2),\si(3), \si(4) ) \; \cr
A_{4;2} (1,2,3,4) &=
-\sum_{\si\in Z_3\{2,3,4\}} A_{4;1} (1, \si(2),\si(3), \si(4) ) \; \cr
&= -{1\over2} A_{4;3}(1,2,3,4)\; , \cr
}\eqn\FourPointLoopDecoupling$$
so all information about the four-point amplitude is contained in
$A_{4;1}$.  It turns out that the decoupling equations also
imply that $A_{5;1}$ contains all information, but for larger numbers of legs
other $A_{n;j}$ besides $A_{n;1}$ are required.

Using the decoupling equations, one can simplify the color-summed
next-to-leading correction to the four-gluon process,
$$\eqalign{
\sum_{\rm colors} \LB\A{4}^*\A{4}^\phstar\RB_{\NLO}
&= 2 g^{6} N^{3}\L N^2 -1\R\Re\sum_{\si\in S_4/Z_4}
\Atree_4(\si) A_{4;1}(\si)\cr
}\eqn\FourPointNLO$$
where we have abbreviated $A_{n;j}(\si(1),\ldots,\si(n))$ by
$A_{n;j}(\si)$.
The leading-order result has the form
$$
\eqalign{
\sum_{\rm colors} \LB\A{4}^*\A{4}^\phstar\RB_{\LO}
&= g^{4} N^{2}\L N^2 -1\R\sum_{\si\in S_4/Z_4}
\LV A^{\rm tree}_4(\si)\RV^2\; .\cr
}\anoneqn
$$
The explicit form of the five-point result can be found in ref.~[\use\Color].

\vskip .3 cm
\noindent
{\it \use\TreeLoopSection.2 Spinor Helicity Techniques}
\vskip .1 cm

One difference between the tree-level and loop-level use of spinor
helicity methods is the appearance of loop momentum.  Since spinor
helicity methods require on-shell momenta, and loop momentum is an
off-shell quantity, until it is integrated out the spinor helicity
method cannot be used to its full power.  As discussed in Section
\use\FieldTheorySection, when following the string-based
organization, integrating out the loop momentum in the string
reorganized form of the $n$-gluon amplitude is an easy step, just as
it is in string theory.  For other amplitudes a direct term-by-term
integration of the loop momentum to obtain a Feynman parametrized form
can be performed.  One systematic approach for integrating out loop
momentum is with the electric circuit analogy [\use\BjD,\use\Lam].
Once the loop momentum is integrated out the full power of the spinor
helicity method can then be used to simplify expressions.

A more significant difference between tree- and loop-level
use of spinor helicity is the apparent incompatibility of spinor helicity
with conventional dimensional regularization (CDR) [\use\Korner,\use\Long].
Dimensional
regularization is by far the most convenient regularization scheme
in practical calculations so this issue must be addressed before spinor
helicity can be used at loop level. The
problem is that spinor helicity
inherently assumes that the
gluon polarization vectors are in four dimensions.  This is in
conflict with the CDR scheme where the external polarization vectors
are continued to $4 - \eps$ dimensions.  However,
this can be repaired by introducing the notion of
`$\epsh$'-helicity [\use\KSpinor].
To do this in the
framework of the spinor helicity basis, one introduces
an additional $\epsh$-helicity, with the following rules
in $4-\eps$ dimensions,
$$
\eqalign{
k\cdot\pol^{(\epsh)}(k';q) &= 0\cr
\pol^{(\pm)}(k;q)\cdot\pol^{(\epsh)}(k';q') &= 0\cr
\pol_1^{(\epsh)}(k;q)\cdot\pol_2^{(\epsh)}(k';q) &=  -\delmeps^{i_1 i_2}\;.
}\anoneqn
$$
In the last expression, $i_1$ and $i_2$ run over the $-\eps$
additional dimensions; in squaring an amplitude (or forming an
interference), one must sum over these additional indices:
$$
\delmeps^{i_1 i_2}\delmeps^{i_2 i_3} = \delmeps^{i_1 i_3},\hskip 1cm
\delmeps^{i_1 i_2}\delmeps^{i_1 i_2} = -\eps\;.
\anoneqn$$
It is convenient to abbreviate $\delmeps^{i_1 i_2}$ to
$\delmeps^{12}$.
Although this scheme allows one to continue using the conventional
scheme (which is quite useful when comparing to previous results
obtained by more traditional methods), the introduction of additional
helicities causes a severe computational penalty in practical
calculations since there is a significant amount of work in calculating
the additional $[\eps]$-helicities.

A more efficient approach is instead to modify the dimensional
regularization scheme so that only plus and minus `observed'
helicities are considered, as advocated in ref.~[\use\Long].
`Observed' gluons are those which are neither virtual, collinear, nor
soft.  The basic idea of these schemes is to retain the polarization
vectors in four-dimensions.  All dimensional regularization schemes
entail continuing the momentum integrals (both the loop integrals and
the integrals over soft and collinear phase space) to $4-\eps$
dimensions in order to render them finite.  There are, however, a
number of versions of dimensional regularization, which differ in
their treatment of the polarization vectors (or helicities) of the
observed and unobserved particles:\hfil\break

\item{(a)} The `conventional' dimensional
regularization (\CDR) used,
for example,
by Ellis and Sexton [\use\Ellis] in which both observed and unobserved
gluon polarization vectors are continued to $4-\eps$ dimensions (so that
all gluons have $2-\eps$ helicity states); and

\item{(b)} the 't Hooft and Veltman scheme [\use\HV] in which
all polarization vectors of unobserved gluons
are continued to $4-\eps$
dimensions (so that unobserved gluons have $2-\eps$ helicity states),
but observed gluon polarizations are kept in four dimensions (so that
observed gluons have $2$ helicity states);

\item{(c)} a four-dimensional helicity scheme (\FDH) which
naturally arises when using the spinor helicity formalism.
In this scheme all helicities
(of both observed and unobserved particles) are treated in four dimensions
(so that all gluons have $2$ helicity states).

\noindent
The defining properties of the various
regularization schemes are summarized in Table 1.

\vskip -.2 truecm
\def\hs{\hskip 4mm}
$$
\vbox{\offinterlineskip
\halign{ &\vrule# & \strut\quad\hfil#\quad\cr
\omit& \omit&\omit&\omit &\omit&\multispan5\hrulefill\cr
\omit&\omit&\omit&\omit &height2pt&\omit&&\omit&&\omit&&\omit&\cr
\omit& \omit &\omit& \omit&& CDR &&
\lower5pt\vbox{\hbox{'t Hooft- \strut}
\setbox0=\hbox{'t Hooft- \strut}
\hbox to \wd0{\hfil Veltman\hfil}} &&
FDH &\cr
\omit&\omit&\omit&\omit&height 2pt&\omit&&\omit&&\omit&\cr
\omit&\multispan9\hrulefill&\cr
height2pt&\omit&&\omit&&\omit&&\omit&&\omit&\cr
& \smash{\lower2pt\hbox{Momentum}}\hfil
&& Unobserved particles && $4-\eps$ \hs &&  $4-\eps$ \hs
&&$4-\eps$\hs &\cr
height2pt&\omit&&\omit&&\omit&&\omit&&\omit&\cr
\omit&\omit &\omit&\multispan7\hrulefill&\cr
height2pt&\omit&&\omit&&\omit&&\omit&&\omit&\cr
& \smash{\raise2pt\hbox{components}}\hfil
&& Observed particles && $4-\eps$ \hs &&  $4$ \hs\hskip 2mm
&& $4$ \hs\hskip 2mm &\cr
height2pt&\omit&&\omit&&\omit&&\omit&&\omit&\cr
\omit&\multispan9\hrulefill&\cr
height2pt&\omit&&\omit&&\omit&&\omit&&\omit&\cr
& \vtop to 0pt{\hbox{\vbox to 7pt{}}\hbox{Helicities}\vss}\hfil
&& Unobserved particles && $2-\eps$ \hs && $2-\eps$ \hs
&& $2$ \hs\hskip 2mm &\cr
height2pt&\omit&&\omit&&\omit&&\omit&&\omit&\cr
\omit&\omit &\omit&\multispan7\hrulefill&\cr
height2pt&\omit&&\omit&&\omit&&\omit&&\omit&\cr
&  && Observed particles && $2-\eps$ \hs && $2$ \hs\hskip 2mm
&& $2$ \hs\hskip 2mm &\cr
height2pt&\omit&&\omit&&\omit&&\omit&&\omit&\cr
\omit&\multispan9\hrulefill&\cr
}}
$$
\vskip -.15 cm \nobreak
{\baselineskip 10 pt\narrower\smallskip\noindent\ninerm
{\ninebf Table 1:} Defining properties of the various dimensional
regularization schemes.
\smallskip}

\vskip .35  cm

The CDR scheme is conceptually the simplest one as all quantities are
uniformly continued to $4-\eps$ dimensions.  (Actually, as a practical
matter the observable external momenta can be effectively taken to be
four-dimensional by taking the momentum components in the $\eps$
dimensions to vanish in a given scattering process.)  In the 't
Hooft-Veltman scheme all observed polarizations are kept in
four-dimensions but the unobserved ones such as the virtual ones in
the loop are continued to $4-\eps$ dimensions.  In the 't
Hooft-Veltman scheme one must carefully distinguish between the
observed and unobserved particles (virtual, soft and collinear) in
order to prevent violations of the optical theorem.  The
four-dimensional helicity scheme involves retaining all states
uniformly in four-dimensions and therefore is a conceptually simpler scheme to
work with.  However, this scheme
has not been well studied beyond the one-loop four-gluon
amplitude.  In terms of computational complexity the CDR scheme is by
far the most complicated to work with when using spinor helicity
because of the additional
$\epsh$-helcities while the 't Hooft-Veltman and FDH schemes are of
comparable complexity.
The calculational differences between the three schemes
are summarized in Table~2.

%\vskip .2 truecm
\def\hs{\hskip .4 cm }
$$
\vbox{\offinterlineskip
\halign{ &\vrule# & \strut\quad\hfil#\quad\cr
\omit&\omit&\omit&\multispan5\hrulefill&\cr
\omit&\omit&height2pt&\omit&&\omit&&\omit&\cr
\omit&\omit && \FDH && 't Hooft-Veltman  && CDR &\cr
\omit&\omit&height2pt&\omit&&\omit&&\omit&\cr
\omit&\multispan7\hrulefill&\cr
height2pt&\omit&&\omit&&\omit&&\omit&\cr
& Continue loop momentum  && Yes \quad &&  Yes \hs
&&Yes \hs &\cr
height2pt&\omit&&\omit&&\omit&&\omit&\cr
\omit&\multispan7\hrulefill&\cr
height2pt&\omit&&\omit&&\omit&&\omit&\cr
& Remove $\eps$ bosonic states && No \quad && Yes \hs
&& Yes \hs &\cr
height2pt&\omit&&\omit&&\omit&&\omit&\cr
\omit&\multispan7\hrulefill&\cr
height2pt&\omit&&\omit&&\omit&&\omit&\cr
& $\pol_i^{(4)}  \rightarrow \pol_i^{(4-\eps)}$
&& No \quad && No \hs
&& Yes \hs &\cr
height2pt&\omit&&\omit&&\omit&&\omit&\cr
\omit&\multispan7\hrulefill&\cr
}}
$$
\nobreak
\vskip -.25 cm \nobreak
{\baselineskip 10 pt\narrower\smallskip\ninerm
\noindent {\ninebf Table 2:}
Modifications needed to construct various versions of
dimensional regularization from the unregularized
amplitude.\smallskip}

\vskip .4  cm

In field theory, with all these schemes one continues the loop
momentum integral from $D=4$ to $D=4-\eps$; this renders the integrals
finite.  The string theory equivalent of this analytic continuation is
obtained by shifting the overall factor in the integrand of
$\tau^{-2}$ to $\tau^{-2 + \eps/2}$ as was done in
eq.~(\use\StringPartFunc).  This change is of the same type as one
would obtain in field theory in a dimensionally regularized Schwinger
proper time formalism after integrating out the loop momentum.

The FDH scheme is similar, but not identical, to Siegel's
regularization by dimensional reduction [\use\Siegel]; in Siegel's
scheme the polarization vectors of the gluons are taken to be in
$D=4-\eps$ dimensions so they represent a total of $2 - \eps$ states,
but there are additional $\eps$-scalars that then brings the total
back to 2 states. The dimensional reduction scheme has been used
together with spinor helicity methods in ref.~[\use\Korner].  One nice
property of dimensional reduction (which is the purpose of the
scheme) is that it maintains space-time supersymmetry.  In
general, the FDH scheme can also be expected to maintain supersymmetry,
since it leaves the number of states at their four-dimensional
value [\use\Future].

A fundamental requirement on any of these schemes is that they
preserve gauge invariance.  String theory provides a useful tool for
quickly verifying the gauge invariance of the diagrams after
regularization.  In field theory the amplitude is described in terms
of a variety of diagrams; it is only their sum which is gauge
invariant.  In string theory (before the field theory limit is taken)
each partial amplitude is described in terms of a single diagram;
under the shift $\pol_i \rightarrow \pol_i + k_i$ this single diagram
should be invariant.  This makes the proof of gauge invariance in
string theory much simpler than in field theory.

In string theory, the substitution of the external momentum $k_i$ for
the corresponding polarization vector
$\pol_i$ formally leads to the vanishing of the
unregulated amplitude, because one obtains the integral
of a total derivative with vanishing boundary terms.
Start with the gluon string vertex operator
$$
V \sim
\; :\pol\c \partial_{\nu} X e^{ik \c X(\nu)} :
\anoneqn
$$
\vskip -.05 cm
\noindent
and set $\pol=k$; the vertex operator becomes
$$
\LP V\RV_{\pol=k} \sim \;
:\partial_{\nu} e^{ik \c X(\nu)} : \; .
\eqn\LongVertex
$$
\vskip -.05 cm
\noindent
If we now compute expectation values using this vertex operator
instead of the usual one for the first external gluon, we obtain an
integrand which is a total derivative; that is with $\pol_1$ replaced
by $k_1$, the integrand of the amplitude is a
total derivative in $\nu_1$.  As discussed above, the various
dimensional regularization schemes modify only the overall factor of
$\tau$, the number of states in the string partition function
and possibly the external polarization vectors; the important
point is that none of these changes alter the fact that the integrand
is a total derivative in $\nu_1$, because they do not affect the
structure of the Green functions.  As a result, the dimensional
regularization schemes do not alter the formal argument.  (There are a
number of subtleties regarding the vanishing of boundary terms, but a
more careful argument shows there are no difficulties [\use\Long].)

Although the FDH scheme maintains gauge invariance and has explicitly
been shown to give identical final results as CDR for the unpolarized
four-gluon amplitude, a complete proof of its consistency, especially
for the case of fermions is lacking.  However, because of the
enormous computational advantage obtained with spinor helicity methods
there is little doubt that regularization schemes (such as the 't
Hooft-Veltman or FDH schemes) which are compatible with spinor
helicity will be used in many future calculations.

\vskip .15 cm
\noindent
{\it \use\TreeLoopSection.3 Space-Time Supersymmetry}
\vskip .1 cm

The supersymmetry identities hold to all orders of perturbation
theory.  However, at loop-level the various states present in
a supersymmetric theory can circulate in the loop modifying
the implication that
can be extracted for QCD.
In particular, the identities (\use\SusyOneMinus) and
(\use\SusyAllPlus) no longer imply that the
corresponding QCD amplitudes vanish.  What it does imply are
relationships between the bosonic and fermionic loop contributions.

Since the states in an $N=1$ supersymmetric Yang-Mills
theory are a gluon and a fermion gluino, each of which can
circulate in the loop, the supersymmetric $n$-gluon amplitude is
$$
\eqalign{
A_{n;j}^{N=1\ \rm susy}(1^\pm, 2^+, 3^+, \cdots, n^+) & =
A_{n;j}^{\rm gluon} (1^\pm, 2^+, 3^+, \cdots, n^+) \cr
\null & \hskip 1 cm
+ A_{n;j}^{\rm fermion} (1^\pm, 2^+, 3^+, \cdots, n^+) \cr
&= 0 \cr}
\eqn\NOneSusy
$$
where the particle labels refer to the states circulating in the loop
and eqs.~(\use\SusyOneMinus) and (\use\SusyAllPlus) were used.  Thus
the contribution to these gluon helicity amplitudes of a fermion loop
is minus that of a gluon loop.  This means that once we have computed
either the fermion or gluon loop for these helicities there is no need
to explicitly compute the other.  In the supersymmetric theory the
fermions are in the adjoint representation, but in QCD the fermions
are in the fundamental representation; this difference is rather minor
since the partial amplitudes are identical in either case but one
would use either eq.~(\use\LoopColorDecomposition) or
(\use\FundamentalColor) to construct the full amplitudes depending on
the color representation.

One can actually do even better by appealing to $N=2$ supersymmetry
[\use\AmplLet,\use\Future].  The basic supersymmetry identities
(\use\SusyOneMinus) and (\use\SusyAllPlus) are still the same, but the
spectrum now consists of two real scalars, two fermions and one gluon
[\use\BasicSusy]. This gives for the one-loop $n$-gluon amplitude
$$
\eqalign{
A_{n;j}^{N=2\ \rm susy} &(1^\pm, 2^+, 3^+, \cdots, n^+)  \cr
\null & =
A_{n;j}^{\rm gluon} (1^\pm, 2^+, 3^+, \cdots, n^+) +
2 A_{n;j}^{\rm fermion} (1^\pm, 2^+, 3^+, \cdots, n^+) \cr
\null & \hskip 2 cm
+ 2 A_{n;j}^{\rm scalar} (1^\pm, 2^+, 3^+, \cdots, n^+) \cr
& = 0 \; .  \cr}
\eqn\NTwoSusy
$$
By combining the $N=1$ identity (\use\NOneSusy) with the $N=2$ identity
(\use\NTwoSusy) we then obtain the result that
$$
\eqalign{
A_{n;j}^{\rm gluon} (1^\pm, 2^+, 3^+, \cdots, n^+) & =
- A_{n;j}^{\rm fermion} (1^\pm, 2^+, 3^+, \cdots, n^+) \cr
& = 2 A_{n;j}^{\rm scalar} (1^\pm, 2^+, 3^+, \cdots, n^+) \; . \cr }
\eqn\FinalSusy
$$
Thus, for these particular gluon helicities, once the scalar contribution to
the $n$-gluon
amplitude is computed we also have the fermion and gluon contributions
to the loop.
The supersymmetry identities for other helicity amplitudes, such as in
eq.~(\use\SusyTwoMinus), relate the
gluon amplitudes to ones with external fermions and
provide useful checks on QCD calculations with external fermions.

It turns out that for
$A_{4;j}(1^\pm, 2^+, 3^+ , 5^+ )$ and $A_{5;j}(1^\pm, 2^+, 3^+, 4^+,
5^+)$ (and most likely for any number of legs) the integrands of each
diagram, whether gluons, fermions or scalars circulate in the loop,
are equal up to the overall constants in eq.~(\FinalSusy).  In
this way the information contained in the supersymmetry identities is
already encoded in the string-based methods.  It also turns out that
the integrands of diagrams in the string-based methods also exhibit
simplifications implicit in $N=4$ supersymmetric theories which go
beyond the above type of supersymmetry identities.  This will be
discussed in the next section where one form of the string-based rules
are presented.

\vskip -.5 cm
\section{Perturbative Rules}
\tagsection\RulesSection

The rules which are presented here are similar to the ones presented
in refs.~[\use\Long,\use\Pascos] except that the kinematic coefficient
is based on the simpler bosonic string.  These rules are the ones
presented in ref.~[\use\Bosonic].  (Other forms exist which avoid
the integration-by-parts step discussed near eq.~(\use\IBPExample)
[\use\Future]. This makes more complicated rules but simpler
Feynman parameter polynomials.)

The starting point of these rules are labeled $\phi^3$ diagrams
excluding tadpoles.  The cyclic labeling of legs of the diagrams must
follow the cyclic ordering of the associated color trace structure.
For partial amplitudes associated with two color traces, the labels
corresponding to the first trace must follow a counterclockwise
ordering while the second trace must follow a clockwise ordering
although the two sets can be ordered arbitrarily with respect to each
other.  The labeling of inner lines of a tree attached to a loop is
determined according to the rule that as one moves from the outer
lines toward the inner lines, one chooses the label of the most
clockwise of the two outer lines at a vertex to label the inner line
as depicted in \fig\TreeLabelingFig .

\vskip .3 cm
\centerline{\epsfxsize 3. truein \epsfbox{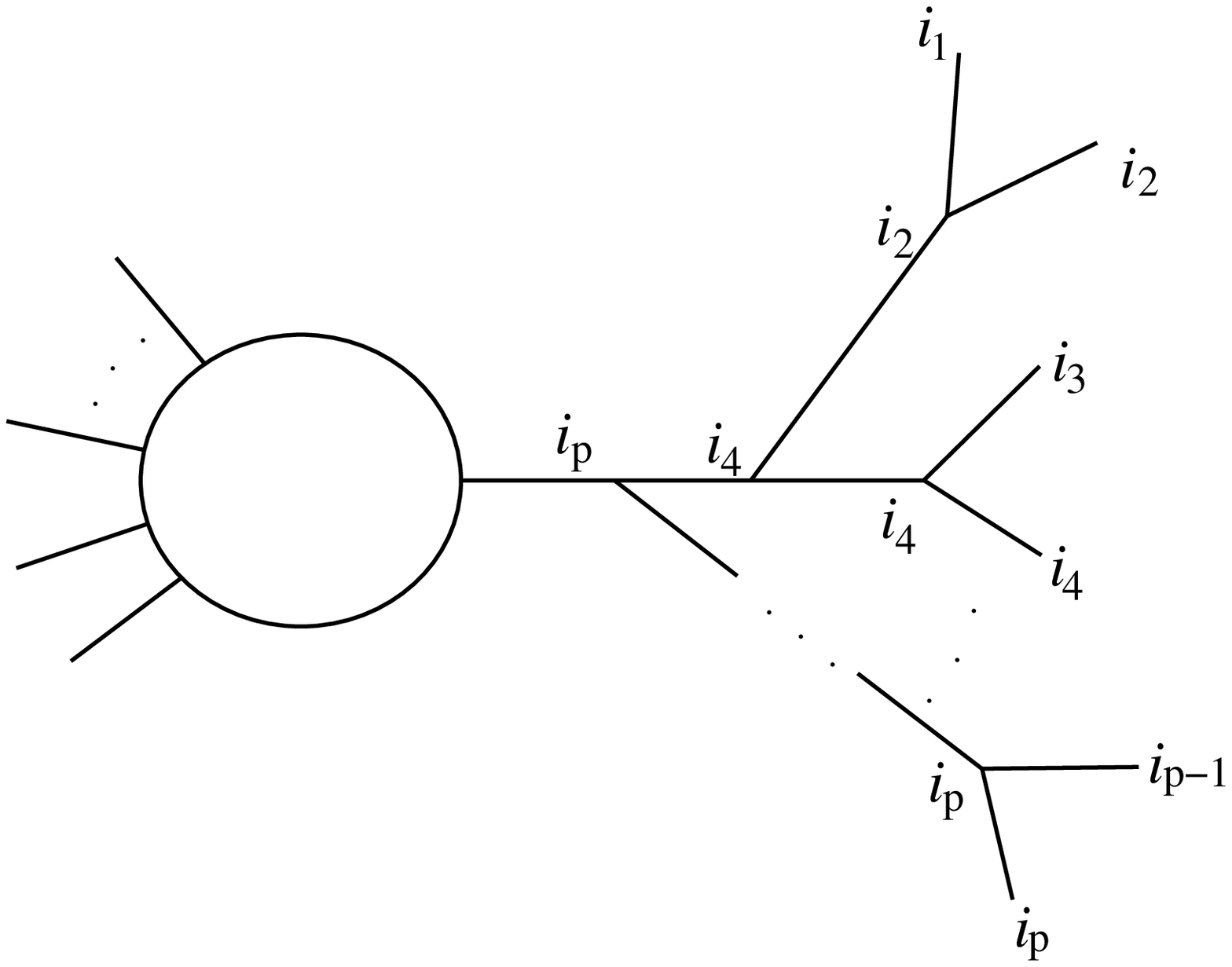} }
 \nobreak
\vskip .15 cm \nobreak
{\baselineskip 10 pt\narrower\smallskip\noindent\ninerm
{\ninebf Fig.~\TreeLabelingFig :} The labeling of the lines making up the
tree parts of the diagrams.
\smallskip}

\vskip .4 cm

For the partial amplitude $A_{n;1}$, the rules presented below for
evaluating a given $\phi^3$ diagram follow directly from the open
string amplitude (\use\StringAmplitude). For $A_{n;j>1}$ the form of
the rules follows the closed string form of the rules which is
what one would obtain by a comparison to the heterotic
string (which is a closed string) form of the rules.
(This distinction between open and closed strings is unimportant up
to the five-gluon calculation since one would not bother computing
$A_{n;j>1}$ directly, but would use the $U(1)$ decoupling equations
to obtain those from $A_{n;1}$.)

According to the rules, each labeled $\phi^3$-like diagram evaluates to
$$
\eqalign{
{\cal D} =
\N \Gamma(n_\ell-2+\eps/2)
&\int_0^1 dx_{i_{n_\ell-1}} \int_0^{x_{i_{n_\ell-1}}} dx_{i_{n_\ell-2}} \cdots
\int_0^{x_{i_3}}  dx_{i_2} \int_0^{x_{i_2}} dx_{i_1} \cr
& \times {K_{\rm red} \over
\Bigl(\sum_{l<m}^{n_\ell} P_{i_l}\c P_{i_m} \x {i_m}{i_l}
(1-\x {i_m}{i_l})\Bigr)^{n_\ell -2 +\eps/2}}  \cr}
\eqn\IntegrationRule
$$
where the ordering of the loop parameter integrals corresponds to the
ordering of the $n_\ell$ legs attached to the loop, $x_{ij} \equiv
x_i - x_j$, and $K_{\rm red}$
is the reduced kinematic factor. With the string-based
rules $K_{\rm red}$ can be obtained in a compact and efficient manner.
The lines attached to the loop carry momenta $P_i$
which need not be on-shell as there may
be trees attached to the loop.  For $K_{\rm red} =(-1)^{n_\ell}$,
${\cal D}$ is an $n_\ell$-point loop in massless $\phi^3$ theory.  The
dimensional regularization parameter $\eps = 4 - D$ handles all
ultra-violet and infrared divergences.
The $x_{i_m}$ are related to ordinary Feynman parameters by
$$
x_{i_m} = \sum_{j=1}^m a_j
\eqn\StandardFeyn
$$
so that the loop parameter integral can alternatively be written as
$$
\eqalign{
{\cal D} =
\N &\Gamma(n_\ell-2+\eps/2)
\int_0^1 \prod_{j=1}^{n_\ell} d a_j  \;
\delta\Bigl( \sum_j a_j - 1\Bigr) \cr
& \times {K_{\rm red} \over
\Bigl(\sum_{l<m}^{n_\ell} P_{i_l}\c P_{i_m}
\Bigl(\sum_{j=l+1}^m a_j \Bigr)
\Bigl(\sum_{j=1}^l a_j + \sum_{j=m+1}^{n_\ell} a_j \Bigr)
\Bigr)^{n_\ell -2 +\eps/2}}  \; .
\cr}
\eqn\integrationRule
$$

The partial amplitudes are then given by the sum over all diagrams whose legs
follow the ordering of the color trace so that
$$
A_{n;j} (1, 2,\cdots, j-1; j, j+1, \cdots , n) =
i (\sqrt{2})^n  \mu^\eps
\sum_{\rm diagrams } (-1)^{j_\ell-1}{\cal D}
\eqn\DiagToAmpl
$$
where $j_\ell-1$ is the number of legs attached to the loop
associated with the first of the two color traces; for $A_{n;1}$
this is always zero.
An additional color combinatoric factor of 2 is required for
$A_{4;3}$; no other combinatoric factors appear.  The parameter $\mu$
is the usual renormalization scale parameter that appears in dimensional
regularization.

The starting point for evaluating $K_{\rm red}$ for any given diagram
is the full kinematic expression given by
$$
\eqalign{
{\cal K} &=
\int \prod_{i=1}^n dx_i \prod_{i<j}^n
\exp\biggl( k_i\c k_j G_B^{i,j} +
(k_i\c\pol_j - k_j\c\pol_i) \, \Gbd^{i,j}
- \pol_i\c\pol_j\, \Gbdd^{i,j} \biggr)
\biggr|_{\rm multi-linear}
\cr}
\eqn\MasterKin
$$
where the exponential should be taylor-expanded to obtain those terms
which are linear in all $n$ polarization vectors. (Since all
powers of the inverse string tension cancel at the end,
all powers of $\alpha'$ have been dropped.)
Although simpler than the kinematic expression obtained from the
heterotic string [\use\Long,\use\Pascos],
this kinematic expression contains identical
information.  This kinematic expression represents all
information contained in a one-loop $n$-gluon amplitude; the
value of all diagrams is encoded in this
kinematic factor.  Although a string theorist may recognize $\Gbd$ and
$\Gbdd$ as derivatives of the bosonic Green function on the world
sheet, a field theorist should view these functions as `Feynman
parameter functions'.
{}From a conventional Feynman diagram point of view the existence
of a universal kinematic function is strange as there is apparently no simple
relationship between the various Feynman diagrams contributing to a
given process.
As discussed in refs.~[\use\Long,\use\Pascos], the fact that no
off-shell momenta
or polarization vectors appear in kinematic expressions of the
type (\use\MasterKin) allows one to use the full power of the spinor
helicity basis [\use\SpinorHelicity,\use\XZC]
on the first line of an explicit computation.

The first step in applying the rules presented here is to integrate by
parts in the kinematic expression (\use\MasterKin) (ignoring surface
terms) so as to remove all $\Gbdd^{i,j}$; this is always possible as
was proven in appendix B of ref.~[\use\Color].  After the integration
by parts has been performed the integrals sitting in front of the
kinematic expression along with the $\prod_{i<j} \exp(k_i \c k_j
G_B^{ij})$ factor should simply be dropped since the rules include the
appropriate factors.  The integration-by-parts step is a matter of
convenience as it simplifies the form of the rules.  (It turns out
that an alternative form of string-based rules exists which avoids the
integration-by-parts step and is of practical significance for the
computations of five-point amplitudes and beyond.  These rules will be
presented elsewhere [\use\Future].)

\vskip .15 cm
\centerline{\epsfxsize 4.5 truecm \epsfbox{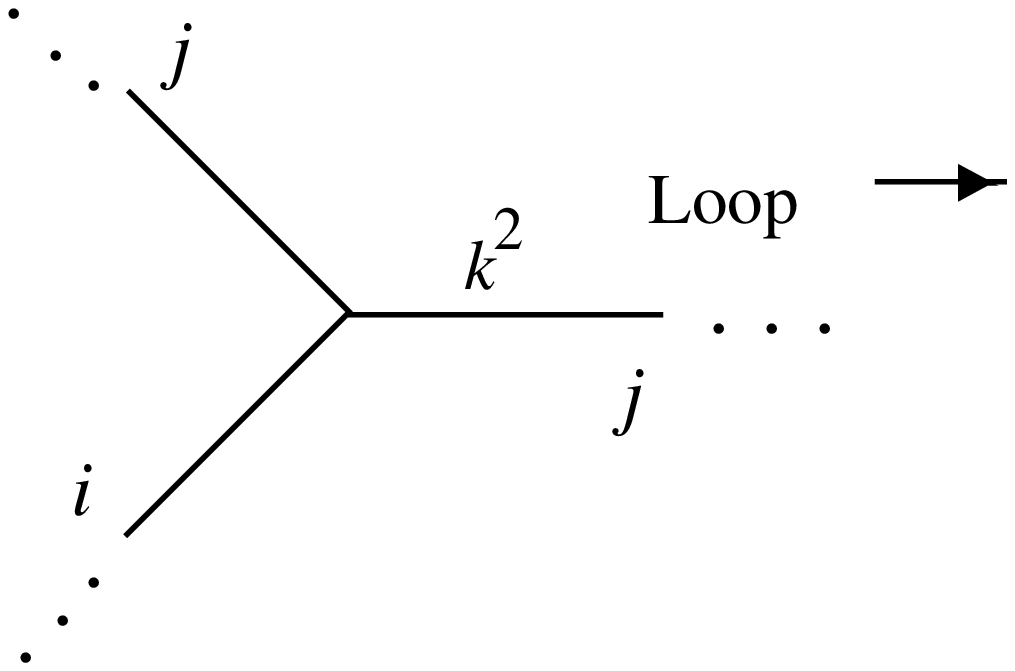} }
\nobreak
\vskip -.25 cm
$$
\eqalign{
& (\Gbd^{i,j})^n \rightarrow
          - {\delta_{n, 1} \over k^2}
            \hskip 1 cm (m , n \ge 0 )  \cr
&i \rightarrow j \hbox{ in remaining factors} \cr}
$$
\nobreak
%\vskip .05 cm \nobreak
\centerline{\ninerm{\ninebf Fig.~\use\TreeRulesFig :}
The tree substitution rules.}

%\vskip .6 cm

Given the integrated-by-parts kinematic factor and a particular
labeled diagram one then applies the tree rules of \fig\TreeRulesFig.
If a given tree contains all the legs associated with
a single color trace the diagram vanishes. (This rule follows
from the closed string color organization of the amplitude and does not
play a role in the calculation of $A_{n;1}$.) In particular,
for a two-point tree with legs labeled by $i$ and $j$ belonging
to a subset of the same color trace and
$i$ appearing before $j$ in the clockwise ordering,
a $\Gbd^{i,j}$ yields a factor of $(-2 k_i \c k_j)^{-1}$.
These tree rules do not depend on the whether gluons, scalars or
fermions circulate in the loop.

The $n$-gluon diagram which has a tree with $(n-1)$-legs so that the
loop is isolated on the remaining leg might seem to be ill-defined as
it contains a `0/0' [\use\Minahan,\use\WaveFunction,\use\Rozhanskii]
after applying the tree rules.  A four-point example of this type of
diagram is given in \fig\ExternalLegBubbleFig .  However, in
dimensional regularization there is an additional factor of zero in
such diagrams of the form $(p^2)^\eps/\eps$ with $p^2=0$ (since the
leg is on-shell).  This is interpreted as a complete cancellation of
ultraviolet and infrared divergences [\use\QCDReview,\use\Long].  (If
one wishes to distinguish between ultraviolet and infrared divergences
then one should resolve the 0/0 according to a prescription such as
the ones given in refs.~[\use\Minahan,\use\Rozhanskii].)

\vskip .5 cm
\centerline{\epsfxsize 4. truecm \epsfbox{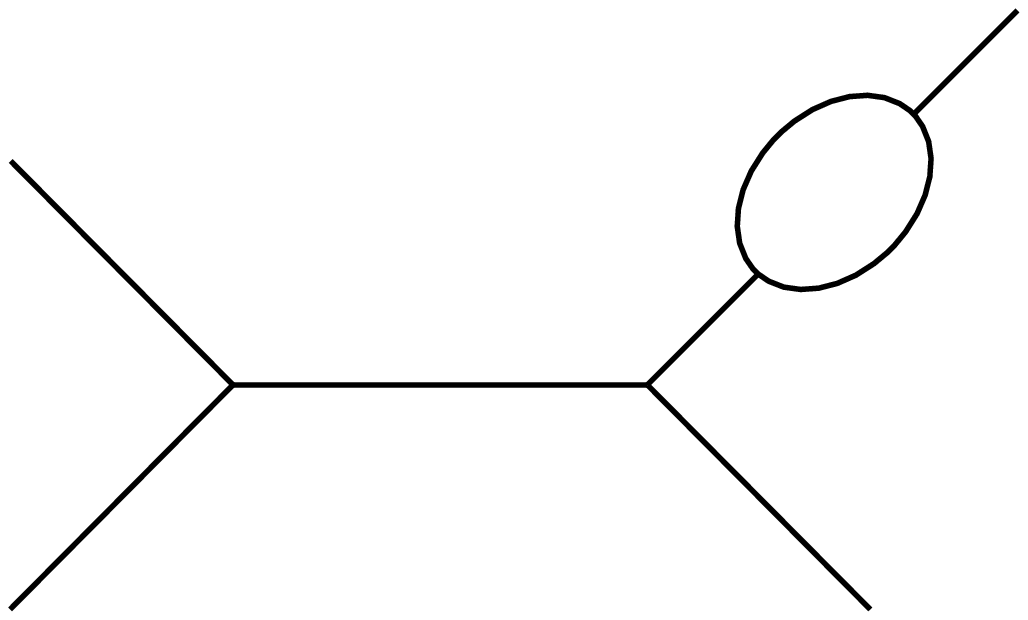} }
\nobreak
\vskip .15 cm
{\baselineskip 10 pt \narrower\smallskip\noindent\ninerm
{\ninebf Fig.~\ExternalLegBubbleFig :} A four-point example of a diagram with
a bubble on an external leg containing a potential 0/0 ambiguity.
\smallskip}
\vskip .4 cm

After the tree rules have been applied to the diagram the loop
substitution rules are then applied. A summary of the loop rules is
provided in \fig\LoopRulesFig .
For the case of gluons (and the
associated ghosts) circulating in the loop, in general every term
generates two types of contributions.

The first type of contribution
is obtained by multiplying the kinematic expression by a factor of
$2 (1- \delta_R \eps/2)$ and substituting
$$
\Gbd^{i,j} \longrightarrow  {1\over 2} (-\sign(\x ij ) + 2 \x ij )  \; .
\eqn\BasicSubst
$$
The regularization parameter is $\delta_R = 1$ in a conventional
or 't Hooft-Veltman type dimensional regularization scheme while
$\delta_R = 0$ in the four-dimensional helicity scheme [\use\Long].

The second type of contribution for gluons arises
if a particular term contains a cycle of $\Gbd$ which follows the ordering
of integration parameters $x_{i_1} \le x_{i_2} \le \cdots \le x_{i_{n_\ell}}$.
A cycle of $\Gbd$'s is defined by
$$
\Gbd^{i_1, i_2} \Gbd^{i_2, i_3} \cdots \Gbd^{i_m, i_1} \; .
\anoneqn
$$
Since the $\Gbd^{i,j}$ are antisymmetric, the indices can be put
into this canonical ordering at the cost of some signs.
For each cycle of $\Gbd$, the loop substitution rules are
$$
\Gbd^{i_1, i_2} \Gbd^{i_2, i_1} \longrightarrow 2
\anoneqn
$$
\vskip -.05 cm
\noindent
and
\vskip -.1 cm
\noindent
$$
\null \hskip 3 cm
\Gbd^{i_1, i_2} \Gbd^{i_2, i_3} \cdots \Gbd^{i_{m-1}, i_m} \Gbd^{i_m, i_1}
\longrightarrow 1 \hskip 1.5 cm  (m>2) \; .
\anoneqn
$$
\vskip -.05 cm
\noindent
Only one cycle at a time may contribute to any given term.  If the
cycle does not follow the ordering of the legs then there is no
contribution.  After these substitution rules have been applied to a
given cycle the substitution rule (\use\BasicSubst) is applied to all
remaining factors in the term of interest.  One then sums over all
cycles in a given term.

%%%%%%%%%%%%%%%%%%%%%%%%%%

\vskip .3 cm
{
\centerline{\epsfxsize 4.5 truecm \epsfbox{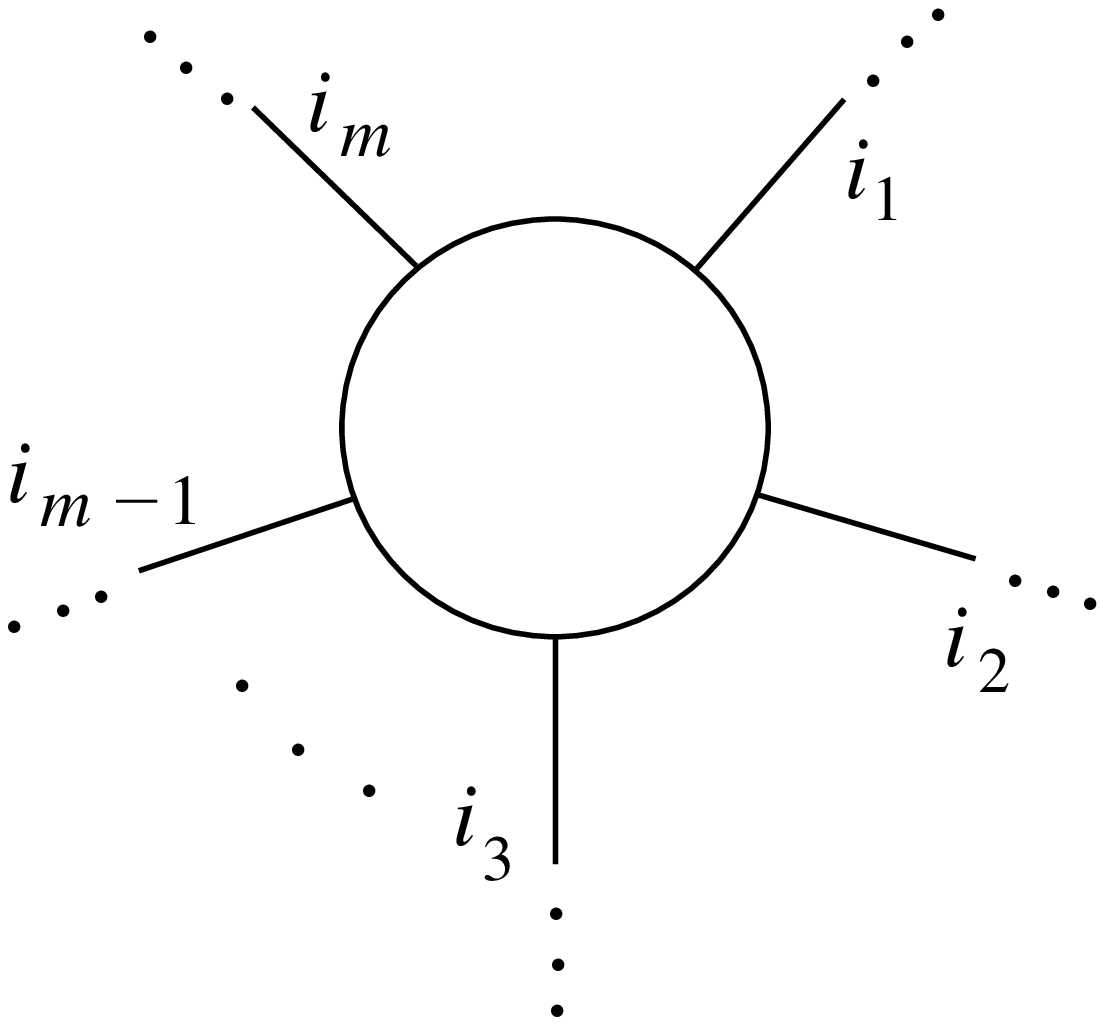} }
\nobreak
\vskip .1 cm \nobreak
\noindent
{\bf Gluon in Loop:}
\vskip -.5 cm
$$
\eqalign{
{\rm (a) \quad}
& \hbox{Overall }2 (1- \delta_R \eps/2) \; ,  \hskip 1.5 cm
\Gbd^{i,j} \longrightarrow  {1\over 2} (-\sign(\x ij ) + 2 \x ij )  \; , \cr
%&\cr
{\rm (b) \quad }& \eqalign{
&\Gbd^{i_1, i_2} \Gbd^{i_2, i_1} \longrightarrow 2 \; , \cr
&\Gbd^{i_1, i_2} \Gbd^{i_2, i_3} \cdots \Gbd^{i_{m-1}, i_m} \Gbd^{i_m, i_1}
\longrightarrow 1 \hskip .65 cm  (m>2) \; , \hskip .8 cm
\hbox{cycle follows leg ordering}\cr} \cr }
$$
\nobreak
%\vskip .1 cm
\noindent
{\bf Real Scalar in Loop:}
\vskip - .15 cm
$$
\hbox{Overall } N_s \; , \hskip 2 cm
\Gbd^{i,j} \longrightarrow  {1\over 2} (-\sign(\x ij ) + 2 \x ij )  \; ,
$$
\vskip .1 cm
\noindent
{\bf Fermion in Loop:}
\vskip -.4 cm
$$
\eqalign{
& \hskip -1.2 cm
\hbox{Overall $-4 N_d$ for Dirac and $-2 N_w$ for Weyl,} \cr
\Gbd^{i_1, i_2} & \Gbd^{i_2, i_3} \cdots \Gbd^{i_{m-1}, i_m} \Gbd^{i_m, i_1}
\longrightarrow \cr
& \Bigl({1\over 2}\Bigr)^m \Bigl[
\prod_{k=1}^m (-\sign(\x{i_k}{i_{k+1}}) + 2 \x{i_k}{i_{k+1}} )
- (-1)^m \prod_{k = 1}^m \sign(\x{i_k}{i_{k+1}}) \Bigr] \; .   \cr}
$$
\nobreak\vskip -.1 cm \nobreak
{\baselineskip 10 pt\narrower\smallskip\noindent\ninerm
{\ninebf Fig.~\use\LoopRulesFig:}
The loop substitution rules for various particle contents.
\smallskip}
}

%\vskip 1.1 cm
%%%%%%%%%%%%%%%%%%%%%%%%%%

As an example of the loop substitution rules for gluons in the loop,
consider the term
$$
\Bigl(\Gbd^{1,2} \Gbd^{2,1} \Bigr) \Bigl(\Gbd^{3,4} \Gbd^{4,5}
\Gbd^{5,3}\Bigr)
\anoneqn
$$
with the ordering $x_1 \le x_2 \le x_3 \le x_4 \le x_5$.
After applying the loop rules a total of three terms are generated
(since it contains two cycles) yielding
$$
\eqalign{
& 2 (1- \eps \delta_R/2) (\half + x_{12})(-\half + x_{21})
(\half + x_{34}) (\half + x_{45}) (-\half + x_{53}) \cr
\null & \hskip 1 cm
+ 2 (\half + x_{34}) (\half + x_{45}) (-\half + x_{53}) \cr
\null & \hskip 1 cm
+ (\half + x_{12})(-\half + x_{21}) \; .  \cr }
\anoneqn
$$

The case of scalars circulating in the loop is simpler and can be obtained
by modifying the string to contain scalars;
the resulting rule for scalars in the loop
is to multiply the kinematic expression by an
overall factor of $N_{\! s}$ representing the number of real scalars
and then apply the substitution rule (\use\BasicSubst).  There are no
further contributions in this case.

Since a bosonic string does not contain space-time fermions in the
spectrum it is not possible to obtain the fermion's contributions
directly from a bosonic string.  However, as was noted in
Section~\use\StringSection, there is a close relationship of
the world-sheet fermions to the world-sheet bosons. A practical
consequence is that all information about the amplitude for any
particle content can actually be extracted from the bosonic Green
functions.  In this way rules for space-time fermions can be
constructed; these reproduce the results of the heterotic string but
are based on the simpler bosonic string kinematic expression.

The first step for
fermions circulating in the loop is to multiply by an
overall factor of $-4 N_d$ where $N_d$ is the number of flavors of
Dirac fermions circulating in the loop. For $N_w$ Weyl fermions the
appropriate factor is $-2N_w$.
Once again cycles must be identified; the main
difference is that in this case all cycles, independent of the
ordering, lead to additional contributions.  The substitution
rule for space-time fermions circulating in the loop is
$$
\eqalign{
\Gbd^{i_1, i_2} & \Gbd^{i_2, i_3} \cdots \Gbd^{i_{m-1}, i_m} \Gbd^{i_m, i_1}
\longrightarrow \cr
& \Bigl({1\over 2}\Bigr)^m \Bigl[
\prod_{k = 1}^m (-\sign(\x{i_k}{i_{k+1}}) + 2 \x{i_k}{i_{k+1}} )
- (-1)^m \prod_{k = 1}^m \sign(\x{i_k}{i_{k+1}}) \Bigr]  \cr}
\eqn\STFermionLoopRule
$$
where $x_{m+1} \equiv x_1$ and the ordering of legs does not matter.
The first term on the right hand side is the same one as one would
obtain by either the scalar loop rules or by the no-cycle gluon
loop rule (\use\BasicSubst), up to the overall constant; this constitutes
the `no-cycle' fermion loop contribution.   The
second term is the additional cycle contribution for a fermion in the loop.
Remaining $\Gbd$ which do
not belong to any cycle should have the substitution rule (\BasicSubst)
applied.

For example, consider
the term $\Gbd^{1,2} \Gbd^{2,3} \Gbd^{1,3} \Gbd^{1,4}$
which contains a (1,2,3) cycle.  Applying the internal fermion
loop rules generates the terms (with $x_1 \le x_2 \le x_3 \le x_4$)
$$
- 4 N_{\! d} \Bigl({1\over 2} \Bigr)^4 \Bigl[(1 + 2\x12) ( 1 + 2\x23)
(1+ 2\x13) - 1\Bigr] ( 1+ 2\x14) \; .
\anoneqn
$$

As mentioned in the previous section the string-based rules go beyond
the standard supersymmetry identities discussed in that section for
simplifying calculations when bosons and fermions are present.
According to the rules, the no-cycle contributions for any particle in
the loop are all equal up to an overall constant.  Furthermore, it is
easy to verify that the two- and three-cycle contributions for a gluon
are $-4$ times that of a Weyl fermion.  This holds for the Feynman
parameter polynomials of {\it all} diagrams in the string-based rules.
For the four-cycle and beyond or products of cycles there is no longer
as simple a relationship between fermion and gluon loop contributions.
However, the most technically complicated terms are the zero-, two-
and three-cycle terms since those generate the most complicated
Feynman parameter polynomials.  This structure then implies that the
contributions of a real scalar, a Weyl fermion and a gluon to the
one-loop $n$-gluon amplitude are given by the generic formulas
[\use\AmplLet,\use\Future] $$
\eqalign{
&A_{n;j}^{\rm scalar} = S \cr
&A_{n;j}^{\rm fermion} = -2 S - F \cr
&A_{n;j}^{\rm gluon} = 2 (1 - \delta_R \eps/2) S  + 4 F + G \cr }
\eqn\GeneralStruc
$$
where $S$ is the no-cycle contribution, $-F$ the terms containing
contributions from
cycles for a
space-time fermion loop and $4F + G$ are the contributions containing
cycles for a gluon loop.
As before, the particle labels refer to the states circulating in the loop.

Thus a good strategy for computing the gluon in the loop is to first
calculate the scalar in the loop to obtain $S$.
This is a universal contribution which appears for all
states circulating in the loop.
Then the cycle parts of fermion in the loop computation can
be computed to determine $F$.
Finally, for the gluon loop contributions $G$ can be obtained by
computing
$$
G = 4 (\hbox{cycle contributions for Weyl fermions})
 + (\hbox{cycle contributions for gluons})
\anoneqn
$$
for each diagram and then summing over diagrams.
In each diagram this quantity vanishes
for all two- and three-cycles leaving behind a much simpler
Feynman parameter polynomial.  In this way $G$ can be directly
computed.  Observe that $S, G$ and $F$ are gauge invariant
since the scalar, fermion and gluon loop contributions are
individually gauge invariant.

The underlying reason for the simple relationship of space-time
boson loops to fermion
loops can be traced back to the essentially equal treatment of either
boson (Neveu-Schwarz) or fermion (Ramond) loops in string theory; only
the world-sheet boundary conditions differ between the two cases.

Observe that from the general structure (\use\GeneralStruc), for $N=4$
super Yang-Mills which contains one gluon, four Weyl fermions and 6
real scalars, the gluon amplitude satisfies [\use\AmplLet] (with
$\delta_R=0$)
$$
A_{n;j}^{N=4\  \rm susy}(1, 2, \cdots, n) = G = {\rm simple}
\anoneqn
$$
independent of the helicity choices.  The string-based rules make
this simple supersymmetry
structure evident at the level of the integrands of each diagram.

Because of this structure once the fermion loop contributions have
been computed, obtaining the gluon loop contributions represents a
small fraction of the work required to obtain the fermion loop
contributions.  It is amusing to make a comparison of the one-loop
diagrams in the gluon-by-gluon scattering computation to the diagrams
of a QED photon-by-photon scattering computation (which makes use of
modern spinor helicity techniques).  (In QCD the conversion of the
matrix elements into quantities which may be compared to experiment is
significantly more complicated but here we are interested in the
comparison of the virtual diagrams, which traditionally are also
far more difficult
in QCD.)  There are three main differences between a one-loop QCD and
QED diagram computation.  In QED one only has massive fermions
circulating in the loop while in QCD one can have gluons, ghosts, and
fermions.  Two other differences are that in QCD there are additional
diagrams with gluon trees and that in QCD masses are generally
negligible but not in QED.  In field theory, the complexity of the
non-abelian vertex indicates that a gluon loop should be significantly
more complicated than a fermion loop.  With the string-based methods,
the computational difference between a gluon loop and a fermion loop
given by $G$ is relatively small.  Furthermore, the diagrams with
lower point loops are generally much simpler to evaluate since the
associated loop integrals are simpler.  The appearance of masses in
QED also complicates the integrals as compared to QCD but lead to less
severe infrared problems.  This leads
to the result that the gluon one-loop diagrams are only moderately
more difficult to compute than photon diagrams, contrary to traditional
field theory expectations.

After the partial amplitudes have been computed,
the full amplitude can then be obtained by summing the partial amplitudes
with appropriate color trace factors; for adjoint representation states
circulating in the loop
the appropriate sum is given in eq.~(\use\LoopColorDecomposition),
while for fundamental states in the loop the appropriate sum is
given in eq.~(\use\FundamentalColor).

Modifying these rules to include masses for the internal fermions or
scalars is simple; the only change that needs to be made is in the
denominator in eq.~(\use\IntegrationRule) where the massless Feynman
denominator is replaced with one corresponding to massive states
circulating in the loop.

\section{Explicit Examples}
\tagsection\ExampleSection

In this section a number of explicit examples are presented of
computations which would be exceedingly difficult with traditional
Feynman diagram techniques but are much simpler with string-based methods.

\vskip .3 cm
\noindent
{\it \use\ExampleSection.1 Four-gluon Amplitudes}
\vskip .1 cm

The first example is the computation of ${\cal
A}(1^-, 2^+, 3^+, 4^+)$.  This example is nice because of its
simplicity.  Since the amplitude turns out to be finite on a
diagram-by-diagram basis there is no need to use dimensional
regularization.

The first step is to insert the
spinor helicity simplifications, which for this case are given
in eq.~(\use\HelicitySimplifications),
into the kinematic expression (\use\MasterKin).
{}From (\use\HelicitySimplifications) we can read off the kinematic
expression for this helicity choice as
$$
\eqalign{
K & = \pol_1 \c k_3 (-\Gbd^{1,3} + \Gbd^{1,2})
\pol_2\c k_4 (-\Gbd^{2,4} + \Gbd^{2,3}) \pol_3\c k_4
(-\Gbd^{3,4}+\Gbd^{3,2}) \cr
\null & \hskip 2 cm \times
\pol_4\c k_3 (-\Gbd^{4,3} + \Gbd^{4,2}) \cr
& =  C (\Gbd^{1,2} - \Gbd^{1,3})
(\Gbd^{2,3} - \Gbd^{2,4})
 (\Gbd^{3,4} + \Gbd^{2,3}) (\Gbd^{3,4} - \Gbd^{2,4}) \cr }
\anoneqn
$$
where
$$
\eqalign{
C & = -\pol_1 \c k_3 \pol_2\c k_4 \pol_3\c k_4 \pol_4\c k_3 \cr
& =  {1\over 4} s^2 t {\spb2.4^2 \over \spb1.2 \spa2.3 \spa3.4 \spb4.1 } \; .
\cr}
\eqn\CDefn
$$
Since there are no $\Gbdd$ factors there is no need to perform the
integration-by-parts step in this particular case.
For simplicity of notation the Mandelstam variables
$$
s = 2 k_1 \c k_2 \; , \hskip 2 cm
t = 2 k_1 \c k_4 \; , \hskip 2 cm
u = 2 k_1 \c k_3 \;
\eqn\Mandelstam
$$
are used.

There are a total of seven diagrams with potential contributions as depicted
in \fig\PotentialFig.  Of these only two diagrams (a) and (b) are non-vanishing
after applying the tree rules.  For example, since there is no
$\Gbd^{14}$ present diagram (c) vanishes.  Diagram (d) vanishes because
one of the factors vanishes when $x_2 \rightarrow x_3$.  Similarly all
other diagrams vanish.

\vskip .6 cm
\centerline{\epsfxsize 3.3 truein \epsfbox{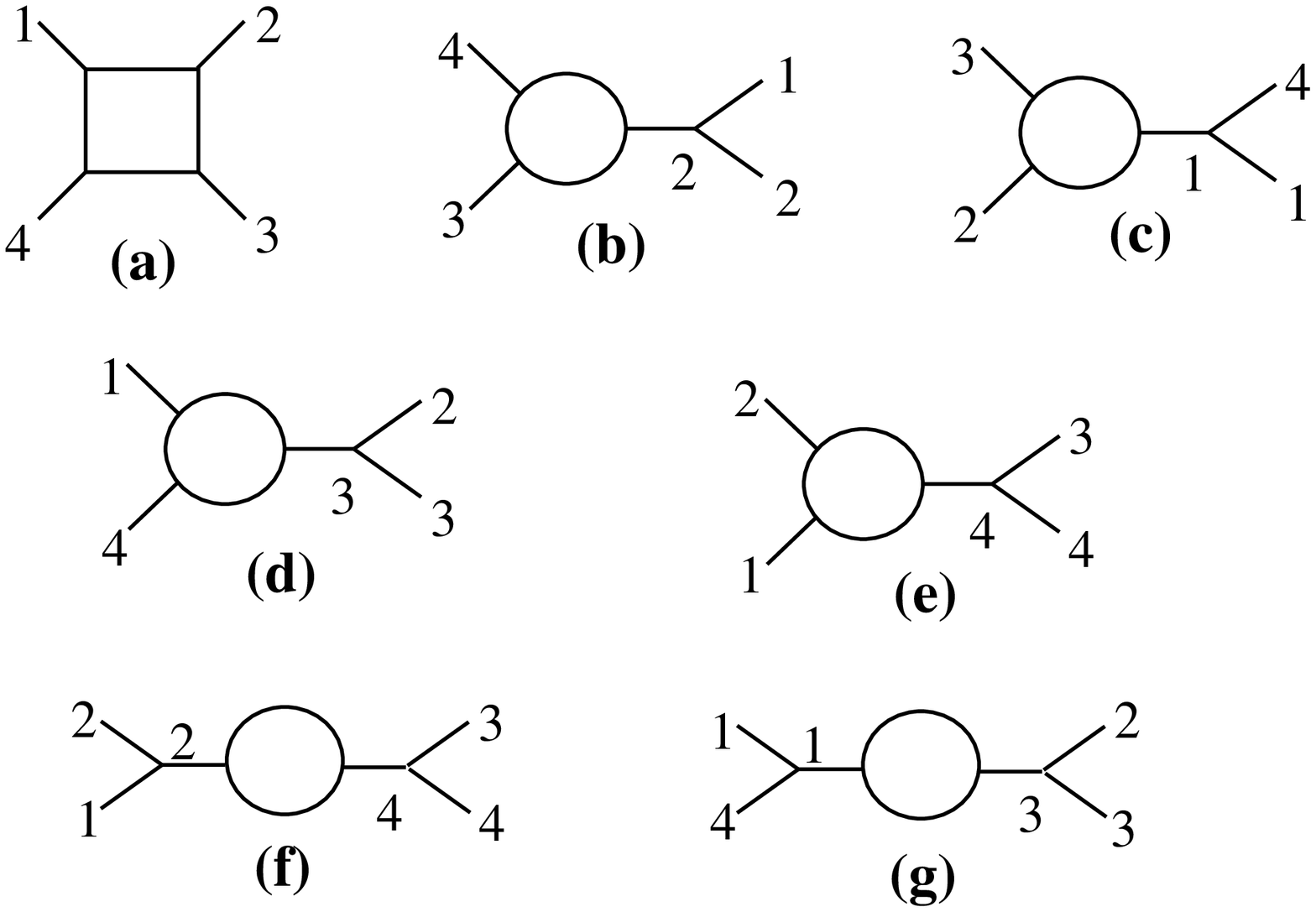} }

\nobreak
\vskip .2 cm
{\baselineskip 10 pt \narrower\smallskip\noindent\ninerm
{\ninebf Fig.~\PotentialFig:} Diagrams which potentially contribute. Diagrams
(c)-(g) trivially vanish after applying the tree rules.
\smallskip}

\vskip .6 cm

The box diagram \use\PotentialFig{a} is computed by applying the
loop substitution rules.  With these rules, the first type of contribution
is obtained by multiplying by 2 (for no dimensional regularization)
and applying the substitution (\use\BasicSubst).
This gives a contribution to
the Feynman parameter polynomial given by
$$
\eqalign{
T_1 & =  2 C (x_{12} - x_{13}) (x_{23} - x_{24}) (1 + x_{34} + x_{23})
(x_{34} - x_{24}) \cr
&=  2 C (x_3-x_2)^2 (1-x_3) x_2 \; . \cr }
\anoneqn
$$

The second type of contribution is given from cycles.  In order
to exhibit the cycles it is best to expand out the kinematic coefficient
$$
\eqalign{
K & = C (\Gbd^{1,2} - \Gbd^{1,3}) [\Gbd^{2,3} (\Gbd^{3,4})^2 -
\Gbd^{2,4} (\Gbd^{3,4})^2 + (\Gbd^{2,3})^2 \Gbd^{3,4}  \cr
\null & \hskip 1 cm
- \Gbd^{2,4} \Gbd^{2,3} \Gbd^{3,4} - \Gbd^{2,3} \Gbd^{3,4} \Gbd^{2,4}
+ (\Gbd^{2,4})^2 \Gbd^{3,4} - (\Gbd^{2,3})^2 \Gbd^{2,4}
+ (\Gbd^{2,4})^2 \Gbd^{2,3}]
\; . \cr}
\anoneqn
$$
The first factor is not expanded out since it is not part of
any cycles.  The cycle substitution rules to be applied to this are
$$
(\Gbd^{i,j})^2 \rightarrow -2\; , \hskip 2 cm
\Gbd^{2,3} \Gbd^{3,4}\Gbd^{2,4} \rightarrow -1
\anoneqn
$$
after which the substitution
(\use\BasicSubst) is performed in remaining factors.  This
gives the cycle contributions
$$
\eqalign{
T_2 & = C {1\over 2} (1 + x_{12}  - 1 - x_{13}) [-2 ( 1+ 2 x_{23}) +
2 (1+ 2 x_{24}) - 2(1+2 x_{34}) + 2]  \cr
& = 0 \; . \cr}
\eqn\CycleVanish
$$
This vanishing of contributions with cycles can be understood from the
space-time supersymmetry identities (\use\FinalSusy), as
explained below.  Observe that this cancellation takes place at the
level of the integrand.

Summing over the two types of contributions and inserting into the
scalar parameter integral gives the value of the first diagram
$$
D_a =  i {C \over 4 \pi^2} \int_0^1 d x_3 \int_0^{x_3} d x_2
\int_0^{x_2}  d x_1  {2 (x_3 - x_2)^2 (1- x_3) x_2 \over
[s x_1 (x_3 - x_2) + t (x_2 -x_1) (1-x_3)]^2 } \; .
\anoneqn
$$
(This can also be rewritten in terms of conventional Feynman parameters
using (\use\StandardFeyn) if one prefers.)
Since
$$
\int_0^{x_2}  d x_1 { 2 (x_3 - x_2)^2 (1- x_3) x_2
\over [s x_1 (x_3 - x_2) + t (x_2 -x_1) (1-x_3)]^2 }
= 2 {(x_3- x_2) \over st}
\anoneqn
$$
the remaining integrals are trivial, yielding
$$
D_a = {i  \over 12 \pi^2} C {1\over st} \; .
\anoneqn
$$

Now we must evaluate the second non-vanishing diagram given in
fig.~\use\PotentialFig{b} with the 1--2 tree.  According to the tree rules
extract the coefficient of $\Gbd^{12}$ and multiply by $-1/k_1 \cdot k_2$,
obtaining the reduced kinematic expression
$$
K^{12}_{\rm red} = - {C \over 2 k_1 \c k_2} (\Gbd^{2,3} - \Gbd^{2,4} )
(\Gbd^{3,4} + \Gbd^{2,3}) (\Gbd^{3,4} - \Gbd^{2,4}) \; .
\anoneqn
$$
Note that the last three factors are the same ones as in the box
diagram.  This means that we can read off the results of applying the
loop rules from the box diagram yielding the Feynman parameter
polynomial
$$
-2 {C\over s} x_2 (1-x_3) (x_3 - x_2)  \; .
\anoneqn
$$
Inserting this into the loop parameter integral yields
$$
D_b = - i C {1\over 4 \pi^2 s } \int_0^1 d x_3 \int_0^{x_3} d x_2
{x_2 (1-x_3) (x_3 - x_2)  \over -s x_2 (x_3 - x_2)} \; .
\anoneqn
$$
Since the denominator cancels against the numerator the integrals are
trivial, yielding
$$
D_b  =  {i  \over 12 \pi^2} C {1\over s^2} \; .
\anoneqn
$$

Summing over the contributions of the two non-vanishing diagrams
and using eq.~(\use\CDefn) yields the amplitude
$$
A^{\rm gluon}_{4;1}(1^-, 2^+, 3^+, 4^+) =
{i \over 48\pi^2 } { (s+t)  \spb2.4^2 \over \spb1.2 \spa2.3 \spa3.4 \spb4.1 }
\anoneqn
$$
which is a result that was first computed with string-based techniques.

Since the contributions with cycles drop out we immediately have that the
contribution of a real scalar in the loop to the four-gluon amplitude
is given by
$$
A^{\rm scalar}_{4;1} (1^-, 2^+, 3^+, 4^+)
= {N_s \over 2} A^{\rm gluon}_{4;1} (1^-, 2^+, 3^+, 4^+) \; .
\anoneqn
$$
It is also not difficult to verify that cycle contributions also drop
out for fermions in the loop.  This then yields
$$
A^{\rm fermion}_{4;1} (1^-, 2^+, 3^+, 4^+)
= - N_w  A^{\rm gluon}_{4;1} (1^-, 2^+, 3^+, 4^+)
\anoneqn
$$
where $N_w$ is the number of Weyl fermions.
Summing over the various contributions yields
$$
A_{4;1} (1^-, 2^+, 3^+, 4^+)
=  (2 + N_s - 2N_w) {i \over 96\pi^2 } { (s+t)  \spb2.4^2 \over
\spb1.2 \spa2.3 \spa3.4 \spb4.1 }
\anoneqn
$$
where all states are in the adjoint representation.
Since this expression vanishes when the number of bosonic states
equals the number of fermionic states, we have that
$$
A^{\rm susy}_{4;1} (1^-, 2^+, 3^+, 4^+) = 0
\anoneqn
$$
in agreement with the supersymmetry identity (\use\SusyOneMinus).
Observe that in the string-based formalism there is no need to
explicitly perform the integrals to obtain this identity
as the supersymmetry identities hold
in the integrands of each diagram.  In particular, the vanishing of the
cycle contributions in (\use\CycleVanish)
is a direct manifestation of the implicit
inclusion of supersymmetry identities in the string-based methods.

The other helicity amplitudes can be evaluated in pretty much the
same manner, except that dimensional regularization is required.
The four-point partial amplitudes which contribute to the next-to-leading
order cross-section are those with two minus and two plus helicities,
since those are the only tree amplitudes which do not vanish.   (The
next-to-leading order correction to the cross-section is obtained
from an interference of tree and one-loop amplitudes.)
By making use of the string-based rules and following the same
type of calculation as discussed above one can obtain [\use\Long]
the dispersive parts of the one-loop partial amplitudes needed for
the next-to-leading corrections (dropping all terms
of ${\cal O}(\eps)$),
$$\eqalign{
&A_{4;1}(\helicitiesA--++)
= i{ {\spa1.2}^4\over\spa1.2\spa2.3\spa3.4\spa4.1}
{\Gamma^2(1-\eps/2)\Gamma(1+\eps/2)\over 8\pi^2\Gamma(1-\eps)}
\L{4\pi\mu^2\over Q^2}\R^{\eps/2}\cr
\times
&\L -{8\over\eps^2} - {22\over 3\eps} + {11\over6}\lQ\moq
+ {2\over\eps} \L \lQ\soq+\lQ\toq \R
- \lQ\soq\,\lQ\toq
+ {11\over 6} \lQ\toq
+ {\pi^2\over2}-{32\over9}-{\delta_R\over 6}
\R\cr
&A_{4;1}(\helicitiesA-+-+)
= i{ {\spa1.3}^4\over\spa1.2\spa2.3\spa3.4\spa4.1}
{\Gamma^2(1-\eps/2)\Gamma(1+\eps/2)\over 8\pi^2\Gamma(1-\eps)}
\L{4\pi\mu^2\over Q^2}\R^{\eps/2}\cr
\times &\L -{8\over\eps^2} - {22\over 3\eps} + {11\over6}\lQ\moq
+ {2\over\eps} \L \lQ\soq+\lQ\toq \R
-{ (u^2 - s t)^2\over 2 u^4}\,\L \lQ\soq + \lQ\toq\R^2\RP\cr
&
+{s^4+t^4\over 2 u^4}\,\L\lQ^2\soq+\lQ^2\toq\R
-{11 u^2 - 3 s t\over 6 u^3}\,\L t\,\lQ\soq+s\,\lQ\toq\R\cr
&\LP
- {s t\over 2 u^3}\,\L s\,\lQ\soq+t\,\lQ\toq\R
-{s t\over 2 u^2}
+{\pi^2\over2}
-{\pi^2\over2}\ThetaHat(u){s t (2 u^2-s t)\over u^4}
-{32\over9}
-{\delta_R\over 6}
\vphantom{{ (s^2 + s t + t^2)^2\over 2 u^4}\,\L \lQ\soq + \lQ\toq\R^2}\R\cr
}\anoneqn$$
where $\mu^2$ is
the renormalization scale, $Q^2$ is a completely arbitrary scale
introduced in order to simplify the comparison to the cross-section
of Ellis and Sexton~[\use\Ellis],
$\lQ(x) = \ln\LV x/Q^2 \RV$,
$\ThetaHat(x\!>\!0)=1$, $\ThetaHat(x\!<\!0)=0$,
$$
\delta_R =
\left\{\eqalign{0,&\qquad\hbox{four dimensional helicity scheme,}\cr
1,&\qquad \hbox{'t\ Hooft-Veltman or conventional scheme,}\cr}\RP
\anoneqn
$$
and where evaluation in the physical region is assumed
(that is, only one of the Mandelstam variables
(\use\Mandelstam) $s, t$, or $u$ may be positive).
The absorptive parts are not included in the above equations but
can be calculated
using the appropriate $i\eps$ prescription.   A modified minimal
subtraction was performed on these amplitudes to subtract out the ultra-violet
divergence;  the remaining divergences are soft and collinear and
cancel against contributions from the five-gluon tree amplitude.

How do we know that the results are correct?  In ref.~[\use\Long] a
number of checks were performed on these expressions including checks
on gauge invariance, unitarity and, best of all, a comparison to a
previous calculation [\use\Ellis] of the next-to-leading order
corrections to the unpolarized cross-section.  Additionally, a mapping
to conventional field theory has been found which verifies
that the string-based methods give the same results for this
calculation as a field theory calculation would.

The checks on gauge invariance were performed in two ways.  One way
was by verifying that a change in the spinor helicity basis does not
modify the result.  From eq.~(\use\SpinorGaugeTrans), a change in the
spinor basis is equivalent to an on-shell gauge transformation which
should leave an on-shell amplitude unchanged.  This indeed works as
expected.  An alternative check is to simply replace a polarization
vector with the momentum of that external line.  When the remaining
legs are all on shell this longitudinal amplitude should just vanish
as indeed it does.

The check on unitarity which was performed was a verification of the
optical theorem.  The optical theorem says that the imaginary or
dispersive part of an amplitude is proportional to a
tree-level cross-section.  In ref.~[\use\Long], the amplitudes
were explicitly shown to satisfy this property.

The best check was against the previous calculation of Ellis and
Sexton [\use\Ellis] who calculated the one-loop corrections to
unpolarized cross-section into two jets.  After including the
$\epsh$-helicities discussed in Section~\use\TreeLoopSection\ in
order to obtain the conventional version of dimensional
regularization,  the unpolarized cross-section is in complete agreement
with their result.  This provides the first complete check of the
Ellis and Sexton cross-section and verifies that the conventional dimensional
regularization prescription used in string theory is identical
to the one of field theory.

\vskip .3 cm
\noindent
{\it \use\ExampleSection.2 Five-gluon Amplitudes}

Using the methods discussed above a computation of the one-loop five
gluon amplitudes has been performed [\use\AmplLet]. Additional
ingredients, beyond those discussed in these lectures, which enter
into this calculation are a simple integral table for the pentagon
parameter integrals [\use\Integrals] and improvements in the spinor
helicity method [\use\Future].  These amplitudes will enter into the
theoretical analysis needed for measurement of $\alpha_s$
at hadron colliders from jets.

The finite helicity amplitudes are
$$
\eqalign{
 A_{5;1}^{\rm 1-loop}&(1^+,2^+,3^+,4^+,5^+)
=\ \Bigl(1+ {1\over 2} N_s^{\rm adj} -
 N_{\! w}^{\rm adj} \Bigr) \cr
& \hskip 1 cm \times
{i\over 48\pi^2}\,
  {  \spa1.2\spb1.2\spa2.3\spb2.3 + \spa4.5\spb4.5\spa5.1\spb5.1
   + \spa2.3\spa4.5\spb2.5\spb3.4
 \over \spa1.2 \spa2.3 \spa3.4 \spa4.5 \spa5.1 } \cr
 A_{5;1}^{\rm 1-loop}&(1^-,2^+,3^+,4^+,5^+)
=\ \Bigl(1+ {1\over 2} N_s^{\rm adj} -
 N_{\! w}^{\rm adj} \Bigr) \cr
& \hskip 1 cm \times
{ i\over 48\pi^2}\,
{1\over\spa3.4^2 }
 \biggl[-{ \spb2.5^3 \over \spb1.2\spb5.1 }
 + { \spa1.4^3\spb4.5\spa3.5 \over \spa1.2\spa2.3\spa4.5^2 }
 - { \spa1.3^3\spb3.2\spa4.2 \over \spa1.5\spa5.4\spa3.2^2 } \biggr]
  \cr }
\anoneqn
$$
where $N_s^{\rm adj}$ and $N_{\! w}^{\rm adj}$ are the number of
adjoint massless real scalars and Weyl fermions.
(For fundamental
representation fermions one would use eq.~(\use\FundamentalColor)
to construct the full amplitude.)
The double trace $A_{5;2}$
and $A_{5;3}$ partial amplitudes follow from the formulae in
ref.~[\Color].  In the string-based formalism the supersymmetry
identities for these amplitudes (\use\FinalSusy)
are satisfied trivially because they
hold for the integrand of each string-based diagram; all cycle contributions
cancel out from the integrands so that the contribution from any
state is identical up to an overall sign determined by statistics.

The infrared divergent ones (which are the ones which interfere with
the tree diagrams to produce the next-to-leading order corrections to
the cross-section) are given in ref.~[\use\AmplLet].  These
amplitudes, have not been obtained with traditional techniques.

The matrix elements with external quark lines also need to be calculated.
These have not been calculated yet, although progress has been
made in a string-based approach [\use\Fermion].

One must then combine the virtual corrections with the singular terms
in the six-gluon tree-level matrix elements arising from the phase
space integration in soft and collinear regions.  One expects all
these divergences to cancel in physical quantities
[\use\LeeNauenberg].  The Giele-Glover formalism
[\use\GieleGlover,\use\Jets] makes use of the color ordering in
construction of universal functions representing the results of the
soft and collinear integrations, and is the most convenient one for
evaluating physical scattering.

\vskip .2 cm
\noindent
{\it \use\ExampleSection.3 A Gravity Example}
\vskip .1 cm

Another application of the string-based technique is to gravity.
Roughly speaking the structure of string theory implies
that
$$
(\hbox {Closed String}) \sim (\hbox{Open String})^2 \; .
\anoneqn
$$
Since closed strings contain gravity and open strings contain gauge theory
one might expect that
$$
(\hbox{Gravity}) \sim (\hbox{Yang-Mills})^2 \; .
\anoneqn
$$
This relationship can be made precise and turned into an extremely
efficient computational tool for perturbative gravity amplitudes.
At tree-level Berends, Giele and Kuijf
[\use\BerendsGrav] have made use of this relationship, as formulated
by Kawai, Lewellen and Tye [\use\KLTtree],
in order to calculate tree-level gravity amplitudes
from known Yang-Mills amplitudes.
At one-loop this relationship can also be made precise [\use\Gravity];
in particular, the calculation of the one-loop
four-graviton amplitude with one minus and three plus helicities is
rather easy by making use of string-based rules modified for the case
of gravity.  The result of such a calculation is given by
$$
A(1^-, 2^+, 3^+, 4^+) = {i \kappa^4 \over (4\pi)^2} {1\over 5760}
 (N_b - N_{\!f}) {s^2 t^2 \over u^2}
 (u^2 - st ) \biggl( {[24]^2 \over [12] \langle
23\rangle \langle 34\rangle [41] } \biggr)^2
\anoneqn
$$
where $\kappa$ is the gravitational coupling,
$N_b$ is the number of physical bosonic states
and $N_{\!f}$ is the number of fermionic states in the particular theory
of gravity under
consideration. The reason why any state gives an identical contribution
up to a sign is in agreement with the supersymmetry identities [\use\Grisaru]
and manifests itself in the string-based formalism as a vanishing of all
cycle contributions.  This is similar to the vanishing of the cycle
contributions in the four-gluon amplitude with the same helicities.

This type of calculation would be exceedingly difficult with
conventional techniques, given the complexity of the gravity three-
and four-point field theory vertices.  This may be compared to the
string-based technique where the calculation of the above helicity
amplitude is reduced to an elementary algebraic exercise.  It is
amusing that the string-based gravity calculation is only slightly
more difficult than the gluon calculation which is in turn only
slightly more difficult than a modern light-by-light scattering
calculation.  It is intriguing that in terms of conventional field
theory the required reorganization is fairly difficult to guess
without some input from string theory.  The case of gravity
will be discussed more fully elsewhere [\use\Gravity].

\section{Field Theory Understanding}
\tagsection\FieldTheorySection

An understanding of the string-based rules in terms of conventional
field theory is important for a number of reasons.  Phenomenologists
tend to be uninterested in string theory, so it is important to be able to
explain the computational advances implied by string theory in terms
of a more conventional field theory language.  With a mapping to conventional
field theory certain string theory subtleties are also
no longer a problem.  In particular, in order to guarantee that the
string-based dimensional regularization scheme is identical to
the field theory scheme a mapping between field theory and string
theory is necessary.  The mapping can also be used to explicitly
demonstrate how the string-based methods bypass many of the large
cancellations inherent in conventional field theory calculations since
it is easy to make comparisons when using the same type of formalism.

As we shall see, the interpretation of the string-based method in
terms of conventional field theory is a collection of ideas combined
in a particular way.  String theory provides the unifying principle
for applying these ideas to a field theory calculation.  Before
turning to the case of loop level we first discuss the tree-level
case.

\vskip .2 cm
\noindent
{\it \use\FieldTheorySection.1 Tree-Level Mapping}
\vskip .1 cm

The string reorganization at tree level can be understood in terms of
three basic ideas [\use\Mapping]: color ordering (which was discussed in
Section \use\TreeSection), a non-linear gauge choice
discovered by Gervais and Neveu [\use\GN] and a systematic evaluation
of the Lorentz contraction algebra generated by the vertices and propagators.

The non-linear Gervais-Neveu gauge makes the comparison of field
theory and string theory tree-level results relatively simple.  Other
gauge choices generally lead to complicated reshuffling of terms
between diagrams, obscuring the connection between field theory and string
theory.  This gauge was originally obtained by Gervais and Neveu
by analyzing the field
theory limit of open string theory.  The terms generated by the
Feynman rules in this gauge are in fairly close correspondence to the terms
generated by tree-level string theory; the main difference is
that with string theory all algebra associated with
contracting momenta is bypassed.

The action in the Gervais-Neveu gauge is given by
$$
S^{\rm GN} = \int d^4 x \Bigl( - {1\over 4} \Tr [F^2] -
{1\over 2} \Tr[(\partial \cdot A + i g A^2/\sqrt{2})^2] \Bigr)
\eqn\GNAction
$$
where
$$
F_{\mu\nu}^a = \partial_\mu A^a_\nu - \partial_\nu A^a_\mu
+ g f^{abc} A_{\mu}^b A_\nu^c
\anoneqn
$$
and we are
ignoring ghosts because here we are only interested in tree level.
The peculiar normalization of the terms in the action (\use\GNAction)
is due to the
unconventional normalization of the group generators $\Tr(T^a T^b) =
\delta^{ab}$.

\nobreak
\vskip .5 truecm
{\baselineskip 14 pt
{\epsfxsize 1.1 truein \epsfbox{feynmanvertfig.ps} } \nobreak
\vskip - 5.8 truecm \nobreak
\noindent \hskip 2.6 truecm $\nu$ \par \nobreak
\vskip 5. truecm \nobreak
\vskip -4.5 truecm \nobreak
\noindent\hskip .05 truecm  $\mu$ \par \nobreak
\vskip 4.5 truecm \nobreak
\vskip -4.6 truecm \nobreak
\noindent\hskip 2.6 truecm $\rho$ \par\nobreak
\vskip 4.6 truecm \nobreak
\vskip -5.9 truecm\nobreak
\noindent \hskip 3.1 truecm
% V^{\rm GN}_{\mu \nu\rho} (k, p , q) =
$ = i \sqrt{2} \Bigl( \eta_{\mu\nu} k _{\rho}
+ \eta_{\nu\rho} p_{\mu}
+\eta_{\rho\mu} q_{\nu} \Bigr) $ \par\nobreak
\vskip 6. truecm \nobreak
\vskip -4.3 truecm \nobreak
\noindent \hskip .2 truecm $\mu$ \hskip 1.9 truecm $\nu$\par\nobreak
\vskip 4.6 truecm \nobreak
\vskip -3.6 truecm \nobreak
\noindent \hskip .25 truecm $\lambda$ \hskip 1.8 truecm $\rho$ \par\nobreak
\vskip 3.6 truecm \nobreak
\vskip - 4.9 truecm \nobreak
\noindent \hskip 3.1 truecm
% V^{\rm GN}_{\mu \nu\rho\lambda}
$= i \eta_{\mu\rho}
\eta_{\nu \lambda} $ \par \nobreak
\vskip 4.9 truecm \nobreak
\vskip -3.5 truecm
\nobreak
{\baselineskip 10 pt\narrower\smallskip\noindent\ninerm
{\ninebf Fig.~\use\GNVertexFig :}
The color ordered Gervais-Neveu gauge vertices.
\smallskip} }

\vskip .6 cm

The color ordered three- and four-vertices
generated by the action (\use\GNAction)
are depicted in \fig\GNVertexFig\ corresponding to the color traces
$\Tr(T^{a_1} T^{a_2} T^{a_3})$ and
$\Tr(T^{a_1} T^{a_2} T^{a_3} T^{a_4})$.
As before, the  convention is to take momenta to be outgoing in three-vertices.
The propagator in this gauge is also
the same simple one as in conventional Feynman gauge
$$
P_{\mu\nu} = -i {\eta_{\mu\nu} \over p^2 + i \eps}
\; .
\anoneqn
$$

These vertices may be compared to the color ordered form of the
conventional Feynman gauge vertices given in fig.~\use\FeynmanColorFig.
The computational simplicity when using the Gervais-Neveu gauge for
tree level calculations is clear; the three- and four-point
Gervais-Neveu vertices have half and a third as many terms as the
corresponding Feynman gauge vertices.  Thus, a diagram in the
Gervais-Neveu gauge has a factor of approximately $2^{n_{\! 3}}
3^{n_{\! 4}}$ fewer terms than a corresponding diagram in color
ordered Feynman gauge, where $n_3$ is the number of three-point
vertices and $n_4$ is the number of four-point vertices.  (We have
ignored simplifications from the on-shell conditions which decrease
the count and the fact that internal line momenta are sums of external
momenta which increase the count.)  This provides an explanation of
how string theory avoids many of the large cancellations inherent in
conventional computations of tree-level amplitudes; at tree level, in
Feynman gauge most of the terms must cancel to reproduce the
simplicity of the Gervais-Neveu gauge.  The use of Gervais-Neveu
gauge provides only a partial explanation of the string-based
rules; by using string-based rules the algebra associated with
sewing the vertices together is completely bypassed.  The
Gervais-Neveu gauge is thus useful for allowing a relatively simple
understanding of the string organization of the tree level amplitude,
although for practical computations it is advantageous to use
recursive methods [\use\Recursive,\use\LightconeRecurrence].

\vskip .3 cm
\noindent
{\it \use\FieldTheorySection.2 Loop Level}
\vskip .1 cm

Given the tree level understanding of the string reorganization of the
amplitude, one might think that this can be directly carried over to
one loop.  However, string theory does not work this way; the field
theory descriptions which most closely resemble the string
reorganizations at tree level and at one loop are rather different.
In particular, a different set of gauge choices are needed at tree and
loop level.  This provides the general notion that one should use
different gauge choices or more generally different field
variables at each order of perturbation theory in order to
maximize efficiency.  String theory
provides a particularly efficient way to accomplish this.

The required field theory ideas which are needed to reproduce much of the
simplicity of the string-based rules for one-loop $n$-gluons
amplitudes are [\use\Mapping]:  background
field Feynman gauge,  color ordering (which was discussed in Section
\use\TreeLoopSection),
systematic organization of the vertex algebra and a second order
formalism for fermions.
The most important new ingredient is the background field
method which we now review.

\vskip .3 cm
\noindent
{\it \use\FieldTheorySection.3 Review of Background Field Method}
\vskip .1 cm

The background field method [\use\Background] is a popular technique
for computing effective actions and $\beta$-functions in field
theories.  Its distinguishing feature is manifest gauge invariance of
the effective action which is a property that does not hold for
conventional Lorentz gauges.

The basic idea of the background field method is to split the gauge
field into quantum and background fields, $A = Q + B$. The quantum
field is then gauge fixed in such a way as
to maintain the gauge invariance of the background
field.  In the background field method the effective action is
computed by considering one-particle irreducible diagrams with
external background $B$ fields.

Although this procedure generates an `effective action', before it can
be used in an amplitude calculation its connection to the usual
effective action which is the Legendre transformation of the connected
diagrams must be understood.  Consider the background field generating
function
$$
Z[B] = \int D Q \; \Delta_{\rm FP} \exp\Bigl[i\int d^4 x \Bigl({\cal L} (Q+B)
- {1\over 2 \alpha} \Tr[(D^B \cdot Q)^2] \Bigl) \Bigr]
\eqn\BFGenerating
$$
where $\Delta_{\rm FP}$ is the Faddeev-Popov determinant and ${\cal L}$ is
the lagrangian for the gauge theory under consideration.  For pure
Yang-Mills theory
$$
{\cal L} = -{1\over 4} \Tr[F^2]
\anoneqn
$$
where $F$ is the field strength.  The generating function (\use\BFGenerating)
is gauge invariant under the background field gauge transformation
$$
\delta B = D^B \lambda \; ,\hskip 2 cm  \delta Q = i [\lambda, Q]
\anoneqn
$$
where $D^B_\mu = \partial_\mu + i g T^a B^a_\mu$
is the derivative covariant with respect to the
background field.  Since $Q$ is a dummy integration field,
$Z[B]$ is gauge invariant with respect to $B$.

But what is $Z[B]$ in terms of the more conventional effective action?
To answer this look at
$$
\tilde Z[J, B] = \int DQ \; \Delta_{\rm FP}  \exp\Bigl[ i \int d^4 x
\Bigl( {\cal L} (Q+B) -
{1\over 2 \alpha} \Tr[(D^B \cdot Q)^2]  + \Tr(J \cdot Q) \Bigr) \Bigr]
\eqn\GeneratingJ
$$
which is a generating functional, but with an additional background field
and a peculiar gauge fixing.  By performing a Legendre transformation
we obtain an effective action defined by
$$
\tilde\Gamma[\tilde Q_{cl}, B] = \tilde W [J, B] -
\int d^4 x \;  J^a \c \tilde Q_{cl}^a
\anoneqn
$$
where
$$
\tilde W = -i \ln \tilde Z \; , \hskip 2 cm
\tilde Q_{cl} = {\delta \tilde W \over \delta J } \; .
\anoneqn
$$

In order to connect this object to the more usual effective action
consider the alternative way of evaluating the path integral
(\use\GeneratingJ) by making
the change of variables $Q \rightarrow Q - B$.
This yields
$$
\eqalign{
\tilde Z [J, B] & = \exp \Bigl[-i\int d^4 x \Tr(J \c B) \Bigr] Z[J]  \cr
& = \exp \Bigl[-i \int d^4 x \Tr(J \c B) \Bigr] \cr
\null & \hskip 1 cm \times
 \int D Q \; \Delta_{\rm FP}
\exp\Bigl[i \int d^4x \Bigl({\cal L} (Q) -
{1\over 2 \alpha} \Tr(D^B \cdot (Q-B))^2  + \Tr(J \cdot Q) \Bigr) \Bigr] \cr }
\anoneqn
$$
where $Z[J]$ is a conventional generating functional, but with
a peculiar gauge fixing which depends on the arbitrary field $B$.

By comparing the two forms of the path integral we have
$$
 \tilde W \equiv -i \ln \tilde Z
= -i\ln Z - \int d^4 x \;  J^a\c B^a  \equiv W -
\int d^4 x \; J^a \c B^a  \; .
\anoneqn
$$
This implies that
$$
\tilde \Gamma = \tilde W - \int d^4 x \; J^a \c \tilde Q_{cl}^a
= W - \int d^4 x \; J^a \c(\tilde Q_{cl}^a + B^a) \; .
\anoneqn
$$
Thus
$$
\tilde \Gamma[\tilde Q_{cl} , B] = \Gamma[\tilde Q_{cl} + B, B]
\anoneqn
$$
which relates the background field effective action $\tilde \Gamma$ to
the conventional effective action $\Gamma$.  The $B$ in the second
argument of $\Gamma$ refers to the dependence of the gauge-fixing
on the background field.
For $\tilde Q_{cl} = 0$ we then obtain the fundamental
equation of the background field method
$$
\tilde \Gamma[0, B] = \Gamma[B, B] \; .
\eqn\BasicBackground
$$
In this equation the object on the left hand side is the background
field effective action (with no other external sources) while the
quantity on the right hand side is the usual effective action, but with
a $B$ dependent gauge fixing.

In an actual calculation with the background field method there
is no need to perform a Legendre transformation since $\tilde \Gamma$ can be
directly computed from the background field Feynman vertices generated
by the path integral (\use\BFGenerating).
The basic background field formula then ensures that one-particle
irreducible diagrams (i.e., the diagrams of the effective action) which
have only external background $B$ fields are meaningful quantities.

The background field method has traditionally been used for effective
action calculations because of its natural interpretation in terms of
effective actions.  Here we are interested in the scattering amplitudes
and not in the effective action.  In order to obtain the scattering
amplitudes from the effective action we need to sew trees onto the
one-particle-irreducible diagrams in order to form the connected
diagrams.  What gauge is this sewing to be performed in?  The fact that
the effective action is gauge invariant leads one to the notion that
it really does not matter what gauge is chosen for the sewing.  Since
the background field effective action differs slightly from the
conventional effective action because of the dependence of the gauge
fixing on the background field, one needs to prove that the $B$ field
in eq.~(\BasicBackground) which comes from the gauge fixing does not
interfere with the sewing procedure.  A proof has been given by
Abbott, Grisaru and Schaefer [\use\AGS].  Thus, the procedure for
constructing scattering amplitudes out of the background field
effective action is to simply sew tree diagrams onto the
one-particle irreducible diagrams of the effective action using some
other gauge, such as ordinary Feynman gauge or Gervais--Neveu gauge.
For practical calculations in field theory the Gervais--Neveu gauge is
advantageous because of its simpler vertices.

\vskip .3 cm
\noindent
{\it \use\FieldTheorySection.4 String Organization of Loop}

The vertices can be generated from the
background field generating function (\use\BFGenerating).
Since we wish to evaluate the term in the amplitude
associated with the color trace $\Tr(T^{a_1} T^{a_2} \cdots T^{a_n})$
it is convenient to color order the background field gauge vertices.
These background field three- and four-point vertices are depicted in
\fig\BFvertexFig .
There are also ghost vertices given by
$$
V^{\rm G}_{\mu}(p, q)  =  {i\over \sqrt{2}} (p - q)_\mu
\eqn\GhostVertex
$$
for the coupling of a ghost and anti-ghost to a single background
field $B$ and
$$
V^{\rm G}_{\mu\nu} =-{  i \over 2} \eta_{\mu\nu}
\anoneqn
$$
for the coupling of a ghost and anti-ghost to two background fields.
Observe that
the ghosts of one-loop background field gauge couple precisely the
same way as complex scalars.  Indeed, at one-loop
in background field gauge the
ghost contribution is minus that of a complex scalar.

\vskip .6 cm
{\baselineskip 14 pt
{\epsfxsize 1.1 truein \epsfbox{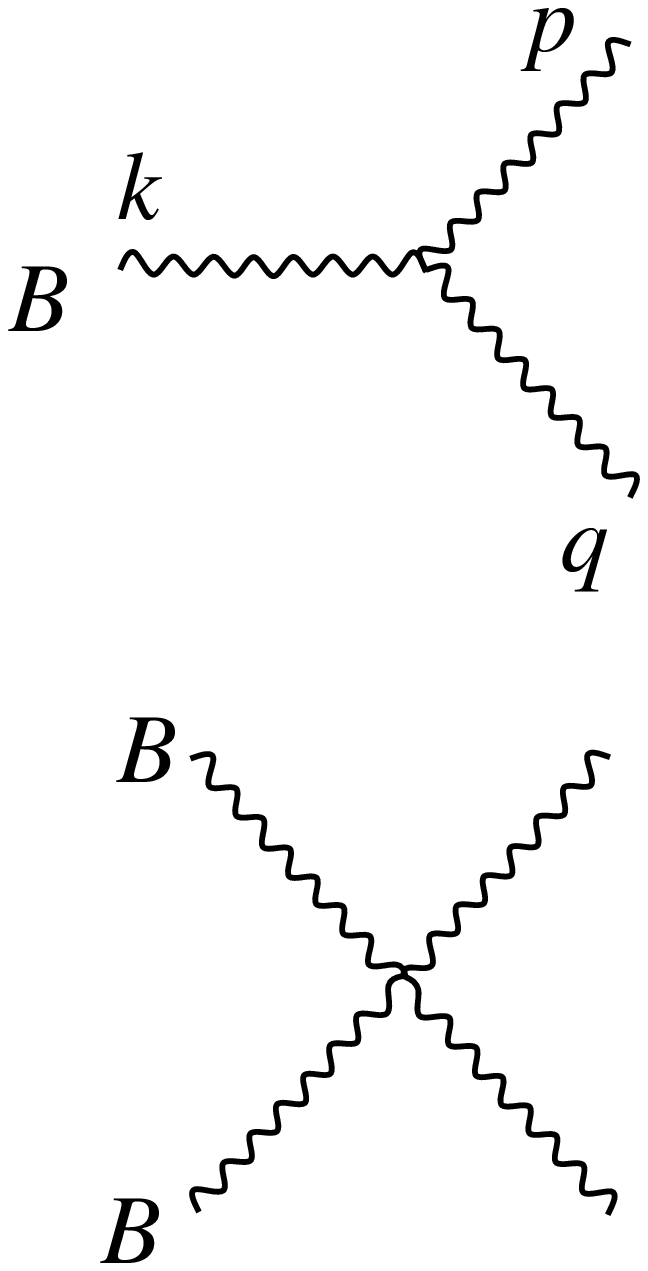} } \nobreak
\vskip - 5.6 truecm \nobreak
\noindent \hskip 2.8 truecm $\nu$\par\nobreak
\vskip 5.6 truecm \nobreak
\vskip -5.1 truecm \nobreak
\noindent\hskip .4 truecm  $\mu$\par\nobreak
\vskip 5.1 truecm \nobreak
\vskip -5.2 truecm \nobreak
\noindent\hskip 2.8 truecm $\rho$\par\nobreak
\vskip 5.2 truecm \nobreak
\vskip -6.6 truecm \nobreak
\noindent \hskip 3.1 truecm
%V^{\rm background}_{\mu\nu\rho} (k, p, q) =
$\displaystyle = {i \over \sqrt{2}}
\Bigl[ \eta_{\nu\rho} (p - q)_\mu  - 2 \eta_{\mu\rho} k_\nu +
2 \eta_{\mu\nu} k_\rho \Bigr]$ \par\nobreak
\vskip 6.6 truecm \nobreak
\vskip -4.8 truecm \nobreak
\noindent \hskip .5 truecm $\mu$ \hskip 1.7 truecm $\nu$\par\nobreak
\vskip 4.8 truecm \nobreak
\vskip -4.1 truecm  \nobreak
\noindent \hskip .5 truecm $\lambda$ \hskip 1.7 truecm $\rho$\par\nobreak
\vskip 4.1 truecm \par \nobreak
\vskip - 5.5 truecm  \nobreak
\noindent \hskip 3.1 truecm
%V^{\rm background}_{\mu\nu\rho\lambda}=
$\displaystyle = -{i\over 2}  \eta_{\mu\nu} \eta_{\rho\lambda}
- i \eta_{\mu\lambda} \eta_{\nu\rho}
+ i \eta_{\mu\rho} \eta_{\nu\lambda}$\par\nobreak
\vskip 5.5 truecm \nobreak
\vskip -4.5 truecm \nobreak
{\baselineskip 10 pt\narrower\smallskip \noindent\ninerm
{\ninebf Fig.~\BFvertexFig:} The color ordered background field
Feynman gauge vertices. The legs labeled by a `$B$' are background field legs
while the others are quantum field legs. \smallskip}
 }

\vskip .6 cm

In order to
to arrange the computation according to the string organization
it is desirable to break the three-point
vertex given in fig.~\use\BFvertexFig\ into three pieces
$$
V_{B} = {i\over \sqrt{2}}
\eta_{\nu\rho} (p-q)_{\mu} \; ; \hskip 0.5 truecm
V_{F}^{+} = i\sqrt{2} \eta_{\mu\nu} k_{\rho} \; ; \hskip 0.5 truecm
V_{F}^{-} = -i\sqrt{2} \eta_{\mu\rho} k_{\nu}
\anoneqn
$$
where $\mu$ is to be contracted against an external polarization
and $k$ is the external background field
momentum as depicted in fig.~\use\BFvertexFig .
With this breakup, the loops containing only three-vertices can be
arranged to follow the string organization.  Observe that the
$V_B$ vertex is identical to the ghost-background field vertex
(\use\GhostVertex) except that it has no $\eta_{\nu\rho}$.

First consider the case with either only $V_B$ or $V_G$
vertices in the $\phi^3$-like loop depicted in \fig\VbExampleFig.
Sewing together the vertices into a loop yields
$$
(\sqrt{2})^n 2 (\delta_\mu^\mu - 2)
\int {d^{4-\eps} p \over (2\pi)^{4-\eps} } \;
{ \prod_{i=1}^n \pol_i \cdot (p+q_i)   \over
\prod_{i=1}^n  ( p+q_i)^2    }
\eqn\FeynmanForm
$$
where $\delta_\mu^\mu$ arises from sewing the $\eta_{\nu\rho}$ of
the $V_B$ vertices around the loop and $-2$ is the statistical
and combinatoric factor associated with the ghost loop.
Rewriting this in terms of Schwinger proper-time variables yields
$$
\eqalign{
( \sqrt{2} )^n 2 {\cal R}
& \int_0^\infty d t_1 \int_0^\infty d t_2 \cdots \int_0^\infty d t_{n-1}
\int_0^\infty d t_n \int
{d^{4-\eps} p \over (2\pi)^{4-\eps} } \;
\cr & \hskip 1.0 truecm  \times
\exp\biggl(-\sum_{i=1}^n t_i (p + q_i)^2 + \sum_{i=1}^n
\pol_i \cdot (p + q_i)
\biggr) \biggr|_{\rm multi-linear}
\cr}
\eqn\MomIntegrals
$$
where
$$
q_i =  k_i + \cdots + k_{n} = - k_1 - k_2 - \cdots - k_{i-1}
\eqn\qDef
$$
and we have observed that
$$
\delta_\mu^\mu - 2 = 2 - \delta_R \eps \equiv  2 {\cal R} \; .
\anoneqn
$$
The parameter $\delta_R$ controls the precise form of dimensional
regularization, as discussed in Section~\use\TreeLoopSection .
This form of the loop momentum integral may be recognized as
the bosonic zero-mode of the string (see Chapter 8 of ref.~[\use\GSW])
providing the connection to string theory.

\vskip .4 cm
\centerline{\epsfxsize 4. truecm \epsfbox{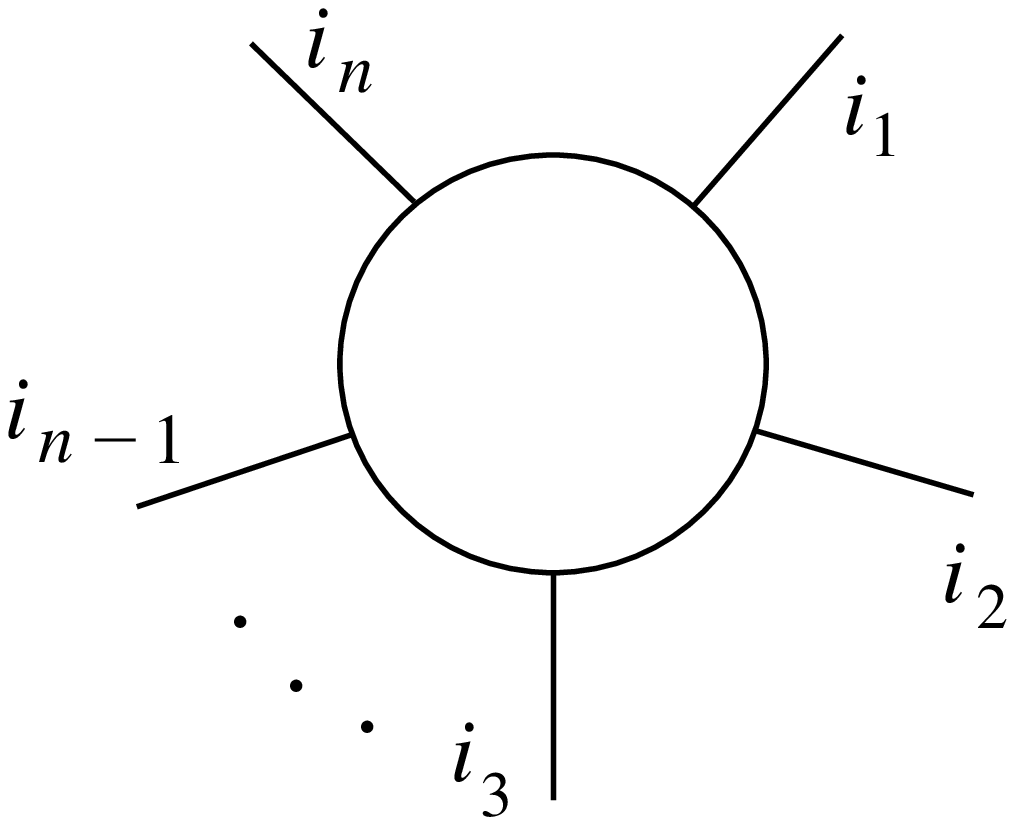} }
\nobreak
\vskip .2 cm
\centerline{\baselineskip 10 pt \ninerm
{\ninebf Fig.~\VbExampleFig :}  A $\phi^3$-like loop diagram with no
attached trees.}
\vskip .6 cm

The loop momentum integral
in eq.~(\use\MomIntegrals) can be performed
by completing the square in the usual fashion
(see for example eq.~(8.1.56) of ref.~[\use\GSW]) yielding
$$
\eqalign{
A_B (\pol_1,k_1; & \cdots ; \pol_n,k_n) = \cr
& i ( \sqrt{2} )^n 2 {\cal R}
{ (4\pi)^{\eps/2} \over 16\pi^2  }
\int \prod_{i=1}^{n-1}  dx_i
 \int_0^\infty dT\;  T^{n-3+\eps/2} \cr
& \null \times \prod_{i<j}^n \exp \biggl\{
\int \prod_{i=1}^{n-1}  dx_i
 \int_0^\infty dT\;  T^{n-3+\eps/2} \cr
& \null \times \prod_{i<j}^n \exp \biggl\{
 k_i\c k_j \tilde G_B^{i,j} +
 (k_i\c\pol_j - k_j\c\pol_i ) \, \dot {\tilde G}_B^{i,j}
         - \pol_i\c\pol_j\, \ddot {\tilde G}_B^{i,j} \biggr\}
\biggr|_{\rm multi-linear}
\cr}
\eqn\KinematicContributions
$$
where $x_i = \sum_{j=1}^i t_j/T$, $T=\sum_{i=1}^n t_i$,
$x_n=1$ and the $\tilde G_B$'s are given by
$$
\tilde G_B^{ij} = T x_{ij} (1 + x_{ij})\; , \hskip 1.5  truecm
\dot{\tilde G}_B^{ij} = {1\over 2} + x_{ij} \; , \hskip 1.5 truecm
\ddot{\tilde G}_B^{ij}  = {1\over 2 T} \; .
\eqn\SchwingPar
$$
As the notation suggests these functions are the same ones as one
obtains from the loop rules applied to the string master formula.  Of
course, in this case the quantities are purely field theory
expressions.  In this way the background field $V_B$ vertices
reproduce the pure $G_B$ parts of the string kinematic expression
evaluated with loop substitution rules without integration by parts
for diagrams of the form in fig.~\use\VbExampleFig.
The integration by parts discussed in Sections \use\StringSection\ and
\use\RulesSection\ obscures the connection between the string-based
rules and field theory.  One can also integrate by parts in field
theory [\use\Mapping] but care must be taken to keep
track of surface terms.  This complicates the match between field
theory and string theory.

Consider now the case with pure $V_F^+$ vertices.  In this case
loop momentum does not appear in the vertices and one obtains
$$
(-1)^n \pol_1\c k_2 \pol_2\c k_3
\cdots \pol_{n-1}\c k_{n} \pol_n \c k_1
\int {d^{4 - \eps}p \over (2 \pi)^{4-\eps}}
{1\over \prod_{i=1}^n  ( p+q_i)^2 }
\anoneqn
$$
where the integral is just a scalar integral.  Replacing any $V_F^+$ with
a $V_F^-$ interchanges $\pol_i \leftrightarrow k_i$ and gives a minus sign.
In this way the pure $V_F^\pm$ terms reproduce all the pure cycles
given by applying the cycle rules to the string kinematic factor.

But what about mixed terms?  Since the $V_B$ vertex contains an
$\eta_{\nu\rho}$ which contracts the two internal indices, there is no
complicated mixing between the $V_B$ and $V_F$ terms.  In this way all
the remaining mixed terms can be reproduced.

At the loop level, the background Feynman gauge three-vertex in
fig.~\use\BFvertexFig\ exhibits a considerable simplicity when
compared to the Feynman gauge three-vertex in fig.~\use\FeynmanColorFig.
Firstly, there are only three terms in the background field vertex
in fig.~\use\BFvertexFig\ as compared to five terms in the Feynman
gauge vertex in fig.~\use\FeynmanColorFig\ (where we have used the on-shell
condition to eliminate one term in each case).  This reduces the
number of terms encountered in a Feynman diagram loop computation,
eliminating many of the cancellations between terms in much the same
way as the Gervais-Neveu gauge does for tree level Feynman diagrams.
At loop level the background field vertex has the additional advantage
over the Gervais-Neveu gauge since it contains loop momentum in only
one of the three terms while the Gervais-Neveu vertex contains loop
momentum in two terms. The simple way that the loop momentum appears
in the vertex allows for the simple structure of the string-based
loop rules and minimizes cancellations between terms.
The background field gauge partially explains the efficiency of
the string organization; one must also organize the algebra in the
particular way discussed above in order to mimic the
simple structure of the string organization of the amplitude.
There is also no need to perform the usual steps of carrying
out the Lorentz contractions, Feynman parameterizations and loop
momentum integrals, as the results of these operations are contained
in the string-based loop rules.

The way that four-point contact terms are matched in string theory is
more complicated because some of these terms
are tied up with the string trees.  In general, the
field theory description of the string tree rules is much more obscure
than the loop rules.  This is especially true when trees become
complicated as contributions can be scattered between diagrams
depending on the precise integration by parts used.  However, there is
no problem in principle to proceed with the match although it becomes
increasingly tedious as pointed out in ref.~[\use\Mapping].  The
$\phi^3$ nature of the integrated-by-parts form of the string-based
rules means that some of the four-point contact terms are to
be found in the collapse of a tree pole by a numerator factor in the
kinematic coefficient as depicted in \fig\CollapseFigure .  One
amusing result of integrating by parts in field theory is that
surface terms cancel against remaining four-point contact interaction
diagrams that are not of the form of the collapsed tree poles
[\use\Mapping,\use\Lam] leaving behind the $\phi^3$-like structure of
the string-based rules.

\vskip .5 cm
\centerline{\epsfxsize 3.2 truein \epsfbox{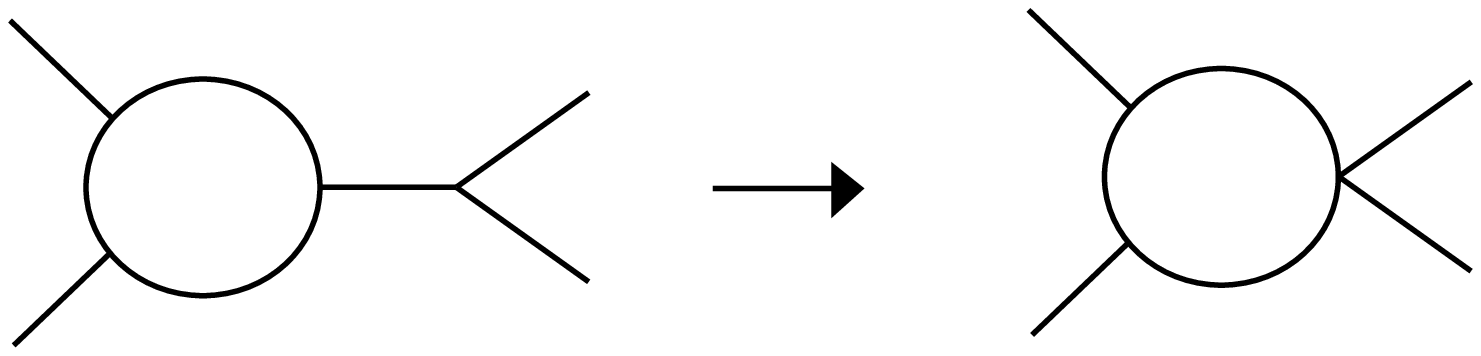} }

\nobreak
\vskip .2 cm
{\baselineskip 10 pt \narrower\smallskip\noindent\ninerm
{\ninebf Fig.~\use\CollapseFigure:}
Contact terms can be generated when a propagator
is cancelled by a momentum invariant in the numerator.
\smallskip}
\vskip .5 cm

It is amusing that other gauge choices yield the same one-loop Feynman
rules; in particular, a background field gauge version of the
Gervais-Neveu gauge yields {\it precisely} the same one-loop effective
action as the background field Feynman gauge.  The gauge fixing term,
$\Tr[(D^A \cdot Q +i Q^2/\sqrt{2})^2]/2$, generates extra terms only in
the vertices with at least three quantum $Q$-fields. The only
vertices contributing to the one-loop effective action have two
$Q$-fields so these extra terms are irrelevant at one loop as are the
extra terms in the Faddeev-Popov determinant.
The additional terms in the Gervais-Neveu background
field gauge, however, might be of interest in understanding the
multi-loop string organization of the amplitude, but this has not been
investigated.

Having identified the mapping between the loops of the string-based
rules and the loops of the background field gauge, one can verify that
the identification of string theory dimensional regularization schemes
in terms of field theory schemes included in the rules of
Section~\use\RulesSection\ is correct.  Although dimensional regularization
schemes can be constructed in string theories [\use\GSB,\use\Long], in
practical calculations it is essential to precisely identify the
corresponding field theory regularization scheme.  With the
term-by-term mapping of string loops into background field gauge loops
a direct comparison of regularizations schemes in string theory and
field theory can be made.  In field theory one can choose variations
of dimensional regularization depending on how the external and
internal states are handled.  The match between external state
prescriptions is rather simple; one simply uses the same prescription
as one would use in field theory.  For internal states, in field
theory one must choose the value of $\eta_\mu^\mu$ resulting from a
trace around a loop.  As discussed above, such traces arise in
background field Feynman gauge when only $V_B$ vertices are used on
all legs. Because of the similarity of the ghost $V_G$ and gluon $V_B$
vertices this quantity always appears in the combination $\eta^\mu_\mu
-2 \equiv 2 {\cal R}$. In a conventional scheme or 't Hooft-Veltman
scheme where $\eta_\mu^\mu = 4 - \eps$ we have ${\cal R} = 1 -
\eps/2$, while in a four-dimensional helicity scheme [\use\Long]
$\eta_\mu^\mu = 4$ so that ${\cal R} = 1$, in agreement with the
string-based rules of Section~\use\RulesSection .

A consequence of the mapping is that the string reorganization can now
be applied to the fully off-shell one-loop effective action.  This
could be of practical interest for higher point amplitudes, since it
allows for use of tree-level recursive techniques
[\use\Recursive,\use\LightconeRecurrence] for the trees sewn onto the
loop.

\vskip .3 cm
\noindent
{\it \use\FieldTheorySection.5  String Form of Effective Action}
\vskip .1 cm

In the string-based rules there is a single master formula which can
be used to describe any particle circulating in the loop.  Can we
mimic this in field theory?  Normally in field theory, the
diagrammatic structure of a scalar, fermion or gluon circulating in a
loop is rather different.  Here we explain how one can reorganize
field theory so that these three cases look pretty much the same.  A
practical consequence is that the space-time supersymmetry relations
between the various amplitudes become much more apparent, just as they
do in string theory.

As we already discussed, the background field method plays a central role
in the one-loop understanding of the string-based rules.
At one-loop the background field Feynman gauge can be summarized by
the one-loop determinant
$$
\Gamma_{\rm gluon}[A] = \ln \det{}^{-1/2} [D^2 \eta_{\mu\nu}
-i g (\Sigma_{\mu\nu})_{\rho\sigma}  F^{\rho\sigma}] \ln \det [D^2]
\eqn\OneLoopAction
$$
where
$$
(\Sigma_{\mu\nu})_{\rho\sigma} = i (\eta_{\mu\rho} \eta_{\nu\sigma}
- \eta_{\mu\sigma} \eta_{\nu\rho} )
\anoneqn
$$
is the generator of Lorentz transformations for a vector field and
$F_{\mu\nu}$ is the adjoint representation field strength matrix
$(F_{\mu\nu})^{bc} = f^{abc} F^a_{\mu\nu}$.   The
first determinant in eq.~(\OneLoopAction)
arises from the gaussian integral over the gauge fields
while the second determinant is the one-loop Faddeev-Popov determinant.
The $D^2$ terms in the determinants generate terms corresponding to the
terms generated by the substitution rule (\use\BasicSubst),
while the $F^{\rho\sigma}$
term generates terms corresponding to the cycle rules.

The scalar contribution, which matches the string form, is the usual
effective action given by
$$
\Gamma_{\rm scalar} [A] = \ln \det{}^{-1/2} D^2 = -{1\over 2} \Tr \ln D^2
\eqn\OneLoopScalarAction
$$
where $D_\mu = \partial_\mu + i g T^a A_\mu^a$.  By expanding out
the trace, the one-loop diagrams representing a scalar in the
loop are reproduced.

What about internal fermions?  Once again the analysis for the
non-cycle contributions of the
$\Gbd$ terms are identical to the case of gluons in the loop.
This indicates that
the fermion determinant should also contain a $D^2$.  Thus, the
expected form of the fermion determinant is the second order form
$$
{\Gamma}_{\rm fermion}[A] = \ln \det{}^{1/2} (D^2  -
{\textstyle {1 \over 2}} \sigma^{\mu\nu} F_{\mu\nu}  + m^2)
\eqn\OneLoopFermionAction
$$
where $\sigma_{\mu\nu} = {i\over 2}[\gamma_\mu, \gamma_\nu]$
is the generator of
spinor Lorentz transformations.  This amounts to a simple rewriting
of the usual fermion determinant
$\det [\; \slash \hskip - .26 cm D + i m  ]$ as $\det{}^{1/2}
[\; \slash \hskip - .26cm D{}^2  + m^2]$.
It is then not difficult to check that additional cycle contribution
in the fermion substitution rule
(\use\STFermionLoopRule) applied to the kinematic
coefficient (\use\KinematicContributions) exactly reproduces those terms
generated by the $\sigma_{\mu\nu} F^{\mu\nu}$ term in the fermion
determinant (\use\OneLoopFermionAction)
after carrying out the trace over the $\gamma$-matrices.

The string organization of gluon, scalar and fermion contributions
to the exponentiated one-loop effective action can then be summarized
by the generic formula
$$
\Gamma_{\rm state}[A] \sim \ln \det{}^{\mp 1/2}
[D^2 - \Sigma_{\rho\sigma} F^{\rho\sigma}]
\eqn\GenericEA
$$
where the operator $\Sigma_{\rho\sigma}$
acts in the representation of the
Lorentz algebra of the state.
The universality of the $D^2$ term in the determinant is a direct
consequence of the fact that the string loop
momentum integral for the $n$-gluon amplitude does not depend
on the choice of states circulating in the loop.
The structure of the remaining
term follows from Lorentz and manifest gauge invariance of the effective action
and the requirement that the three-vertex contains only a single
power of momentum.

\vskip .3 cm
\noindent
{\it \use\FieldTheorySection.6 Applications to General Gauge Theories}
\vskip .1 cm

The above field theory ideas can be applied to any gauge theory
calculation which involve non-abelian vertices [\use\Mapping].  In the
future extensions of the rules to include external fermions, weak
interactions and multi-loops can be expected, but in the meantime,
the above ideas can be directly used by anyone wishing to do loop
level Feynman diagram computations in non-abelian gauge theory.

The following strategy incorporates the ideas which were extracted
from the mapping between field theory and string theory
greatly improving the calculational efficiency over traditional
Feynman diagram computations.
First background field Feynman gauge [\use\Background,\use\AGS]
should be used in calculations where
a non-abelian vertex appears in the loop.  In this way a gauge invariant
effective action is produced.  For sewing trees onto the one-particle
irreducible loop diagrams
the Gervais-Neveu gauge [\use\GN] is
a particularly efficient gauge
because of the simplicity of the three- and four-point vertices.
One should use color ordered [\use\TreeColor] vertices
in order to minimize the number of diagrams which must be explicitly
computed. For
internal fermions it is best to use the second order formalism because
then the gauge boson and fermion contributions are quite similar
so that a good fraction of the work does not have to be duplicated.

Spinor helicity methods [\use\SpinorHelicity,\use\XZC] are also
important to help minimize the amount of required algebra.  Since spinor
helicity methods do not handle off-shell loop momentum efficiently it
should be integrated out early in the calculation to obtain a
representation in terms of Feynman parameters.  In order to minimize
the number of terms which appear, spinor helicity should be applied on
a term-by-term basis in the numerator as one integrates out loop
momentum.  An alternative approach which implicitly and systematically
integrates out the loop momentum is the electric circuit analogy
discussed by Lam [\use\BjD,\use\Lam].  In order to maintain the gains in
efficiency obtained with the spinor helicity method one should use
either the 't Hooft-Veltman or four-dimensional helicity scheme, as
the conventional scheme would undo much of the gain implicit in the
spinor helicity method due to the extra $\epsh$-helicities.
The resulting Feynman parameter integrals can then be
conveniently integrated using the method of ref.~[\Integrals].

Although, in this way one can expect to greatly improve the efficiency
over a traditional Feynman diagram computation, in general this type
of approach cannot be expected to be as efficient as a more direct
string approach.  Gravity is a concrete example where further input
from string theory provides further large improvements in
computational efficiency [\use\Gravity].  It is also difficult to
understand within a conventional field theory context the relatively
compact multi-loop structure implied by string theory.

\vskip .3 cm
\noindent
{\it \use\FieldTheorySection.7 First Quantized Formalism}
\vskip .1 cm

Starting from the one-loop determinants (\use\OneLoopAction),
(\use\OneLoopScalarAction) and (\use\OneLoopFermionAction), a one-loop
first quantized formalism can be obtained
[\use\FirstQuantized,\use\Matt] which mimics the structure of string
theory.  This provides a complementary description of the loop parts
of the rules within a field theory context.

For the case of the scalar in the loop this is given by
$$
\Gamma_{\rm scalar} [A] = \ln \det{}^{-1/2} D^2 =
{\cal N} \int_0^\infty {dT \over T} \int
D X \exp \Bigl[ - \int_0^T d \tau
\Bigl( {1\over 2} \dot X^2(\tau) - i g A(\tau) \cdot \dot X(\tau) \Bigr)\Bigl]
\anoneqn
$$
where $A$ is the gauge field and ${\cal N}$ is an appropriate normalization.
This path integral looks very much like the Polyakov string path integral,
except that here the path integral is over world-lines instead of
world-sheets.

By functionally differentiating with respect to $A$ $n$-times, setting
$A\rightarrow 0$ and then Fourier
transforming the vertex operator form can be recovered
$$
\Gamma_n =
{\cal N} \int_0^\infty {dT \over T} \int D X \exp \Bigl[ - \int_0^T d \tau
{1\over 2} \dot X^2(\tau) \Bigr] V_1 V_2 \cdots V_n
\anoneqn
$$
where the vertex operator is
$$
V_j = \int d \tau_j \; \pol_j \cdot \dot X(\tau_j) e^{i k_j \cdot X(\tau_j)}
\anoneqn
$$
and $\pol_j$ is the polarization vector.
One can then construct field theory Green functions on the circle
which reproduce the field theory loop substitution rules.

The structure of space-time fermion and gluon loop contributions
has also been worked out by Strassler in a first quantized formalism
[\use\Matt] based on the superstring construction.

The main advantage of this first quantized formalism is its simplicity
in deriving the loop part of the rules. However, the non-abelian
contact terms are not conveniently described by the master formula as
they are in the full string-based formalism. Furthermore, tree parts
of the rules have as yet not been obtained with first quantized
methods.  This makes the first quantized formalism useful for studying
effective actions but not amplitudes.  It may also provide a possible
alternative path for extensions to multi-loops.

\section{Summary and Conclusions}
\tagsection\ConclusionSection

In these lectures a new method, based on superstring theory, for
evaluating one-loop $n$-gluon amplitudes in perturbative QCD was
discussed [\use\Long,\use\Pascos].  The method was originally derived
by taking the field theory limit of an appropriately constructed
[\use\Beta] four-dimensional string theory [\use\KLT].  It was first
applied to reproduce, in a relatively simple way, the two-jet cross-section
of Ellis and Sexton [\use\Ellis]. As a by-product a first
calculation of all the four-gluon one-loop helicity amplitudes was
also performed.  More recently, using the string-based method,
together with a simple integral table [\use\Integrals] and
improvements in the spinor helicity method, a first calculation of all
five-gluon helicity amplitudes has been performed [\use\AmplLet].  These
amplitudes are required for the analysis of three-jet events at
Fermilab.

The string-based method meshes naturally with spinor
helicity methods since loop momentum does not appear in the initial
expression, but to take full advantage of the gains one should use a
version of dimensional regularization which leaves observed
polarization vectors in four-dimensions, such as the 't Hooft-Veltman
or four-dimensional helicity schemes [\use\Long].
Another important ingredient is the color decomposition of the
amplitude into smaller gauge-invariant partial amplitudes.
The string also provides a systematic and compact expression for the
$n$-point partial amplitude, eliminating many of the large
cancellations inherent in traditional Feynman diagram computations.

In the string-based method the diagrams are obtained by applying
certain substitution rules to the string `master formula'.  In this
way one obtains a set of Feynman parameter polynomials which are far
more compact than one would obtain by traditional Feynman diagram
methods.  The master formula is the usual kinematic expression one
obtains in string theory for the $n$-gluon amplitude.  In the
string-based method the master formula contains all information about
all field theory diagrams and particle contents.  Because the
contribution of any type of particle is contained in the master
formula, relationships between fermion and boson contributions become
apparent within the integrands of each diagram.  This can be used to
obtain even further simplifications; once the fermion loop
contribution to the $n$-gluon amplitude has been computed, calculating
the gluon loop contribution is relatively simple
[\use\AmplLet,\use\Future].

The collection of conventional field theory ideas which describe many
of simplifications of the string-based method are [\use\Mapping] color
ordering [\use\TreeColor,\use\ManganoReview,\use\Color], use of
background field Feynman gauge [\use\Background] for the one-particle
irreducible parts of diagrams, systematic organizations of the vertex
algebra and a second order formalism for fermions.  One can also use
the Gervais-Neveu gauge [\use\GN] for the tree parts of calculations since this
gauge has particularly simple vertices. (Within the background field
method the different choices of gauge are made for the loop and tree
parts of a diagram [\use\AGS].)  The spinor helicity
method can also be used [\use\SpinorHelicity,\use\XZC] to provide
further simplifications.
These ideas can be applied to more general
calculations within a conventional field theory framework
[\use\Mapping], such as ones which include external fermions, massive
gauge bosons, or multi-loops.

Although one can obtain improvements in computational efficiency in
field theory in this way, string theory goes beyond this naive
application of known field theory ideas.  String theory provides a
guiding principle for finding compact organizations; as yet, no
corresponding principle has been found within conventional field
theory.  In particular, obtaining a compact string-like organization
for gravity is fairly straightforward by directly using string theory,
but rather obscure in field theory due to the non-trivial field
redefinitions needed to mimic the simple string reorganization
[\use\Gravity].  String theory also implies that the compact structure
of the one-loop `master formula' should continue to hold to
multi-loops; it is not clear how one would obtain this in conventional
field theory (without looking at string theory to some extent).

There are a number of other extensions of the string motivated
techniques.  One important area which was not discussed in these
lectures is external fermions.  In string theory, external fermions
are technically more involved than the case of external bosons.
Progress has however been made for this case and an explicit test
calculation of $q \bar q \rightarrow g g$ has been performed based on
string theory [\use\Fermion].  The methods can also be applied to
certain weak interaction processes [\use\Unpublished].  Some progress
has also been made on the extension of these methods to multiloops
[\use\Roland].

In summary, there is every reason to believe that string theory will
continue to be as helpful for gauge theory perturbative computations
as it has been in the past.

\vskip .5 cm

I would like to thank Lance Dixon, David C. Dunbar and David A.
Kosower for collaborating on many of the ideas presented here.  I
would also like to thank them for contributing to the writing of these
lecture notes.  I also thank Al Mueller for discussions on QCD, Kaj
Roland for discussions on multi-loop extensions and Tokuzo Shimada for
discussions on gravity.  Finally I thank the Institute for Advanced
Study at Princeton for its hospitality where these lecture notes
were completed.  This work was supported by the Texas National
Research Laboratory Commission grant FCFY9202.

\vfill
\vfill\eject\immediate\closeout\rfile%\parindent=20pt
\centerline{{\bf References}}\bigskip\frenchspacing%
\input refs.tmp\vfill\eject\nonfrenchspacing

\break
\bye